\documentclass[aps,prd,showpacs,nofootinbib,showkeys,superscriptaddress]{revtex4-2}

\usepackage{bm}
\usepackage{booktabs}

\usepackage{epsfig,graphicx,amsfonts,amsmath,amssymb,bbm}
\usepackage{bbold}
\RequirePackage{mathrsfs}
\usepackage{color}
\usepackage{epsf}
\usepackage{epstopdf}

\def\fig#1{{Fig.~(\ref{#1})}}
\def\eq#1{{Eq.~(\ref{#1})}}
\newcommand{\Le}{\left(}
\newcommand{\Ra}{\right)}

\newcommand{\beq}{\begin{equation}}
	\newcommand{\eeq}{\end{equation}}
\newcommand{\beqar}{\begin{eqnarray}}
	\newcommand{\eeqar}{\end{eqnarray}}
\newcommand{\nn}{\nonumber}
\newcommand{\D}{\partial}
\newcommand{\g}{{\rm g}}
\newcommand{\el}{{\cal L}}

\newcommand{\al}{\alpha}
\newcommand{\ep}{\varepsilon}

\newcommand{\de}{\delta}
\newcommand{\De}{\Delta}

\newcommand{\ph}{\varphi}
\newcommand{\si}{\sigma}
\newcommand{\Si}{\Sigma}

\newcommand{\Ga}{\Gamma}
\newcommand{\om}{\omega}
\newcommand{\Om}{\Omega}

\DeclareMathOperator{\diag}{diag}

\newcommand{\Madm}{M_{\rm ADM}}
\newcommand{\Rreg}{R_{\rm reg}}
\newcommand{\Rcore}{R_{\rm core}}
\newcommand{\RtL}{R_{\rm 3L}}
\newcommand{\rhocr}{\rho_{\rm cr}}
\newcommand{\rhocrust}{\rho_{\rm crust}}

\newcommand{\rhoP}{\rho_{P}}

\newcommand{\RS}{R_{S}}

\newcommand{\Tloc}{T_{\rm loc}}
\newcommand{\Tinf}{T_{\infty}}

\newcommand{\abs}[1]{\left| #1 \right|} 

\newcommand{\toremove}[1]{\textcolor{blue}{#1}}

\DeclareMathOperator{\sech}{sech}
\DeclareMathOperator{\arctanh}{arctanh}
\DeclareMathOperator{\arcsinh}{arcsinh}

\begin{document}
	\title{On the structure of black hole interior in a model of scalar quasi-particles within the GR framework}
	\author{S. Bondarenko}
	\affiliation{Ariel University, Ariel 4070000, Israel}

	
\begin{abstract}

 We propose an effective, singularity-free model of the black hole interior described entirely by a scalar field 
with a non-linear self-interaction potential. 
The interior consists of three layers -- a core, a transition layer,
and a crust -- each fixed by the local quasi-particle density and the corresponding extremum
of the potential of the field. The crust is a layer of massive, positive-energy thermal excitations above
the zero-potential well, beneath a genuine Schwarzschild horizon at $r = 2GM_{\rm ADM}$. The
core is an AdS-type region of negative energy density, it simulates a condensate of quasi-particles which carry zero classical
kinetic energy and form a negative potential well governed by a negative inverse temperature parameter. The two regions
are joined through Israel matching across the transition layer, which sits at an approximately
null-gravity hypersurface of maximal regular matter density and where both the sign of the
energy density and the type of thermal excitations change. Solving the static Einstein equations, we obtain
the metric and mass functions of each layer, the edge equation of state of the crust, the linear
stability condition of the null-gravity surface for ordinary and effectively negative mass matter, and the
two-temperature thermodynamics linking the kinetic excitations to the negative temperature
ground state. The framework unifies core and crust within a single field
description and highlights the role of the negative energy AdS core and the associated
negative temperature notion in describing black hole interior.
In this picture the formation and Hawking evaporation of a black hole appear as a
quasi-static cycle of an almost adiabatic thermodynamic engine, with the negative energy
core as the working substance. The model offers a natural setting for 
addressing the entropy and information problems of evaporating black holes and for connecting horizon scale thermodynamics to a regular, singularity free interior
region.

\end{abstract}
\maketitle

\section{Introduction}\label{SecA}

  The thermodynamics of a black hole, as perceived by an external observer, is well defined theoretically, see
\cite{Beken1,Beken2,Hawk1,Hawk2,Hawk3,Strom1,Frol1,Frol2,Page1,Page2,Page3} for a few examples from a very long list. It is not, however, explained from first principles based on the structure of the black hole (BH) interior. Exploration of the interconnection between the interior of a black hole and its external characteristics is, without doubt, a task of great importance. In \cite{BondBH} we proposed a model of the BH interior\footnote{There is a long list of papers dedicated to different approaches to the BH interior construction, see a very short list of   \cite{Int,Int1,Int2,Int3,Int4} papers concern the subject.} 
intended to address this problem, basing the interior structure on a few simple first-principle statements about BH thermodynamics and its desired properties.

The picture of the interior introduced in \cite{BondBH}, which we continue to explore in the present manuscript, rests on an interpretation of the idea of an infinite-density limit of matter. Borrowing from the physics of neutron-star interiors, see \cite{Nov1,Nov2,Neut1,Neut2,Neut3,Neut4} and references therein, we regard a black hole as a "factory" of matter-energy packing -- an object that concentrates the maximum amount of matter (energy) in a minimal volume. In this picture the BH is built from a core, a crust, and a transition layer between them, with the highest attainable matter density reached in the core. This highest density does not mean an infinite density, but rather the creation of a special matter state inside the core. In \cite{BondBH} the core was proposed to be a condensate of quasi-particles that carry only potential energy and no kinetic energy. These quasi-particles have the maximal possible binding of their constituents; the binding energy was assumed to exceed the rest-mass energy of the quasi-particles thereby producing a negative potential well in the core. In the model this construction replaces the notion of an infinitely large density.\footnote{Strictly speaking, we speculate that at very large density and temperature one effectively obtains quasi-particles that provide a negative energy density and act gravitationally as a repulsive matter relative to ordinary matter.} The much thinner crust in \cite{BondBH} was likewise described in terms of regular quasi-particles of a different kind, which possess kinetic energy and are built from ordinary matter constituents. The transition layer is the region where the transmutation of the two types of quasi-particles and the change of sign of the energy density take place. In this construction the central singularity disappears; the core region appears in \cite{BondBH} as a flat-band type of space-time, see \cite{FB,FB1,FB2,FB3}, free of a singularity at $r\,=\,0$.

The state of the core matter proposed in \cite{BondBH} is not entirely new. A similar idea can be found in \cite{Zeld1,Zeld2}, see also \cite{Wil1,Wil2,Wil3,Wil4}, where a state of matter built from quarks interacting through attractive channels and paired as bosons in antisymmetric color combinations was studied, as well as the discussion in \cite{Nozi}. As in the previous paper, we therefore do not address the microscopic physics of the formation of these quasi-particles, and we treat the appearance of the core's quasi-particles as the result of an attractive interaction between the fundamental constituents of matter. The number and kind of the quasi-particle constituents are not essential in our framework, see again \cite{Nozi} for a similar discussion in the case of a boson liquid. The quasi-particles are produced in a regime of initially very high density and temperature of the crust matter, which makes the binding of fermions a preferable process from the point of view of packing matter inside the black hole. Once the core density is very high, even some cooling of the state cannot reverse the situation, because there is not enough phase space for the dissociation of the bound states. Further growth of the density then leads to the condensation of the Bose quasi-particles in physical space. These quasi-particles fill the geometrical (physical) volume of the core and create the new matter state we discuss; as noted, their vanishing momentum and the negative energy density of the core are a manifestation of the maximally attained matter density. The physical picture is thus an analogue of Fermi-particle condensation, the difference being that here we have a genuine condensation of Bose particles in coordinate space.

The equation of state (EOS) of the core matter proposed in \cite{BondBH} may be called a flat-band EOS, see \cite{FB,FB1,FB2,FB3}. It is characterized by negative energy density and negative pressure, and is, overall, an unusual equation of state because in general relativity energy and pressure may carry different signs, but not both negative simultaneously. This EOS can be shown to follow from two factors. The first is the particular regularization of the integral over momentum phase space proposed in \cite{BondBH}; the second is the non-relativistic origin of the model explored there. Performing the calculation within the general-relativistic framework therefore requires us to revise this set-up of the EOS. Moreover, since the \cite{BondBH} framework involves two distinct subsystems of quasi-particles, a unified description of the whole interior in terms of a single fundamental field considerably simplifies the task and the calculations. Accordingly, in the present manuscript we introduce a scalar field with a non-linear interaction potential as the main degree of freedom describing the entire interior. The potential we use has a first minimum corresponding to the crust and a negative-valued region with a second minimum corresponding to the core, with a transition layer connecting the two, see Fig.~\ref{Poten} below.

We thus employ two complementary points of view in describing the quasi-particles of the interior. From the microscopic standpoint, the regular matter in the crust undergoes a "reconstruction", i.e. a rebinding, which produces new matter constituents -- the quasi-particles -- that have zero kinetic energy and a negative potential energy. In this language the quasi-particles are new matter constituents created from regular matter at asymptotically large crust density, and, once formed, they condense in the core, a picture similar in some respects to a Bose-Einstein condensate. This description of the core formation, being non-relativistic and not Lorentz invariant, is not really suitable for the description of GR processes. For that reason in this paper we instead describe all the interior layers and their evolution in terms of a scalar field with a non-linear interaction potential. In the crust the scalar field mimics the behavior of regular matter with attractive interactions between the particles; the attracting scalar particles then collapse toward the center, reaching larger densities and creating a negative-density core. In this sense the field description of these processes is not identical to the one introduced in \cite{BondBH}; here we bypass the notion of quasi-particles by using the complicated form of the interaction potential, reproducing the main stages and states of the interior in terms of the scalar field alone. We nonetheless assume that, at the microscopic level, the description of quasi-particle condensation into the core given in \cite{BondBH} remains correct, and we match these two formulations of the same problem.\footnote{We will use henceforth the notion of quasi-particles in the sense of \cite{BondBH}, assuming that the scalar field provides an effective description of the behavior of those quasi-particles.}

\begin{figure}[htbp]
\centering
\includegraphics[width=0.65\columnwidth]{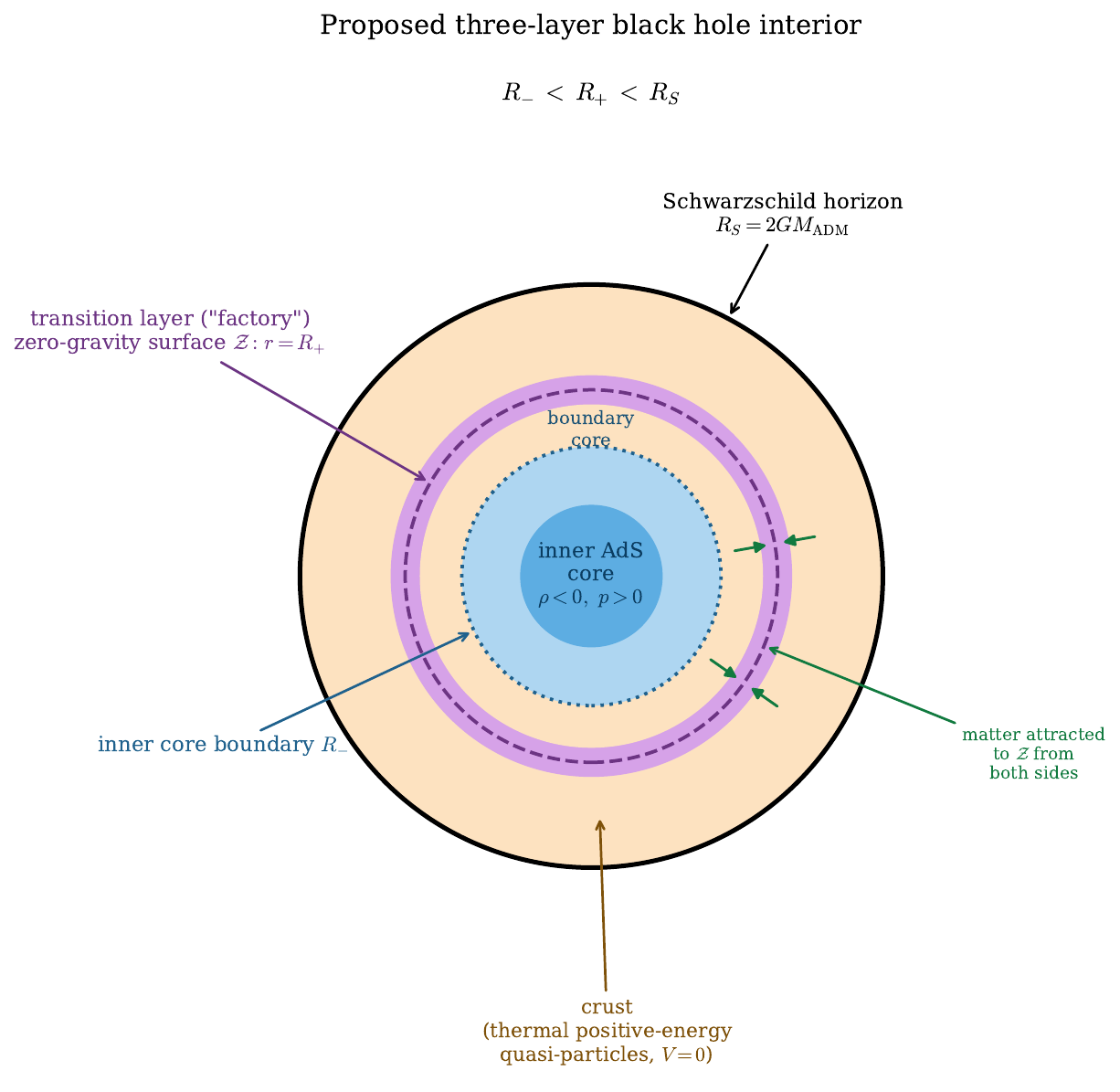}
\caption{Schematic structure of the proposed black hole interior. The radial 
ordering is $R_{-}\,<\,R_{+}\,<\,R_{S}$: the inner AdS-like core with negative energy density 
($\rho\,<\,0,\,p\,>\,0$) and its boundary-core shell are enclosed within $R_{-}$; the 
transition layer (the "factory"), which approximately coincides with the zero-gravity 
hypersurface $\mathcal{Z}:\,r\,=\,R_{+}$, attracts the regular matter from both sides and 
repels the negative-energy bubbles; the thermal positive-energy crust ($V\,=\,0$) fills the 
region up to the Schwarzschild horizon at $R_{S}\,=\,2\,G\,M_{\rm ADM}$.}
\label{Interior}
\end{figure}

In summary, in this paper we discuss a three-layer model of a black hole described in terms of a scalar field with a non-linear interaction potential. Besides the time scales, the direction of the system's evolution is set by the change in the density of the quasi-particles in the interior, which drives the changes between the system's states. The formation of the BH is assumed to proceed through the following sequence of macroscopic stages. First, collapsing matter accumulates in the interior near the future horizon, forming an initial dense layer that produces quasi-particles without any specific core; these quasi-particles provide a positive energy density, similar to the regular matter state. As the inward flux of matter/energy continues, the quasi-particle density grows and the layer rearranges internally, generating a small core, which in the model is assumed to be an AdS-like vacuum interior with negative energy density. This process goes through a gravitational restructuring accompanied by the attractive self-interaction of the quasi-particles, and takes place in a special layer that we call the "factory", where the transition to a negative energy density of the quasi-particle matter occurs. At this stage the internal interaction between the quasi-particles is already turned on, and the core grows as further matter falls into the BH, while the near-horizon layer remains in an approximate steady state. Evolving, the whole system reaches the following configuration: a large core with negative energy density; a thin transition layer between the crust and the core, which contains the "factory"; and a near-horizon crust, built mainly from quasi-particles (i.e. regular matter) with positive energy density and regarded as an internal boundary of the horizon. The final element of the construction is the external horizon at $r\,=\,2\,M$. The "factory", where the rebinding of matter and the transition to a negative energy density occur, is approximately a null-gravity hypersurface. By construction this null-gravity hypersurface carries the largest density of regular matter attained in the interior; it attracts regular matter from both sides and repels the negative-energy quasi-particles. The BH grows while matter continues to fall inward, and begins to evaporate once the outward fluxes exceed the inward ones. A schematic picture of the proposed three-layer interior is shown in \fig{Interior}.
The scalar-field model we discuss has a limitation, in the present manuscript we solve a static problem in which the matter/energy fluxes between the different layers are not introduced; for each particular part of the potential we consider a free field Lagrangian. In this simplified set-up we therefore obtain snapshots of the interior, without deriving the mechanisms governing the growth of the quasi-particle density; that means we consider a static solution for the interior with fixed values of the extrema of the scalar-field potential. Without external currents in the scalar-field Lagrangian, and the corresponding non-static solutions, the evolution of the system together with the growth of the density is a problem that cannot be solved in terms of a scalar field that does not interact with its environment. This issue is important, of course: if the extrema of the potential are fixed, we do not have a truly dynamical picture of the interior evolution. We propose then the following way to overcome this difficulty without performing the full calculation.\footnote{The model's calculations with the external currents included will be published in a separate paper; in Appendix~\ref{AppenA} the simplest form of the required currents and the corresponding changes of the equations are presented.}

We note that the construction contains not one but two different thermodynamics. The first is the thermodynamics of thermal excitations on top of the lowest energy level; these excitations can be defined as analogues of phonons and/or rotons above the quasi-particle condensate, or above the vacuum state of the scalar quasi-particles. They spread heat over the system's levels and mediate the interactions between the different layers of the interior. The second is the thermodynamics of the ground state of the core, which cannot be described in terms of an ordinary temperature in any formulation, because it has none. We therefore treat the ground state of the quasi-particles in the core as a condensate of negative-energy particles with zero or very small kinetic energy, whose thermodynamic properties are defined not through a regular temperature but through the negative inverse-temperature parameter introduced in \cite{BondBH}. Operationally, this parameter is the inverse-temperature variable $\beta\,\equiv\,\partial S/\partial U_{pot}$ conjugate to the potential energy of the condensate; because the condensate spectrum is bounded, $\beta$ takes negative values, in contrast to the ordinary positive inverse temperature $1/T\,\equiv\,\partial S/\partial U_{kin}$ of the kinetic excitations. We give the precise definitions and properties of both in Subsection~\ref{SecAA5}. In this set-up the evolution of the core can be determined through the change in the number of quasi-particles in the core, treated by the regular thermodynamic approach formulated in terms of the new temperature parameter, without an actual solution of the non-stationary Einstein equations. The interconnection between the regular temperature of the excitations and the new temperature parameter of the ground state then clarifies the stages of the system's evolution through universal thermodynamic principles. The two ways of describing the ground state of the BH core, in terms of scalar field well and in terms of quasi-particles, are dual to each other; together they clarify the issues under discussion, and we use both formulations to elucidate the dynamics of the core.

The evaporation mechanism assumed in the model is Hawking radiation at zero inward matter/energy infall. Regarding the black hole as a heat engine with two reservoirs -- an internal one, the core, and an external one -- the system undergoes a quasi-static evolution in the direction opposite to the formation of the black hole, through the following stages. First, the horizon shrinks as the black hole mass decreases. The core shrinks as well and feeds quasi-particles into the crust, maintaining it in an approximate steady state at roughly constant density and slowly decreasing temperature. The whole system passes through these steady-state (quasi-equilibrium) cycles, set by the Hawking radiation rate. When the core shrinks toward zero, only the regular-matter BH remains; this remnant returns to the core-free configuration, the same as at the initial stage of formation. Further emission brings the BH to some final configuration, which may be either full evaporation or a cold remnant. Because of the length of the article, we do not consider this problem here and postpone it for a separate publication.

The proposed framework thus suggests regarding a black hole as a non-stationary, almost adiabatic thermodynamic engine. In this picture the formation and evaporation of a BH are analogous, to some extent, to an almost adiabatic thermodynamic cycle. The collapsing matter/energy transforms heat and matter into a structured configuration consisting of a core, a transition layer, a crust, and a horizon, through a sequence of quasi-static cycles. The loss of heat through Hawking radiation recovers this energy by reversing the cycle, again through a sequence of reversed quasi-static cycles. The intermediate states are then almost-equilibrium configurations of the layers, traversed quasi-statically. In this analogy the core plays the role of the working substance, while the transition layer and the near-horizon layer play the role of the reversible heat-exchange interface of the engine. The external heat reservoir is identified with the Hawking radiation, with the surface gravity $\kappa$ playing the role of the inverse temperature of the reservoir. The first law of black hole thermodynamics is then the analogue of the energy-conservation statement of the thermodynamic cycle.

The requirement of reversibility should, in turn, fix the structural features of the model. The three-part interior, with its steady-state near-horizon layer, is precisely a configuration that allows a quasi-static evolution. The rate of heat transfer between the core and the crust is assumed to match the Hawking luminosity, the process being enforced by the equilibrium condition, while the scaling of the black hole interior entropy is fixed by the requirement that the total entropy be conserved. Concerning the Hawking radiation and the black hole entropy, we require in this model, first, that the radiation carry calculable correlations encoding the information, and, second, that the black hole possess an interior structure carrying, by construction, a macroscopic entropy. In general, the evaporation is quasi-static with respect to the internal equilibration time of the layers, and its evolution is determined by the structure of the proposed interior configuration. An evaporating black hole, in this view, is not a singular object that destroys information, but an ordinary thermodynamic system running a slow adiabatic cycle, with the Hawking radiation reflecting the internal thermalization and cooling of the crust.

The manuscript is organized as follows. In the next Section~\ref{SecAA} we discuss a set-up of the model. There we introduce the scalar field we use through the corresponding interior description and clarify general ideas and construction of the model of the BH interior without the solution of the corresponding Einstein equations. The Section~\ref{SecAB} is dedicated to the GR description of the inner, i.e. located at $r\sim\,0$, core region. We solve Einstein equations which describe the region with corresponding EOS and discuss the properties and characteristics of this part of the BH core. After that, in Section~\ref{SecBB}, we discuss the boundary core region of the BH which is located at the boundary of the core. Then we also calculate the solution of the Einstein equations with corresponding EOS close to the boundary and 
determine the form of the whole boundary solution by interpolating the found metric with the metric from Section~\ref{SecAB}, in both cases we talk about the static metric solutions.
In the Section~\ref{SecCC} we describe the GR construction of the BH crust. There we discuss the form of the crust's metric, the EOS of scalar field there, the Israel matching between the outer Schwarzschild vacuum solution and calculated inner crust's metric and other questions concerning the crust's construction and properties. In the Section~\ref{SecDD} we close the interior's description,
calculating a metric of the transition layer and its matching with core and crust metrics inside the interior. In this Section we also consider the zero-gravity hypersurface that appears in the model and describe its properties. The last Section~\ref{SecEE} is a section where we summarize the obtained results and propose next possible directions of the model's development. The main details of the calculations concern different technical issues we put in the
Appendixes~\ref{AppenA}--\ref{AppenE} with corresponding references through the Sections.


\section{Scalar field quasi-particles and inner structure of BH: introductory picture}\label{SecAA}

\begin{figure}[htbp]
\centering
\includegraphics[width=0.65\columnwidth]{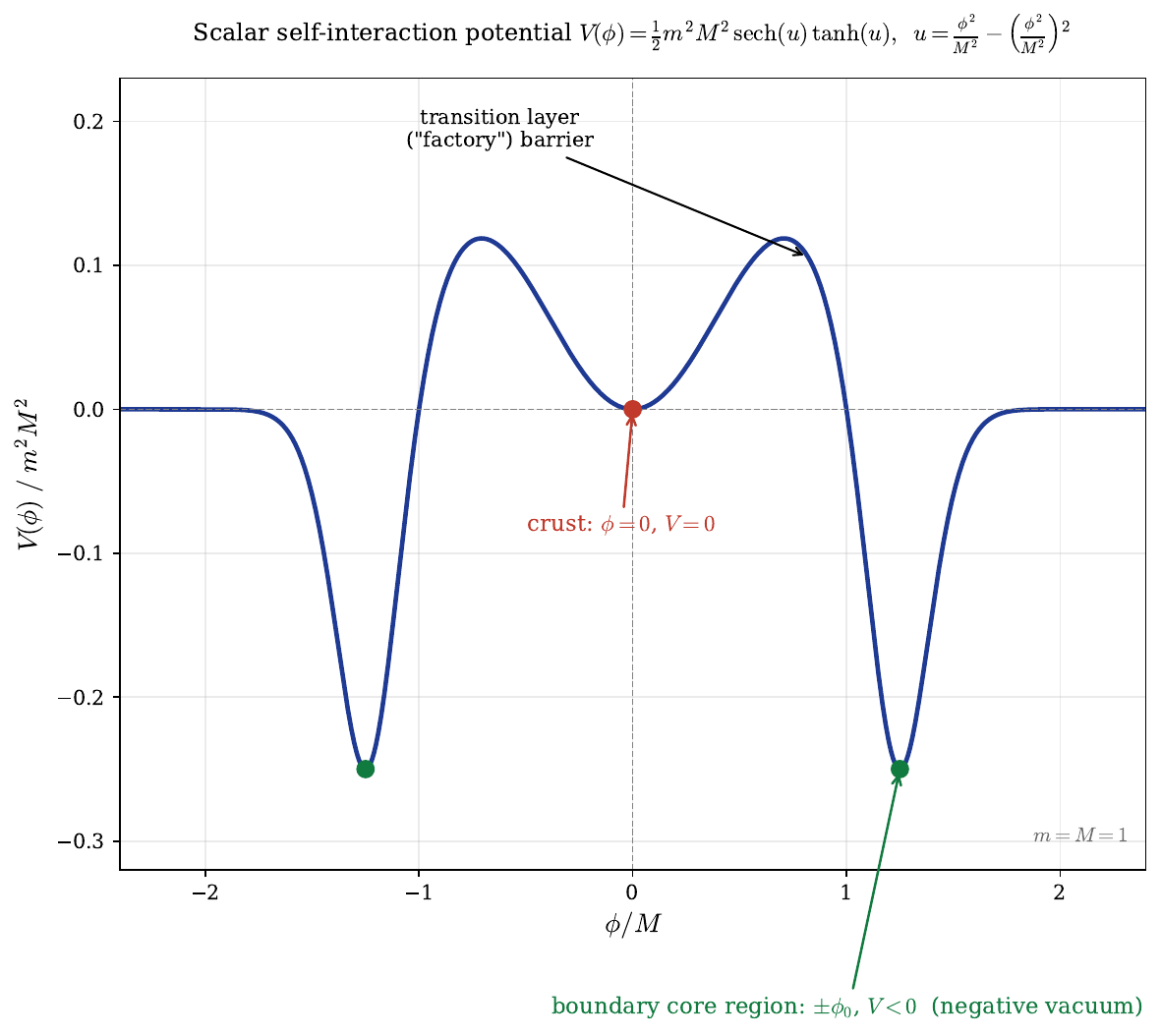}
\caption{The scalar self-interaction potential $V(\phi)$ of \eq{AA3}, plotted at $m\,=\,M\,=\,1$ 
(with $\phi$ in units of $M$ and $V$ in units of $m^{2}M^{2}$). The quasi-stable minimum at 
$\phi\,=\,0$ with $V\,=\,0$ corresponds to the crust; the two symmetric negative minima at 
$\pm\phi_{0}$ with $V\,<\,0$ correspond to the negative-energy boundary core region; the 
barrier separating the wells is associated with the transition layer ("factory").}
\label{Poten}
\end{figure}

 As a model for the interior interacting quasi-particles, 
we consider the following Lagrangian:
\beq\label{AA1}
\el_{0}\,=\,\frac{1}{2}\,\sqrt{-g}\,\g^{\mu \nu}\,\D_{\mu}\phi_{i}\,\D_{\nu}\phi^{i}\,-\,\frac{1}{2}\,\sqrt{-g}\,m^2\,M^{2}\,
\sech\Le \frac{\phi_{i}\,\phi^{i}}{M^{2}}\,-\,\Le\frac{\phi_{i}\,\phi^{i}}{M^{2}}\Ra^{2}\Ra\,
\tanh\Le \frac{\phi_{i}\,\phi^{i}}{M^{2}}\,-\,\Le\frac{\phi_{i}\,\phi^{i}}{M^{2}}\Ra^{2}\Ra\,;\,\,\,\,
i\,=\,1,\,2\,.
\eeq
Introducing 
\beq\label{AA2}
\phi_{i}\,\phi^{i}\,=\,\phi^{2}\,;\,\,\,\, 
\eeq
we discuss a scalar field with the
\beq\label{AA3}
V(\phi)\,=\,\frac{1}{2}\,m^{2}\,M^{2}\,
\sech\Le \frac{\phi_{i}\,\phi^{i}}{M^{2}}\,-\,\Le\frac{\phi_{i}\,\phi^{i}}{M^{2}}\Ra^{2}\Ra\,
\tanh\Le \frac{\phi_{i}\,\phi^{i}}{M^{2}}\,-\,\Le\frac{\phi_{i}\,\phi^{i}}{M^{2}}\Ra^{2}\Ra\,
\eeq
non-linear potential, the shape of this potential, for the representative parameter values 
$m\,=\,M\,=\,1$, is shown in \fig{Poten}.
The energy momentum tensor we use for the whole interior is a standard one defined through
\beq\label{SAA1}
T_{\mu\nu}\,=\,\frac{2}{\sqrt{-g}}\,\frac{\delta \el}{\delta g^{\mu \nu}}\,=\,
\Le \delta^{\alpha}_{\mu}\,\delta^{\beta}_{\nu}\,-\,\frac{1}{2}\,g_{\mu \nu}\,g^{\alpha\beta}\Ra\,\D_{\alpha}\phi_{i}\,\D_{\beta}\phi^{i}\,+\,
g_{\mu \nu}\,V(\phi)\,.
\eeq
For the chosen form of the potential, we have the following special regimes.

  At the first minimum, or quasi-stable state at $\phi\,=\,0$, 
we have
\beq\label{AA4}
\frac{\phi^{2}}{M^{2}}\,=\,\frac{\epsilon^{2}}{M^2}\,\sim\,0
\eeq
obtaining the following leading order answer for the potential:
\beq\label{AA5}
V(\phi)\,=\,\frac{1}{2}\,m^{2}\,\epsilon^{2}\,-\,
\frac{1}{2}\,\frac{m^{2}}{M^2}\,\epsilon^{4}\,.
\eeq
At the limit
\beq\label{AppenA5001}
M\,\rightarrow\,\infty
\eeq 
the potential is simplified to
\beq\label{AA7}
V(\phi)\,=\,\frac{1}{2}\,m^{2}\,\epsilon^{2}\,.
\eeq
We see then that there are attractive $\epsilon^{4}$ field interactions at some mass and self-interaction vertex values, which at large field strength leads to
a non-stable behavior of the system.

  The two additional minimum of the potential are located at 
\beq\label{AA8}
\pm\,\phi_{0}\,=\,\pm\,\frac{M}{\sqrt{2}}\,\Le 1\,+\,\sqrt{1\,+\,4\arctanh(1/\sqrt{2})}\Ra^{1/2}\,=\,\pm\,\frac{M}{\sqrt{2}}\,\Le 1\,+\,
\sqrt{1\,+\,4\arcsinh(1)}\Ra^{1/2}\,=\,\pm\,M\,\beta\,
\eeq
values of the field. Defining
\beq\label{AA9}
\phi\,=\,\left (\begin{array}{c} 
\zeta\\
\phi_{0}\,+\,\epsilon
\end{array}\right)\,;\,\,\,\zeta\,,\epsilon\,\sim\,0\,;\,\,\,\phi_{0}\,>\,0\,,
\eeq
we obtain
\beq\label{AA10}
V(\phi)\,=\,-\,\frac{1}{4}\,m^{2}\,M^{2}\,+\,\alpha\,\frac{\phi_{0}^{2}}{M^{2}}\,m^{2}\,\epsilon^{2}\,+\,
\alpha\,\frac{\phi_{0}}{M}\,\frac{m^{2}}{M}\,\epsilon^{3}\,+\,\frac{1}{4}\,\frac{m^{2}}{M^{2}}\,\epsilon^{4}\,+\,
\frac{1}{4}\frac{m^{2}}{M^{2}}\,\zeta^{4}\,+\,
\frac{1}{2}\,\alpha\,\frac{m^{2}}{M^{2}}\,\epsilon^{2}\,\zeta^{2}\,+\,
\alpha\,\frac{\phi_{0}}{M}\,\frac{m^{2}}{M}\,\epsilon\,\zeta^{2}\,
\eeq
where
\beq\label{AA11}
\alpha\,=\,1\,+\,4\,\arcsinh(1)\,.
\eeq
At the limit
\beq\label{AA12}
M\,\rightarrow\,\infty
\eeq 
the potential is simplified to
\beq\label{AA13}
V(\phi)\,=\,-\,\frac{1}{4}\,m^{2}\,M^{2}\,+\,\alpha\,\beta^{2}\,m^{2}\,\epsilon^{2}\,
\eeq
expression. The reduction from the \eq{AA5} value of the potential to \eq{AA13} answer can be considered as a process of formation of a black hole with negative energy density core, see also
\cite{Interior} paper. This transition from positive value of the potential to the negative one takes place at the $\phi\,=\,M$ scale that determines a physical 
meaning of the parameter.

  Another vacuum-like state is achieved at
\beq
\phi^{2}\,\rightarrow\,\infty
\eeq
or
\beq\label{AA14}
\frac{\phi^{2}}{M^{2}}\,=\,\frac{\epsilon^{2}}{M^2}\,\gg\,1\,.
\eeq
value of the field.
We have then
\beq\label{AA15}
V(\phi)\,\rightarrow\,-\,m^{2}\,M^{2}\,e^{-\phi^{4}/M^{4}}
\eeq
value of the potential.

 In general, as mentioned in the introduction, we can consider the Lagrangian with the addition of, for example, source terms of the following form:
\beq\label{AA16}
\el_{J}\,=\,g^{\mu\nu}\,\phi_{i}\,\Le D_{\nu}J_{\mu}^{i}\,+\,J_{\mu \nu}^{i} \Ra
+\,g^{\mu\nu}\,\phi_{i}\phi^{i}\,\Le D_{\nu}\mathcal{J}_{\mu}\,+\,\mathcal{J}_{\mu \nu}\Ra\,
\eeq
which can be interpreted as  source terms responsible for the heat (matter) transfer in the system. On the classical level, the additional terms
redefine the position and the value of the extrema of the potential.
A brief discussion of the redefinitions of the energy-momentum tensor and the corresponding non-stationarity of the system caused by the first term in the expression is presented in the  
Appendix~\ref{AppenA}.

\subsection{Three layer structure of the interior: simplest analysis of AdS core}\label{SecAA1}

 As described in the Introduction, the proposed model of the interior consists of three main parts, see \cite{BondBH} for additional details. Now we want to match
the proposed interior structure with the potential of the scalar field.
The first part of the interior, the crust, we roughly can relate with the first minimum of the potential at $\phi\,=0\,$. The crust
consists of quasi-particles at low density which provide a positive energy density
contribution to the energy-momentum tensor, the quasi-particles there
mimic the behavior of regular matter located at internal vicinity of the BH horizon. The crust is located at $R_{+}\,<\,R_{S}$, where the
$R_{+}$ is a radial distance of the crust lower boundary from the center, the $R_{s}$ is a Schwarzschild radius. Additionally the crust is characterized by some width 
$l_{+}\approx\,R_{S}\,-\,R_{+}$. Next layer is the transition one, in the potential it relates to the regions with positive and zero value of the energy, 
i.e. maximum and zero value of the potential approximately. It is as well characterized by some width $l_{0}$, we can consider it as a kind of domain wall. In this region the energy density changes sign with a growth of the quasi-particle density. On the language of \cite{BondBH}, there, negative-energy bound quasi-particles are created and absorbed thereafter into the core.  We can consider the layer in the model as 
a "factory" which provides the growth of the negative-energy core because of an influence of an inward heat/matter flux into the crust.
The third region of the interior is the core located inside a radial distance $R_{-}$ where $l_{0}\,=\,R_{+}\,-\,R_{-}$. The core, in turn, consists of two parts. The first one is an inner region of the core, i.e. it is located 
at some distance from the core's boundary, we can call it as a deep core region. The second part of the core is located at vicinity of the core's boundary, 
it includes the second minimum of the potential and can be named as a boundary core region. In general, the boundaries between the regions are dynamical ones and depend on the process of BH evolution. An exterior of the model is, of course, a usual Schwarzschild solution exterior.

 An immediate consequence of the construction with a negative energy density core and regular matter crust is an appearance of a zero-gravity hypersurface, which we can then identify with the  introduced transition layer. Indeed, the interplay between the repulsion and attraction leads to an appearance of the hypersurface to which the usual matter from below and above will be attracted but which will not be affected by any gravitational forces if the matter is located on the hypersurface itself. As a consequence, the maximal matter density will be achieved on the hypersurface and we can consider it as mentioned above
"factory". We discuss this issue further as well.

 The proposed structure of the interior, even without microscopic description, provides two useful scales. The first one is a potential difference of the energy captured by the scalar potential:
\beq\label{AA17}
\Delta V(\phi)\,\propto\,\rho_{-}\,-\,\rho_{+}\,=\,-|\rho_{-}|\,-\,\rho_{+}\,=\,-L\,,
\eeq
here the $\rho_{+}$ and $\rho_{-}$ are values of the energy density in the layer and core interior regions correspondingly.
Another scale is a scale of energetic cost of core's quasi-particles creation, i.e. so called lattice energy, see \cite{Latt1,Latt2}.
Namely, it is assumed that the "factory" dissociate a regular matter and create a new quasi-particles, the cost is
\beq\label{AA18}
\Delta E_{f}\,=\,E_{bind}^{-}\,-\,E_{bind}^{+}\,,
\eeq
here the $E_{bind}^{\pm}$ is a binding energy of the matter in the regions of different energy density signs correspondingly. 
In general, as well we can introduce a latent heat provided by the possible phase transitions in the system, but, for sake of simplicity, we assume that any 
processes of heat creation or absorption are encoded in the \toremove{$\Delta E_{0}$}$\Delta E_{f}$. We, as well assume, that 
\beq\label{AA19}
E_{bind}^{-}\,>\,E_{bind}^{+}\,,
\eeq
i.e. the process is endothermic, see again discussion in \cite{BondBH}, and therefore the "factory" works and core grows only when an external energy/matter flux exists.
Otherwise the whole system stays in a stable adiabatic state, or, if we turn on a Hawking radiation, the state becomes unstable and evolves slowly, by the evaporation, toward a new quasi-stable state till full evaporation.
In this picture, one of the conditions for the "factory" activation  is that the temperature of the crust must achieve 
\beq\label{AA20}
E_{act}\,\sim\,\Delta E_{f}
\eeq
value, see also discussion in Subsection~\ref{SecDD3}. This barrier between the two layers of the interior regulates, i.e. suppresses, the classical and quantum-mechanical transitions between the   
different regions of the interior. In general, of course, this is a dynamical quantity and as well depends on the particular configurations of the core and crust.

 Now consider the core assuming that it has a simple structure with homogeneous value of the negative internal vacuum energy
without any exited states in, i.e. we consider $N\,\gg\,1$ number of quasi-particles in the core with the following equation of state:
\beq\label{AA21}
\rho\,<\,0\,,\,\,\,p\,>\,0\,,\,\,\,\rho\,+\,p\,=\,0\,,
\eeq
this picture is approximate and does not account for the boundary part of the core, see further, but still it provides a good qualitative understanding of the processes in the interior. In general, then,
this is a picture of false vacuum bubbles, see \cite{FalB,FalB1,FalB2}, is reproduced at this approximation\footnote{The negative energy density EOS with different
possibilities of the energy density to pressure ratio were also considered in \cite{NEM} paper.}. So, approximating roughly the core by the bubble structure, the total energy of the core (bubble) is
\beq\label{AA22}
E_{\rm bub}(R,N)\,=\,\frac{4\pi}{3}\,R^{3}\,\rho(N,R)\,+\,4\pi\,R^{2}\,\sigma_{\rm wall}\,.
\eeq
Here $\sigma_{\rm wall}$ is the surface tension of the bubble wall, i.e. the surface-energy density of the transition layer separating the negative-energy core from the positive-energy exterior.
The effective energy of a quasi-particle inside the bubble follows by
differentiating at fixed $R$,
\beq\label{AA23}
\varepsilon_{\rm in}\,=\,\frac{\D E_{\rm bub}}{\D N}\bigg|_{R}\,=\,
\frac{4\pi R^{3}}{3}\,\frac{\D \rho}{\D N}\,<\,0\,,
\eeq
which is negative. Namely, adding a quasi-particle deepens the negative
energy density well and therefore lowers the total energy, it is a dynamical process of course.
Outside, in the ordinary--matter phase, the matter's particle has positive
energy
\beq\label{AA24}
\varepsilon_{\rm out}\,>\,0\,,
\eeq
comprising its rest mass and any kinetic energy in the exterior phase.
The change in energy when a single quasi-particle leaves is therefore
\beq\label{AA25}
\Delta E_{\rm leave}\,\simeq\,\varepsilon_{\rm out}\,-\,\varepsilon_{\rm in}\,+\,\Delta E_{\rm wall}\,>\,0\,,
\eeq
where $\Delta E_{\rm wall}$ is the cost of crossing the domain wall. All three contributions are positive, therefore
\beq\label{AA26}
E_{\rm bind}\,\simeq\,|\varepsilon_{\rm in}|\,+\,\varepsilon_{\rm out}\,+\,\Delta E_{\rm wall}\,>\,0\,.\;
\eeq
This is the binding energy of a quasi-particle in the bubble. Decay of the bubble  is classically forbidden then unless an external source provides $E_{\rm bind}$ to
each escaping particle. So, if we want a pump of a heat from core to the crust then
that is also a point where the second minimum of the potential must appear. It has a thermal positive excitations above the minimum and 
can provides a heat transfer from core to crust, which is impossible otherwise without additional special assumptions.

 Next we note that bubble wall is under tension, and the bulk has nonzero pressure
$p_{\rm in}\,=\,|\rho_{-}|\,>\,0$, where we used the core EOS \eq{AA21} so that $p_{\rm in}\,=\,|\rho_{-}|$. Mechanical balance is the Young--Laplace, \cite{YounLaplaceEq},
condition
\beq\label{AA27}
p_{\rm in}\,-\,p_{\rm out}\,=\,\frac{2\,\sigma_{\rm wall}}{R}\,\simeq\,|\rho_{-}|\,-\,|\rho_{+}|\,+\,\Delta E_{f}/V_{l_{0}}\,=\,\Delta \rho_{eff}
\eeq
which gives the equilibrium radius
\beq\label{AA28}
R_{eq}\,=\,\frac{2\,\sigma_{\rm wall}}{|\Delta \rho_{eff}|\,}\,=\,R_{crit}\,.
\eeq
Examining the energy \eq{AA22} as a function of $R$ at fixed $N$,
\beq\label{AA29}
\frac{\D E_{bub}}{\D R}\,\sim\,4\pi\,R\,\Le R\,\rho\,+\,2\,\sigma_{\rm wall}\Ra\,,
\eeq
we see it vanishes at $R\,=\,R_{crit}\,$.
The second derivative at this point,
\beq\label{AA30}
\frac{\D^{2}E_{bub}}{\D R^{2}}\bigg|_{R_{crit}}\,=\,-\,8\pi\,\sigma_{\rm wall}\,<\,0\,,
\eeq
shows that $R_{crit}$ provides a maximum of the bubble's energy, as it must be
in the case of the critical bubble structure of false vacuum decay, see \cite{FalB,FalB1,FalB2}. There are then two usual
regimes we have.
The first is when we have $R\,<\,R_{crit}$, then the wall tension dominates, the bubble shrinks and collapses.
The second is provided by $R\,>\,R_{crit}$, the bulk energy dominates, the bubble is unstable to growth and not decay.
The condition for a large "bubble" (core) in the present case is therefore
\beq\label{AA31}
R\,\gg\,R_{crit}\,=\,\frac{2\,\sigma_{\rm wall}}{|\rho|}\,.
\eeq
In this regime the bubble does not classically collapse and this simple core construction becomes stable against the collapse in the classical sense.

  When decay of the core is energetically forbidden classically, it can proceed through some
thermal fluctuations. The rate at which a single quasi-particle escapes
across the wall is governed simply by a Boltzmann factor
\beq\label{AA32}
\Gamma_{escape}\,\sim\,\exp\!\Le -\,\frac{E_{bind}}{k_{B}\,T_{wall}}\Ra\,,
\eeq
where $T_{\rm wall}$ is the effective temperature of the wall. Again there are two different
regimes we have.
The first one is a cold wall regime $k_{B}\,T_{\rm wall}\,\ll\,E_{\rm bind}$. In this case  the rate
 $\Gamma_{escape}$ is exponentially suppressed and the bubble is stable on some timescale.
The second regime is a hot wall situation, $k_{B}\,T_{\rm wall}\,\gtrsim\,E_{\rm bind}$. In this case the
thermal escape is efficient; the bubble loses quasi-particles continuously, $\rho$ rises toward zero, and the configuration evaporates if the opposite flux of the matter is absent. We notice then that in the second case the crust heating is accomplished by the "factory" activation, see \eq{AA18}, and therefore the crust in turn is cooling by the quasi-particles creation, i.e. by the "factory" work. The mechanism, discussed as well in \cite{BondBH}, leads the proposed interior
system to an equilibrium if the system is truly adiabatically isolated. In the case of an equilibrium, if we assume that the crust and the wall
temperature is set by the Bekenstein--Hawking temperature\footnote{We take here and further $c\,=\,1$, $k_{B}\,=\,1$\, and $\hbar\,=\,1$.}
\beq\label{AA33}
T_{H}\,=\,\frac{1}{\kappa\,M}\,,
\eeq
which is small 
compared to any plausible $E_{\rm bind}$, we realize the cold wall regime appears
generically for an equilibrium or quasi-equilibrium situations. Important to notice of course, the Hawking radiation changes the picture 
leading to the quasi-stable sates of the system. Namely, whatever the rate of this outward flux, it is shrink the whole interior leading to the decrease of the
$E_{bind}$ and to a reestablishing of a new quasi-equilibrium state. The process, if inward fluxes are not appearing, leads to some scenario of the full system evaporation through this chain of the quasi-equilibrium states.

\subsection{Two regions of the core: boundary core heat reservoir}\label{SecAA2}

 The core we construct has the two different regions, we denote them as inner core and boundary core region. These two regions of the core have quite different properties. So, firstly we discuss the boundary core region located near the $r\,\sim\,R_{-}$ and
next we go back to the inner core region located $r\,\sim\,0$ whose classical properties we discussed in the previous Subsection. We note once and for all that the excitation spectra in this Section are computed in flat space-time, without external fields present; this is a leading-order approximation valid to the first order with respect to the metric we use for the interior, and the curvature corrections are deferred to the GR Sections below.

 In the scalar field model, the matter component of the boundary field region consist of two excitations, massive and massless on the top of the constant negative background, see \eq{AA10} expression.
As illustration of the proposed construction, consider dispersion relations for the Goldstone and massive boson at the case of a flat space-time, we have:
\beq\label{AA34}
\omega_{\zeta}^{2}(\vec{k})\,=\,\vec{k}^{2}\,,
\qquad
\omega_{\epsilon}^{2}(\vec{k})\,=\,\vec{k}^{2}\,+\,m_{\epsilon}^{2}\,.
\eeq
Both have positive $\omega^{2}$ for all $\vec{k}$, the spectrum of the excitations then is represented by the
massless relativistic boson and a massive relativistic
boson, on top of the $V_{0}^{B}\,=\,-\,m^{2}\,M^{2}/4$ negative background. Introducing the Boltzmann temperature $T$ of the boundary core region,
we can write an energy density of this internal layer as
\beq\label{AA34001}
\rho^{B}(T)\,=\,V_{0}^{B}\,+\,\int\frac{d^{3}k}{(2\pi)^{3}}
\Big[\omega_{\zeta}\,n_{B}(\omega_{\zeta},T)\,+\,\omega_{\epsilon}\,n_{B}(\omega_{\epsilon},T)\Big]\,;\,\,\,\,
n_{B}(\omega,T)\,=\,(e^{\omega/T}\,-\,1)^{-1}\,.
\eeq
In this definition, the background $V_{0}^{B}$ provides a negative constant contribution to the thermodynamical quantities. To some extent
it is similar to what we have for description of phonon gas in crystals, the $V_{0}^{B}$ in this case is an analog of the crystal atoms contributions.
The excitation, in turn, is an analog of the phonons/rotons in the crystals thermodynamic.  They contribute to the regular heat
capacity and define what the regular temperature $T$ the region has. Yet, there exists a dual temperature related to the value of the 
$V_{0}^{B}$, see again \cite{BondBH} for the definition of it in terms of quasi-particles. In the model, then, the massless modes behave as a gas of photons or a low-temperature gas of phonons\footnote{The result, of course, is a consequence
of a linear dependence between the energy and momentum.} with only one polarization included, i.e. we have for the energy density, corresponding heat capacity and pressure:
\beq\label{AA35}
\rho^{B}_{\zeta}(T)\,=\,\frac{\pi^{2}}{30}\,T^{4}\,;\,\,\,C_{\zeta}\,=\,\frac{2\pi^{2}}{15}\,T^{3}\,;\,\,\,
P\,\simeq\,\frac{\pi^{2}}{90}\,T^{4}\,.
\eeq
The result is that the massless modes excitations behave as a low-temperature quantum Bose liquid system.
For the massive mode, assuming that in the spatially limited system at low temperatures we have
\beq\label{AA35001}
m_{\epsilon}\,\gg\,|\vec{k}|
\eeq
and correspondingly
\beq\label{AA35002}
\omega_{\epsilon}(\vec{k})\,\approx\,m_{\epsilon}\,+\,\frac{\vec{k}^{2}}{2\,m_{\epsilon}}\,\,.
\eeq
Then we obtain an analog of the gas of rotons which provides in this case:
\beq\label{AA36}
\rho^{B}_{\epsilon}(T)\,\simeq\,\frac{2\,m_{\epsilon}\,\vec{k}^{2}\,}{(2\pi)^{3/2}}\,\sqrt{m_{\epsilon}\,T}\,e^{-m_{\epsilon}/T}\,=\,m_{\epsilon}\,N;\,\,\,
C_{\epsilon}\,\simeq\,N\,\Le \frac{3}{4}\,+\,\frac{m_{\epsilon}}{T}\Ra\,;\,\,\,
P\,\simeq\,\frac{2\,\vec{k}^{2}\,}{(2\pi)^{3/2}}\,\Le m_{\epsilon}\,T\Ra^{3/2}\,e^{-m_{\epsilon}/T}\,\,,
\eeq
in the case opposite to \eq{AA35001} inequality, we will obtain a contribution similar to the massless one.
These two types of terms contribute differently at low and higher temperatures; there is some intermediate region at $T\,\sim\,m_{\epsilon}$ where both contributions must be taken into account. In general, the excitations have a positive energy density, therefore there is the core gravity influence which push them outward from the core toward the boundary, see further discussion about null gravity hypersurface. Of course, these positive energy density excitations modify the metric of negative energy density core.

 We see now, that because the heat capacity of the boundary core excitations is positive in respect to the regular temperature, there is a heat pump which
transfers heat from the core to the crust in the ordinary thermodynamic sense. The overall energy of the core is decreasing then if an opposite heat flux is absent.
As we already underlined, the heating of the crust in turn leads to a flux of the additional quasi-particles into the core
and we again face some quasi-equilibrium state but with an additional stabilization channel added. We stress that this positivity of the excitation heat capacity is precisely what distinguishes the excitation thermodynamics, which is ordinary with $C\,>\,0$, from the condensate thermodynamics, which is governed by the negative inverse-temperature parameter $\beta\,<\,0$; this distinction pre-figures the two-temperature structure analysed in Subsection~\ref{SecAA5}.

\subsection{Two regions of the core: inner core heat reservoir}\label{SecAA3}

 At the inner core region, a classical field value is large $\phi_{cl}\,\gg\,M$, see also \eq{SAB4} further.
The second derivative of the potential provides the negative value of the effective "mass" then:
\beq\label{AA37}
m_{eff}^{2}(\phi_{cl})\,\equiv\,V''(\phi_{cl})\,
\approx\,-\,\frac{16\,m^{2}\,\phi_{cl}^{6}}{M^{6}}\,e^{-\,\phi_{cl}^{4}/M^{4}}\,=\,
-\,16\,\frac{\varepsilon^{4}(t)}{M^{2}}\,\ln^{3/2}\Le \frac{m\,M}{\varepsilon^{2}(t)}\Ra^{2}\,<\,0\,,
\eeq
here $\varepsilon^{4}(t)$ is an absolute value of the energy density deep in the core.
The mass is negative that means an instability of the system, it can evolve either toward smaller values of the field density or toward 
$\phi\rightarrow\,\infty$ depending on the heat and matter fluxes affect the core. The characteristic scale we have then is a stationary  maximum of the mass.
Taking derivative of the \eq{AA37} with respect to the $\phi_{cl}$, we obtain that the stationary maximum is located at
\beq\label{AA38}
\phi_{cl,m}^{4}\,=\,3\,M^{4}/2\,\rightarrow\,m_{eff,m}^{2}(\phi_{cl,m})\,\sim\,m^{2}\,e^{-3/2}\,.
\eeq
For the growing core system, the value of the mass and corresponding instability decrease exponentially in correspondence to \eq{AA37} expression.

 For the further clarification of the inner core region behavior, consider a small fluctuation $\delta \phi$ around the introduced $\phi_{cl}$.
There is a simple quadratic action for fluctuations $\delta\phi$ around $\phi_{cl}$ we can write:
\beq\label{AA39}
S_{\delta}\,=\,\int d^{4}x\,\Big[\,\frac{1}{2}\,(\partial\,\delta\phi)^{2}\,
+\,\frac{1}{2}\,|m_{eff}^{2}|\,\delta\phi^{2}\,\Big]\,.
\eeq
The instability appears in the Fourier modes of the fluctuations through the wrong mass sign in the dispersion relations.
Namely, for 
\beq\label{AA40}
\delta\phi_{\,\vec k}(t)\,\sim\,e^{i\,\vec{k}\cdot\vec{x}\,-\,i\,\omega t}
\eeq
and
\beq\label{AA41}
\omega^{2}(\vec k)\,=\,\vec{k}^{2}\,-\,|m_{eff}^{2}|\,.
\eeq
we obtain 
\beq\label{AA42}
\delta\phi_{\,\vec k}(t)\,\sim\,e^{i\,\vec{k}\cdot\vec{x}\,-\,i\,t\,\sqrt{\vec{k}^{2}\,-\,|m_{eff}^{2}|}}
\eeq
and the result now depends on the value of the momentum we have. There is a critical momentum
\beq\label{AA43}
|\vec k|\,=\,|m_{eff}|
\eeq
and
the spectrum splits at this value of it.

 We see that, first of all, there is a stable sector of the excitations exists.
When
\beq\label{AA44}
|\vec{k}|\,>\,|m_{eff}|\,\rightarrow\,\omega^{2}\,>\,0\,,
\eeq
there are modes which propagate with group velocity
\beq\label{AA45}
v_{g}\,=\,d\omega/dk\,=\,k/\omega\,<\,1 
\eeq
and dispersion law
\beq\label{AA46}
\omega(\vec k)\,=\,\sqrt{\vec{k}^{2}\,-\,|m_{eff}^{2}|}\,
\,\,\,\textrm{for }|\vec k|\,>\,|m_{eff}|\,.
\eeq
These are ordinary positive-energy excitations defined with an IR cutoff at $|\vec k|\,=\,|m_{\rm eff}|$,
below which no propagating mode exists. When, in turn, we have
\beq\label{AA47}
|\vec{k}|\,<\,|m_{eff}|\,\rightarrow\,\omega^{2}\,<\,0
\eeq
we obtain modes which do not oscillate but grow
exponentially with rate
\beq\label{AA48}
\gamma(\vec k)\,=\,\sqrt{|m_{eff}^{2}|\,-\,\vec{k}^{2}}\,
\,\,\,\textrm{for }|\vec k|\,<\,|m_{eff}|\,,
\eeq
the maximum rate $\gamma(0)\,=\,|m_{eff}|$ is at zero momentum.

 These growing modes encode the dynamical instability of the inner core region
configuration. We see that the small fluctuations of long wavelength grow
exponentially, it means that this field configuration is not stable
and will roll. In our particular case, the field at any point
eventually rolls either back to the second minimum at
$\phi\,=\,\phi_{0}$, see \eq{AA9} or further out toward $\phi\,\to\,\infty$.
As already underlined, the direction depends on overall thermodynamic processes of the whole system.
In general, if the inner core region which has a finite spatial size inside the whole core,
the growth is cut off by the spatial scale of this region. Namely, if the
region has linear size $l_{I}$, then unstable modes with $\vec k$
smaller than $\sim\,1/l_{I}$ are excluded by the boundary and then the
growth rate of the surviving unstable modes behaves at most as
\beq\label{AA49}
\gamma(\vec k)\,\sim\,
\sqrt{|m_{eff}^{2}|\,-\,(1/l_{I})^{2}}\,.
\eeq
If, in turn, $l_{I}\,<\,1/|m_{\rm eff}|$ then 
no unstable modes are allowed and the configuration is
dynamically stable purely by virtue of being small enough.

  From the thermodynamic point of view, the presence of the unstable sector means that there is no
equilibrium thermal state for the inner core region. Namely, a Gibbs
state cannot be defined because the unstable
modes have $\omega^{2}\,<\,0$, the corresponding Hamiltonian is not bounded from below and that leads to the divergence of the partition function.
Yet, a quasi-thermal state of the stable sector of the inner region can
be defined, valid on timescales shorter than the instability growth
time $\gamma^{-1}$. We restrict the trace to states populating only
the modes with $|\vec k|\,>\,|m_{eff}|$:
\beq\label{AA50}
\rho^{I}_{quasi}\,=\,\frac{1}{Z}\,e^{-\,H_{I}/T}\,,
\qquad H_{I}\,=\,\int_{|\vec k|\,>\,|m_{\rm eff}|}\frac{d^{3}k}{(2\pi)^{3}}\,
\omega(\vec k)\,a_{\vec k}^{\dagger}\,a_{\vec k}\,.
\eeq
This expression we can use as a thermodynamic framework valid in the inner core region. The
stable modes are in thermal equilibrium, the unstable modes are
assumed quenched or otherwise unexcited. It is important, of course, to mention that we discussed the state in a flat space-time, in the case of a curved background 
the gravity corrections will enter in the picture.

  Next, treating the stable modes as a massive relativistic boson gas with
mass $|m_{\rm eff}|$ and IR cutoff at $|\vec k|\,=\,|m_{\rm eff}|$, we define
the thermal energy density as
\beq\label{AA51}
\rho^{I}_{stable}(T)\,=\,V_{0}^{I}(\phi_{cl})\,+\,
\int_{|m_{\rm eff}|}^{\infty}\frac{d^{3}k}{(2\pi)^{3}}\,
\frac{\sqrt{\vec{k}^{2}\,-\,|m_{\rm eff}^{2}|}}{e^{\sqrt{\vec{k}^{2}\,-\,|m_{\rm eff}^{2}|}/T}\,-\,1}\,,
\eeq
with $V_{0}^{I}(\phi_{cl})\,=\,-\,m^{2}\,M^{2}\,e^{-\,\phi_{cl}^{4}/M^{4}}$
the background contribution. Then we have two regimes depending on the regular temperature value in this inner core region.
The first one is when 
\beq\label{AA52}
T\,\sim\,\vec{k}\,\gg\,|m_{\rm eff}|\,.
\eeq
In this case the IR cutoff is unimportant and we obtain an answer similar to the \eq{AA35} expressions, i.e.
the spectrum looks effectively the same as for the massless mode of the boundary core region. When, in turn,
\beq\label{AA53}
|\vec{k}|\,-\,|m_{\rm eff}|\,\sim\,|m_{\rm eff}|\,\geq\,T
\eeq
we obtain an answer similar to the roton gas \eq{AA36} contributions with corresponding replacement of the mass in the expressions.
The IR cutoff in this case freezes out the all modes 
and the heat capacity is exponentially suppressed so the inner region becomes thermodynamically inert
at temperatures below the effective mass scale.
The transition between the contribution can be considered as a crossover  at $T\,\sim\,|m_{eff}|$ approximately.
Therefore, for a deep
inner-core configuration, when $\phi_{cl}\,\gg\,M$, the crossover lies far
below the scale $m$ of the boundary core region, so the inner core remains thermally active
at temperatures where the boundary core region massive mode is already frozen out.

  It important to note again, that so far we considered the thermodynamic of excitations of the core region. Additionally, we can treat the constant 
value of the potential energy of
the whole core as a source of a thermodynamic based on the temperature parameter related to the interaction potential of the quasi-particles
and not to the regular heat excitations, see \cite{BondBH}.
In this case overall framework is a two temperature thermodynamic, the analog of that construction can be found in a plasma or crystal systems,
see \cite{TwoT1,TwoT2} for example. Therefore, the equilibrium states for the core system must be defined not only through the thermal spectrum of the excitation but also through a value of the second "temperature" defined by the value of the potential energy. We discuss this question further in the corresponding 
Subsection ~\ref{SecAA5}.

\subsection{The core as a heat machine}\label{SecAA4}
 
 An observation we make is that the inner core region is less deep in
potential energy than the boundary core region:
\beq\label{AA54}
V_{0}^{B}\,=\,-\,\frac{m^{2}\,M^{2}}{4}\,,
\qquad V_{0}^{I}(\phi_{cl})\,=\,-\,m^{2}\,M^{2}\,e^{-\,\phi_{cl}^{4}/M^{4}}\,\to\,0^{-}\,.
\eeq
The difference
\beq\label{AA55}
\Delta V(\phi_{cl})\,\equiv\,V_{0}^{I}\,-\,V_{0}^{B}\,=\,\frac{m^{2}\,M^{2}}{4}
\,-\,m^{2}\,M^{2}\,e^{-\,\phi_{b}^{4}/M^{4}}\,
\xrightarrow{\phi_{b}\,\to\,\infty}\,\frac{m^{2}\,M^{2}}{4}
\eeq
is positive and approaches a fixed value $m^{2}\,M^{2}/4$ deep into
the inner core. So, when, due to evaporation, for example, the density decreases and 
the field rolls from the inner core back toward the boundary core region, this energy is released per unit volume converted:
\beq\label{AA56}
\delta \rho_{I}\,\approx\,\frac{m^{2}\,M^{2}}{4}\,.
\eeq
The inner core is therefore an energy reservoir which slowly drains as the field density decreases.
Now we can roughly estimate an evaporation time of the inner core reservoir. The evaporation decreases the field density
from some $\phi_{cl}$ value to some $\phi_{0}$ value where the energy density inverts sign, this $\phi_{0}$ determines the spatial size of the core approximately.
Then, taking into account an energy conservation
\beq\label{AA57}
\dot\phi^{2}\,=\,-\,2\,V(\phi)\,,
\eeq
we obtain an approximate total time for the homogeneous field to roll from the $\phi_{cl}$ to $\phi_{0}$:
\beq\label{AA58}
\tau_{roll}(\phi_{cl})\,=\,\int_{\phi_{0}}^{\phi_{cl}}\,\frac{d\phi}{\sqrt{-\,2\,V(\phi)}}\,=\,\frac{1}{m\,M\,\sqrt{2}}
\int_{\phi_{0}}^{\phi_{cl}}\,d\phi\,e^{\phi^{4}/2M^{4}}\,;\,\,\,\,\phi_{0}\,\sim\,m_{eff}\,,\,\,\,\,
\phi_{cl}\,=\,M\,\ln^{1/4}\Le \frac{m\,M}{\varepsilon^{2}}\Ra^{2}\,;
\eeq
or
\beq\label{AA58001}
\tau_{roll}(\phi_{cl})\,=\,
\frac{1}{m\,\sqrt{2}}\int_{1}^{\ln^{1/4}\Le \frac{m\,M}{\varepsilon^{2}}\Ra^{2}}\,dx\,e^{x^{4}/2}\,,
\eeq
see \eq{SAB4} further.
We obtain therefore to the first order precision with respect to the large logarithm:
\beq\label{AA59}
\tau_{roll}(\phi_{cl})\,\approx\,\frac{1}{2\sqrt{2}\,m}\,\frac{1}{\ln^{3/4}\Le \frac{m M}{\varepsilon^{2}}\Ra^{2}}\,\Le\frac{m M}{\varepsilon^{2}}\Ra^{2}\,,
\eeq
the smaller the $\varepsilon$ the larger the time of evaporation; recall that the $\varepsilon^{4}$ is an energy density of the inner core region.
This is of course a rough estimate, in general $\varepsilon\,=\,\varepsilon(t)$ and for a core which shrinks because of an evaporation, an additional 
term will appear in the r.h.s. of the \eq{AA57}, i.e. in general the evolution of the core due the evaporation is non-linear. Nevertheless, accounting for the per-volume amount of heat in \eq{AA55}, we can conclude that the rate of the reservoir depletion is approximately
\beq\label{AA60} 
P\,\sim\,\frac{\delta \rho_{I}}{\tau_{roll}}\,=\,\frac{m^{2} M^{2}}{4\,\tau_{roll}}\,\sim\,m\,\varepsilon^{4}\,\ln^{3/4}\Le \frac{m M}{\varepsilon^{2}}\Ra^{2}\,,
\eeq
the rate is maximal when $\frac{m M}{\varepsilon^{2}}\approx\,const\,>\,1\,$, i.e. not deep in the inner core region. In general, the whole evaporation evolution of the core is evolving through a shrinking of the both inner core and boundary core regions consequently. The thermodynamics of this process has to account also, additionally to the heat excitations,  the change of the
potential well of the system, see \cite{BondBH}. At the last stage, the only heat remnant of the system is a crust which is evaporating through 
the same Hawking radiation mechanism.

 The whole evaporation mechanism we describe is the following then. In the quasi-static cycle picture we discuss, 
the Hawking radiation cools the external layer by removing some amount
energy from it. Then there is a small thermal gradient between the layer and the
core, see further, and there are two distinct contributions to the
core quasi-particles release we have then.
The first one is set by ordinary thermal physics and that
is a thermal-gradient transfer between the outer core and
layer, governed by a thermal conductance.
This is a "fast" process which provides a restoration of the quasi-equilibrium within each cycle built from a "portion" of a Hawking radiation.
The second process is provided by a rolling release from inner core to outer one,
governed by the $|m_{\rm eff}|$ scale. This is exponentially slow and
provides an almost steady trend over many cycles.
Therefore, we can roughly describe the cycles of the heat engine as  following.
On the one cycle timescale, the
outer core and crust reach quasi-equilibrium via the first channel mentioned above.
On the much longer rolling timescale, the inner core feeds
the outer core via second channel, refilling the thermal reservoir.
As the inner core drains, the rolling rate decreases, i.e. whole core is shrinking through the decreases of quasi-particle density.
The feed becomes weaker and the evaporation continues toward the inner core dissociation.
At the end the inner core is exhausted, the system stays with
a crust layer and boundary core only. Further, the continued evaporation as well drains
the outer core's thermal energy and we face a remnant BH system.\footnote{Simplifying, we have a core as a quasi-equilibrium heat bath which provides, through the interactions of the thermal excitations of the field in the first and second minimums,  some quasi-equilibrium temperature of the crust. The temperature of the core aka heat bath in general is different from the temperature of the crust, 
see next Section ~\ref{SecAA5}.}

 An additional detail concerns the core is that it is described by a negative temperature parameter, in scalar field description this parameter is related to the depth and dynamics of the
negatively valued potential well. This negative temperature system, as well known, being in contact with an another regular temperature subsystem, can not be stable and transfers heat to
the subsystem. So, the core system instability is underlined also by its negative temperature, see discussion in \cite{BondBH}.

\subsection{Two temperature description of the interior}\label{SecAA5}

 In connection to the general description of the three layer BH construction we discuss, it is important to clarify the 
roles of two different notions of the temperatures introduced in \cite{BondBH}. In terms of the quasi-particles, the core aka condensate sector 
of the framework, is composed of identical bosonic
quasi-particles all possessing the same single--particle energy
\beq\label{AA61} 
\varepsilon_q \,=\, -\,U_{0},\qquad \,U_{0}\,>\,0,
\eeq
with zero classical kinetic energy. In any realistic quantum-liquid analog (super-fluid
$^4$He being the canonical example), the condensate is accompanied by
a thermal cloud of collective excitations which are phonons at small
momentum and rotons at intermediate momentum, see discussion above and further. The excitations carry kinetic
energy and entropy on top of the condensate ground state. Treating
the condensate as the precise state of the core's content is appropriate only
in the $T\to 0$ limit, this is an approximation we consider further in the quasi-particles set-up, at any finite $T$ the bulk has
non-zero excitation density of the quasi-particles as well.
Accordingly, the total energy of the bulk can be represented as sum of two parts:
\beq\label{AA62}
U \,=\,U_{pot} \,+\, U_{kin}\,,\qquad
U_{pot}\,=\, -N_{cond}\,U_{0}\,,\qquad
U_{kin} \,=\,\sum_{k}\hbar\omega(k)\,n_k\,,
\eeq
where $N_{cond}$ is the condensate occupation and $n_k$ the
thermal phonon/roton population at wavevector $k$. The entropy then splits
similarly 
\beq\label{AA63}
S \,=\, S_{cond}(N_{cond}) \,+\, S_{exc}(U_{kin})\,,
\eeq
with the condensate entropy from the BEC bookkeeping over phase-space
cells and the excitation entropy from the standard thermal occupation
of $\omega(k)$ modes. Thus, the system we discuss is similar to the thermodynamics of crystals, to some extent, with only 
difference that the non-dependent on $T$ part of the total energy is negative here and changes dynamically in general.

 Considering now the energy splitting, we notice that there are two independent temperature parameters exist:
\begin{align}
\beta &\;\equiv\; \left.\frac{\partial S}{\partial U_{pot}}\right|_{V,\,U_{kin}}\,,
\qquad \text{(potential--reservoir inverse temperature)}\,;
\label{AA64}
\\
\frac{1}{T} &\;\equiv\; \left.\frac{\partial S}{\partial U_{kin}}\right|_{V,\,U_{pot}}\,,
\qquad \text{(kinetic--reservoir inverse temperature)}\,.
\label{AA65}
\end{align}
These are conjugate to different extensive quantities and they
are not the two values of the same temperature. The obtained picture is similar 
to the coexisting of
electron and ion temperatures of a non--equilibrium plasma, see for example \cite{TwoT1,TwoT2}. Therefore, since this system consists of two sub-systems, 
the dynamics of the kinetic heat (phonons exchanged with a bath, for instance) is
governed by $T$ whereas the dynamics of the potential energy exchange (quasi-particle
condensation/dissociation) is governed by $\beta$ temperature parameter. There is no
single thermal hierarchy on which $\beta$ and $T$ can be compared without a consideration of non-static processes in the whole system of interior.
The BEC construction with two separated reservoirs has the following properties:
\beq\label{AA66}
\left.\frac{\partial U_{pot}}{\partial N_{cond}}\right|_V\,<\,0\,,\qquad
\left.\frac{\partial S_{cond}}{\partial N_{\rm cond}}\right|_V\,>\,0\,,
\eeq
hence
\beq\label{AA67}
\frac{1}{\beta}\,=\,\frac{\partial U_{\rm pot}/\partial N_{\rm cond}}{\partial S_{\rm cond}/\partial N_{\rm cond}}\bigg|_V\,<\,0\,,
\eeq
i.e.\ $\beta\,<\,0$. By contrast, the phonon/roton sector has an
unbounded continuum spectrum $\omega(k)\geq 0$ and obeys ordinary
positive--temperature statistics:
\beq\label{AA68}
\frac{1}{T}\;>\;0.
\eeq
Thus we notice that the two parameters sit on opposite sides of zero. The $\beta$
is conjugate to a bounded spectrum reservoir which allows
population inverted behavior and the $T$ is conjugate to an
unbounded spectrum reservoir with ordinary thermal behavior.

 Because the system's split, the condition of the full thermodynamic equilibrium now is defined by the entropy maximization
in the following form:
\beq\label{AA69}
\left.\frac{\partial S}{\partial U_{pot}}\right|_{V,U_{kin}}\,=\,\left.\frac{\partial S}{\partial U_{kin}}\right|_{V,U_{pot}}\,=\,\beta_{\rm eq}\,.
\eeq
In general, the condition would force the two derivatives to coincide. Nevertheless, we have to clarify separately how the relaxation could work. The two possibilities exist for the relaxation scenario, it is important
what and how is passing to an equilibrium firstly.
If the inter-reservoir relaxation channel, mentioned above and which is
quasi-particle dissociation into kinetic excitations and
the reverse process, is faster than every other relevant timescales then the equilibrium of the whole system could be achieved in meaning of \eq{AA69}. 
When, in turn, it is slower, then each reservoir/sub-system internally equilibrates independently and
the two sides of the system carry different temperature parameters for as long as the overall
relaxation has not had time to run. This is the
two temperature regime we consider in the application to the core of BH.
Since $\beta\,<\,0$ and $1/T\,>\,0$ for as long as the condensate is intact,
the two parameters are on opposite sides of zero and cannot coincide. Indeed, 
the negative $\beta$ system is tied to the bounded
spectrum of the condensate, i.e. every quasi-particle sits at the single
level $-\,U_{0}$, with no lower state available. Adding occupation does
not lower energy further, the same algebraic feature is what
underlies negative temperature ensembles in the spin systems with negative temperatures. The
unbounded phonon/roton spectrum, by contrast, has ordinary
$T\,>\,0$ behavior and the two structures coexist as long as no microscopic process
mixes them. The mixing of the subsystems is provided by a
thermal dissociation, when a phonon (or a combination of phonons)
with kinetic energy $U_{0}$ can convert into the unbinding of a single
condensate quasi-particle and the production of free kinetic
excitations. This process becomes thermally accessible at order-unity
probability when
\beq\label{AA70}
T \,\sim\, U_{0} \,.
\eeq
Below $T\,\ll\,U_{0}$ thermal dissociation is exponentially suppressed, i.e. it is proportional to
$\sim e^{-\,U_{0}/T}$, and the reservoirs remain decoupled on relevant
timescales. Above $T\,\sim\,U_{0}$ the rate becomes fast, the bounded
spectrum dissolves and the negative $\beta$ subsystem collapses into
the positive $T$ thermal branch.

 Equivalently, we can clarify the picture through the writing of the forward and backward rates of inter-reservoir
exchange. For the forward process, a phonon (or phonon cluster) with kinetic energy $\sim \,U_{0}$ ionizes a condensate quasi-particle out of its
bound level. The rate of the process could be determined as
\beq\label{AA71}
\Gamma_{\to} \,\propto\, e^{-\,U_{0}/T}.
\eeq
A backward process is when a free excitation is captured into the condensate and, because the negative value of the potential energy, it
releases the $U_{0}$ amount of energy into the phonon/roton bath. The rate of the process can be roughly defined as
\beq\label{AA72}
\Gamma_{\leftarrow}\,\propto\;T^{\,k}\,e^{\beta\,U_{0}}
\eeq
with some power $k$ set by the matrix element and density of states.
Detailed balance between the two,
$\Gamma_{\to}=\Gamma_{\leftarrow}$, closes only when the
exponential suppression of the forward channel becomes mild, which
again requires $T\,\sim \,U_{0}$. At smaller $T$ the system is locked into
slow inter-reservoir exchange and two distinct temperature
parameters persist.
Next we notice that at the moment the condensate dissolves, the BEC inverse temperature
parameter goes singular. Using the relation from \cite{BondBH}, we can define the $\beta$ parameter as
\beq\label{AA73}
\frac{1}{\beta}\,=\,-\frac{\,U_{0}}{\ln(1+1/\eta)},\qquad \eta \,=\, \rho_v/\rho_p\,,
\eeq
here $\rho_v$ and $\rho_p$ are density of the states in coordinate and momentum phase space correspondingly.
The dissolution corresponds to $\rho_v\to 0$ limit when the condensate density
drains into the kinetic continuum, hence $\eta\to 0$ and
$\ln(1+1/\eta)\to\infty$, providing
\beq\label{AA74}
\frac{1}{\beta}\,\longrightarrow\,0^{-}.
\eeq
The BEC bookkeeping therefore signals condensate dissolution at
$\beta\,\to\,-\,\infty$ or $1/\beta\,\to\, 0^-$ value of the parameter. The parameter approaches its
zero limit from below, never crossing to positive values; instead it
ceases to refer to any extant condensate. Above this crossover, only
$T$ remains as a meaningful temperature.
This behavior, therefore, is structurally identical to the way nuclear--spin
negative--temperature ensembles relax. There the inverse temperature
parameter passes through its zero limit (corresponding to
$T\to\pm\infty$) as the population inversion is removed, after which
the system passes in the ordinary positive-temperature regime.

  We conclude then that the two temperature structure of the core behaves as following. As in the regular two temperature systems, the heat is always flowing from negative temperature to the positive one, i.e. the negative $\beta$ branch of the system is hotter than the branch with regular $1/T$. Therefore, a real
equilibrium of the BH core particularly and whole BH interior system in general can be achieved only if we consider a BH as a fully adiabatically isolated system. When, in turn, the external inward fluxes of heat/matter stop working, but the crust is cooling by the Hawking radiation, then the system is not isolated and then the evaporation of the whole system begins by the dissociation of the core's main state till it disappears. 
Important that it is a usual fate of such negative temperature systems. The core is thermodynamically unstable, the details of its evolution 
depend on the characteristic time scales of the processes in core and crust but when the only Hawking radiation present it always ends through the full evaporation.
This time interval during which the system evaporates can be considered as the characteristic relaxation time of the entire system in the sense of \eq{AA69}.


\section{GR description of the inner core region}\label{SecAB}

\subsection{Metric of the core}\label{SecAB1}

The main property of the core is that there the energy density is negative and so the mass function:
\beq\label{SAA2}
\mu(t,r)\,<\,0\,.
\eeq
Then, the general form of the interval we can introduce for the core is the following one:
\beq\label{SC38}
ds^2\,=\,e^{\Phi(t,r)}\,\chi(t,r)\,dt^2\,-\,\chi^{-1}(t,r)\,dr^{2}\,-\,r^{2}\,d^{2}\Omega\,,
\eeq
with
\beq\label{SC39}
\chi(t,r)\,=\,1\,+\,\frac{2\,|\mu(t,r)|}{m_{p}^{2}\,r}\,.
\eeq
The metric for the core we have then is 
\beq\label{SC40}
g_{\mu \nu}\,=\,\left(
\begin{matrix}
e^{\Phi(t,r)}\,\chi(t,r)\,&\,0\,&\,0\,&\,0 \\
0\,&\,-\,\chi^{-1}(t,r)\,&\,0\,&\,0 \\
0\,&\,0\,&\, -r^{2}  \,&\,0 \\
0\,&\,0\,&\,0 \,&\, -r^{2}\sin^{2}\theta \\
\end{matrix}\right)\,
\eeq
with correspondingly defined 
\beq\label{SC41}
g^{\mu \nu}\,=\,\left(
\begin{matrix}
e^{-\Phi(t,r)}\,\chi^{-1}(t,r)\,&\,0\,&\,0\,&\,0 \\
0\,&\,-\,\chi(t,r)\,&\,0\,&\,0 \\
0\,&\,0\,&\, -1/r^{2}  \,&\,0 \\
0\,&\,0\,&\,0 \,&\, -1/r^{2}\sin^{2}\theta \\
\end{matrix}\right)\,.
\eeq
This metric we also write in the monad form
\beq\label{SC42}
g_{\mu \nu}\,=\,\tau_{\mu}\tau_{\nu}\,-\,h_{\mu \nu}\,;\,\,\,\tau_{\mu}\tau^{\mu}\,=\,1\,;\,\,\,\tau_{\mu}h^{\mu \nu}\,=\,0\,
\eeq
with
\beq\label{SC43}
g^{\mu \nu}\,=\,\tau^{\mu}\tau^{\nu}\,-\,h^{\mu \nu}\,;\,\,\,\tau^{\mu}h_{\mu \nu}\,=\,0\,.
\eeq
see \cite{Zelmanov,Vladinirov}.
Here
\beq\label{SC44}
\tau_{\mu}\,=\,\Le \tau_{0},\,\tau_{1},\,\tau_{2},\,\tau_{3}\Ra\,=\,\Le \tau_{0},\,\tau_{r},\,\tau_{\theta},\,\tau_{\phi}\Ra\,.
\eeq
We have then
\beq\label{SC45}
\tau^{\mu}\,=\,\frac{g_{0\alpha} g^{\alpha \mu}}{\sqrt{g_{00}}}\,
\eeq
that provides
\beq\label{SC45001}
\tau^{0}\,=\,\frac{g_{0\alpha} g^{0 \alpha}}{\sqrt{g_{00}}}\,=\,e^{-\Phi(t,r)/2}\,\chi^{-1/2}(t,r)\,;\,\,\,\tau^{r}\,=\,\tau^{\theta}\,=\,\tau^{\phi}\,=\,0\,.
\eeq
Correspondingly
\beq\label{SC46}
\tau_{\mu}\,=\,\frac{g_{0 \mu}}{\sqrt{g_{00}}}\,
\eeq
and
\beq\label{SC47}
\tau_{0}\,=\,e^{\Phi(t,r)/2}\,\chi^{1/2}(t,r)\,,\,\,\,\,\tau_{r}\,=\,\tau_{\theta}\,=\,\tau_{\phi}\,=\,0\,.
\eeq
The expression for the components of the induced $h$ metric can be found as well, it has the following form:
\beq\label{SC48}
h_{\mu \nu}\,=\,\tau_{\mu}\tau_{\nu}\,-\,g_{\mu \nu}\,;\,\,\,\mu,\nu\,=\,0,\,r,\,\theta,\,\phi\,
\eeq
and correspondingly the non-zero components we have are:
\beq\label{SC49}
h_{i k}\,=\,\left(
\begin{matrix}
\,\chi^{-1}(t,r)\,&\,0\,&\,0 \\
\,0\,&\, r^{2}  \,&\,0 \\
\,0\,&\,0 \,&\, r^{2}\sin^{2}\theta \\
\end{matrix}\right)\,\,;\,\,\,\,i,k\,=\,r,\,\theta,\,\phi\,.
\eeq
As well we have
\beq\label{SC50}
h^{i k}\,=\,
\left(
\begin{matrix}
\chi(t,r)\,&\,0\,&\,0 \\
\,0\,&\, 1/r^{2}  \,&\,0 \\
\,0\,&\,0 \,&\, 1/r^{2}\sin^{2}\theta \\
\end{matrix}\right)\,.
\eeq
The components of the energy-momentum tensor now are defined as:
\beq\label{SC4}
\rho\,=\,\tau^{\alpha}\tau^{\beta} T_{\alpha \beta}\,;\,\,\,p^{\mu}\,=\,-\,h^{\mu \alpha}\tau^{\beta} T_{\alpha \beta}\,=\,-\,h^{\mu i} T_{i 0}\,.
\eeq
We have then that $p^{\mu}\,=\,0$ for the particular case of diagonal energy-momentum we have.

\subsection{Classical field solutions and corresponding energy-momentum tensor at inner core region}\label{SecAB2}

 In this case we can consider a generally non-static and homogeneous solution of the equations of motion for the $\phi$ field, i.e. we consider a system's state when
\beq\label{SAB1}
\phi\,=\,\phi(t)\,
\eeq
and we have therefore
\beq\label{SAB2}
\g^{\mu \nu}\,\D_{\mu}\phi_{i}\,\D_{\nu}\phi^{i}\,=\,\dot{\phi}_{i}\,\dot{\phi}^{i}\,=\,\dot{\phi}^{2}\,.
\eeq
We notice that we are looking for an asymptotic value of the potential at asymptotically large value of the field which is not an extremum of the Lagrangian. Therefore,
we consider a classical field's configuration which satisfies the following identity:
\beq\label{SAB3}
V(\phi_{cl})\,\approx\,-\,m^2\,M^{2}\,e^{-\phi^{4}_{cl}(t)/M^4}\,=\,-\varepsilon^{4}(t)\,\sim\,-\,0
\eeq
that provides
\beq\label{SAB4}
\phi^{2}_{cl}(t)\,=\,\sqrt{2}\,M^{2}\,\ln^{1/2}\Le \frac{m M}{\varepsilon^{2}(t)}\Ra
\eeq
Using the \eq{SAA1} definition we can write the components of the energy-momentum tensor:
\beqar\label{SAB5}
&\,&
T_{\mu \nu}\,=\,0\,,\,\,\,\,\mu\neq\nu\,;
\\
&\,&
T_{00}\,=\,\frac{1}{2}\,\dot{\phi}^{2}_{cl}\,+\,e^{\Phi(t,r)}\,\chi(t,r)\,V(\phi_{cl})\,;
\label{SDD4001}
\\
&\,&
T_{rr}\,=\,\frac{1}{2}\,e^{-\Phi(t,r)}\,\chi^{-2}(t,r)\,\dot{\phi}^{2}_{cl}\,-\,\chi^{-1}(t,r)\,V(\phi_{cl})\,;
\label{SDD4002}
\\
&\,&
T_{\theta \theta}\,=\,r^{2}\,\Le e^{-\Phi(t,r)}\,\chi^{-1}(t,r)\,\frac{1}{2}\,\dot{\phi}^{2}_{cl}\,-\,V(\phi_{cl})\Ra\,;
\label{SDD4003}
\\
&\,&
T_{\phi \phi}\,=\,r^{2}\,\sin^{2}(\theta)\,\Le e^{-\Phi(t,r)}\,\chi^{-1}(t,r)\,\frac{1}{2}\,\dot{\phi}^{2}_{cl}\,-\,V(\phi_{cl})\Ra\,;
\label{SDD4004}
\\
&\,&
T\,=\,g^{\mu \nu}\,T_{\mu \nu}\,=\,-\,e^{-\Phi(t,r)}\,\chi^{-1}(t,r)\,\dot{\phi}^{2}_{cl}\,+\,4\,V(\phi_{cl})\,.
\label{SD4005}
\eeqar
Now, with the help of \eq{SC4} and \eq{SC42} expressions, we have then for the energy density:
\beq\label{SD5}
\rho\,=\,e^{-\Phi(t,r)}\,\chi^{-1}(t,r)\,\frac{1}{2}\,\dot{\phi}^{2}_{cl}\,+\,V(\phi_{cl})\,.
\eeq
Correspondingly, using \eq{SC4}, we have for the momentum:
\beq\label{SDD6}
p^{\mu}\,=\,-\,h^{\mu i} T_{i 0}\,=\,0\,.
\eeq
For the pressure we can use a standard definition:
\beq\label{SDD7}
p\,=\,\frac{1}{3}\,h^{\mu\nu}\,T_{\mu\nu}\,=\,e^{-\Phi(t,r)}\,\chi^{-1}(t,r)\,\frac{1}{2}\,\dot{\phi}^{2}_{cl}\,-\,V(\phi_{cl})\,
\eeq
obtaining
\beq\label{SDD8}
\rho\,=\,p\,+\,2\,V(\phi_{cl})\,
\eeq
equation of state.

\subsection{Static solution with negative energy and zero sum of energy and pressure EOS case}\label{SecAB3}

 In the simplest static limit we have the following expressions for the quantities of interests:
\beq\label{SBB1}
\Phi\,=\,0\,,\,\,\,\chi\,=\,\chi(r)\,,\,\,\,\mu\,=\,\mu(r)\,,\,\,\,\varepsilon^{4}(t)=\varepsilon^{4}\,,\,\,\,\dot{\phi}^{2}_{cl}\,=\,0\,
\eeq
with
\beq\label{SBB1001}
p\,=\,-\,\rho\,>\,0\,
\eeq
equation of state (EOS) of the quasi-particles.
The general form of the Einstein equations with cosmological constant included\footnote{The convention we use here is a positive sign of the $\Lambda$ in the Einstein-Hilbert Lagrangian.}
has the following form:
\beq\label{SBB2}
R_{\mu\nu}\,-\,\frac{1}{2}\,g_{\mu\nu}\,R\,=\,\frac{8\pi}{m_{p}^{2}}\,T_{\mu\nu}\,+\,g_{\mu\nu}\,\Lambda
\eeq
where
\beq\label{SBB3}
R\,=\,-\,\frac{8\pi}{m_{p}^{2}}\,T\,-\,4\,\Lambda
\eeq
In the static case, the energy-momentum tensor obtained from the scalar field reduces then to 
\beq\label{SBB4}
T_{\mu\nu}\,=\,g_{\mu\nu}\,V(\phi_{cl})\,,
\eeq
see above. Therefore,
defining $\Lambda$ simply as
\beq\label{SBB6}
\Lambda\,=\,\frac{8\pi}{m_{p}^{2}}\,V\,=\,\kappa\,V(\phi_{cl})\,\equiv\,-\,\kappa\,\varepsilon^{4}\,,
\eeq
the Einstein equation can be written as with cosmological constant included but with zero energy momentum tensor:
\beq\label{SBB7}
G_{\mu}^{\nu}\,=\,g_{\mu}^{\nu}\,\Lambda\,.
\eeq
Provided by only diagonal components of $G_{\mu}^{\mu}$ with $\Phi^{'}\,=\,0$ condition, the solution of the equations is trivial, we obtain:
\beq\label{SBB8}
\chi(r)\,=\,1\,-\,\frac{\Lambda}{3}\,r^{2}\,=\,1\,+\,\kappa\,r^{2}\,\frac{\varepsilon^{4}}{3}\,=\,1\,+\,\frac{8\pi}{3\,m_{p}^{2}}\,r^{2}\,\varepsilon^{4}\,=\,
1\,+\,\frac{2}{m_{p}^{2}}\,\frac{\mu(r)}{r}\,=\,1\,+\,r^{2}/L^{2}\,.
\eeq
The obtained solution has the form of an AdS solution with curvature radius of the AdS space-time defined by the value of the $\varepsilon^{4}$:
\beq\label{SBB9}
L^{2}\,=\,\frac{3\,m_{p}^{2}}{8\pi\varepsilon^{4}}\,,
\eeq
which is standard result of course, see \cite{FalB,FalB1,FalB2}.

 Next we can discuss a condition of hydrostatic equilibrium which is the covariant conservation of the
energy-momentum tensor in the radial direction, $\nabla_{\mu}T^{\mu}{}_{r}\,=\,0$.
For the given static solution, a direct calculation gives
\beq\label{SBB10}
\nabla_{\mu}\,T^{\mu}{}_{r}\,=\,-\,\frac{dp}{dr}\,-\,\frac{1}{2}\,\Le \rho\,+\,p\Ra\,\frac{\chi'(r)}{\chi(r)}\,=\,0\,,
\eeq
which is the Tolman--Oppenheimer--Volkoff (TOV) equation in the form
\beq\label{SBB11}
\frac{dp}{dr}\,=\,-\,\Le \rho\,+\,p\Ra\,\frac{\chi'(r)}{2\,\chi(r)}
\eeq
Here, as usual, the factor $\chi'/(2\,\chi)$ is the gravitational redshift gradient. The combination $\rho\,+\,p$
playing the role of the inertial mass density is the standard relativistic
generalization of the Newtonian $\rho\,$.
In the case of \eq{SBB1001} EOS, this inertial mass density vanishes
identically,
\beq\label{SBB12}
\rho(r)\,+\,p(r)\,=\,0\,,
\eeq
so the TOV equation \eq{SBB11} is satisfied trivially:
\beq\label{SBB13}
\frac{dp}{dr}\,=\,0\,,
\eeq
which is consistent with $p\,=\,-\,V\,=\,$const everywhere inside the
core or in some internal region of the core. This
is the standard property of the  introduced $\Lambda$-vacuum: it does not "weigh"
anything, gravitational effects arise entirely through its contribution
to space-time curvature and not through a hydrostatic pressure gradient

 To quantify the gravitational field experienced inside the bubble, we compute
the four-acceleration of a static observer at radius $r$, and the geodesic
acceleration of a freely-falling test particle initially at rest. Both
quantities are well-defined and unambiguous.
A static observer has four-velocity
\beq\label{SBB14}
\tau^{\mu}\,=\,\Le \frac{1}{\sqrt{\chi(r)}},\,0,\,0,\,0\Ra\,,
\eeq
see \eq{SC45},
and four-acceleration $a^{\mu}\,=\,\tau^{\nu}\,\nabla_{\nu}\,\tau^{\mu}$. A direct
computation yields the only non-zero component
\beq\label{SBB15}
a^{r}\,=\,\frac{\chi'(r)}{2}\,=\,\frac{r}{L^{2}}\,,
\eeq
where we used $\chi(r)\,=\,1\,+\,r^{2}/L^{2}$. The proper magnitude of the
acceleration, $|a|\,=\,\sqrt{-\,a_{\mu}\,a^{\mu}}$ for the spacelike vector
$a^{\mu}$, is
\beq\label{SBB16}
|a(r)|\,=\,\frac{|\chi'(r)|}{2\,\sqrt{\chi(r)}}\,=\,\frac{r/L^{2}}{\sqrt{1\,+\,r^{2}/L^{2}}}\,.
\eeq
This is the magnitude of the force per unit mass that, as a clarifying example, a rocket would have to
exert outward to maintain the observer at fixed $r$. The fact that the proper force is directed outward means
that the gravitational pull on the observer is inward: gravity inside
the region is attractive toward the center, and two limits exist then.

 The first one concerns the position near the center, $r\,\ll\,L$, the proper acceleration reduces there to
\beq\label{SBB17}
|a(r)|\,\approx\,\frac{r}{L^{2}}\,=\,\frac{8 \pi \varepsilon^{4}}{3 m_{p}^{2}}\,r\,=\,\frac{\kappa}{3}\,\varepsilon^{4}\,r\,,
\eeq
the expression is linear in $r$. This is a harmonic restoring acceleration toward the origin,
identical in form to the Newtonian gravitational field inside a uniform
sphere of constant density but with the relativistic
enhancement due to the contribution of the pressure to the active
gravitational mass.

 The second one is when the particle is placed far from the center when $r\,\gg\,L$, the proper acceleration saturates there:
\beq\label{SBB18}
|a(r)|\,\approx\,\frac{1}{L}\,=\,\varepsilon^{2}\,\sqrt{\frac{\kappa}{3}}\,,
\eeq
the expression is the characteristic surface-gravity scale of AdS, set by the inverse AdS
radius, a static observer at
large $r$ requires only a finite thrust to get accelerated.

Considering a test particle released from rest at radius $r$, we see that this feels no proper acceleration
but follows a time-like geodesic. The radial geodesic equation of motion for a
particle initially at rest is
\beq\label{SBB19}
\frac{d^{2}\,r}{d\,\tau^{2}}\,+\,\Gamma^{r}_{tt}\,\Le \frac{dt}{d\tau}\Ra^{2}\,=\,0
\eeq
or
\beq\label{SBB20}
\frac{d^{2}\,r}{d\,\tau^{2}}\,=\,-\,\Gamma^{r}_{tt}\,\Le \frac{dt}{d\tau}\Ra^{2}\,
=\,-\,\frac{\chi(r)\,\chi'(r)}{2}\,\cdot\,\frac{1}{\chi(r)}\,=\,-\,\frac{\chi'(r)}{2}\,,
\eeq
where the normalization $g_{00}\,(dt/d\tau)^{2}\,=\,1$, was used, see \eq{SC38}. Substituting \eq{SBB9} in it we obtain:
\beq\label{HS15}
\frac{d^{2}\,r}{d\,\tau^{2}}\,=\,-\,\frac{r}{L^{2}}\,=\,-\,\frac{\kappa}{3}\,\varepsilon^{4}\,r\,
\eeq
This is exactly the equation of motion of a harmonic oscillator with angular
frequency
\beq\label{HS16}
\omega\,=\,\frac{1}{L}\,=\,\varepsilon^{2}\,\sqrt{\frac{\kappa}{3}}\,\,.
\eeq
A test particle released anywhere inside the vacuum interior of the core oscillates radially
through the center with this frequency, never reaching infinity. This is the
defining feature of time-like geodesic confinement in AdS: the negative
vacuum energy acts as a harmonic trap, and no singularity appears, of course. This is the standard property of global AdS space-time, there the time-like geodesics are oscillatory with period $2\pi\,L$ independent of the oscillation amplitude; the core we obtained thus realizes this confinement with the AdS radius $L$ set by the negative vacuum energy density \eq{SBB9}.

 It is worth mentioning then that for this type of the core we obtained, the negative vacuum energy with vacuum equation of state produces an
attractive gravity inside the bubble, but repulsive, with respect to ordinary matter.


\section{Boundary core and an interpolation between two regions of the core}\label{SecBB}

\subsection{Boundary core regions set-up}\label{SecBB1}

 The general form of the interval for the boundary core region in the case of a static metric is the same as in the previous case we discussed 
in  Section ~\ref{SecAA}:
\beq\label{BBA1}
ds^{2}\,=\,e^{\Phi}\,\chi(r)\,dt^{2}\,-\,\chi^{-1}(r)\,dr^{2}\,-\,r^{2}d\Omega^{2}\,,
\eeq
see \eq{SC38}.
The differences from the inner core region case is that the $\chi(r)$ function for the boundary region of the core is defined as 
\beq\label{BBA2}
\chi_{B}(r)\,=\,1\,+\,\frac{r^{2}}{L_{B}^{2}}\,
\eeq
with an another AdS radius
\beq\label{BBA3}
L_{B}^{2}\,=\,\frac{12}{\kappa\,m^{2}M^{2}}\,
\eeq
and that in this region of the core the thermal excitations exist above the potential energy minimum.
Denoting $a\,\in\,\{\zeta,\epsilon\}$ as the massless and massive modes of the region, see \eq{AA34} and Section~\ref{SecAA2} discussion, we write  the relevant
quadratic action of the modes:
\beq\label{BBA4}
S^{(a)}\,=\,\int\!d^{4}x\,\sqrt{-g}\,\el^{(a)}\,,\qquad
\el^{(a)}\,=\,\frac{1}{2}\,g^{\mu\nu}\,\D_{\mu}\phi^{(a)}\,\D_{\nu}\phi^{(a)}\,
-\,\frac{1}{2}\,m_{a}^{2}\,(\phi^{(a)})^{2}\,,
\eeq
with $m_{\zeta}\,=\,0$ and $m_{\epsilon}\,\neq\,0$, see \eq{AA10}, further in this Section
we use results of the Appendix~\ref{AppenB} calculations.
The background energy density of this boundary region is different from the one of the inner core, see \eq{SBB8}, where we denote the AdS radius of the 
inner core region as
\beq\label{BBA5}
L_{I}^{2}(\phi_{cl})\,=\,\frac{3\,m_{p}^{2}}{8\pi\varepsilon^{4}}\,,
\quad
\eeq
see \eq{SBB9} definition.
We obtain thus that a hierarchy of the radii of the two core regions exists in general:
\beq\label{BBA6}
L_{I}\,\gg\,L_{B}\,,
\eeq
i.e. the deep inner core has a  much weaker AdS curvature (larger AdS radius) than the boundary core. In the limit
$\varepsilon\,\to\,0$ we have $L_{I}\,\to\,\infty$ and the deep inner
core approaches Minkowski space. 

 Therefore, the background metric for the whole core is not a
single pure AdS solution but must interpolate between these two AdS patches
with two different curvature radii:
\beq\label{BBA7}
\chi_{\rm I-B}(r)\,=\,\begin{cases}
\chi_{I}(r)\,=\,1\,+\,r^{2}/L_{I}^{2}\,,\quad
&0\,\leq\,r\,<\,r_{*}\quad\text{(deep core, weak AdS)}\,,
\\
\chi_{B}(r)\,=\,1\,+\,\dfrac{r^{2}}{L_{B}^{2}}
\,+\,\dfrac{r_{*}^{3}}{r}\!\Le\dfrac{1}{L_{I}^{2}}\,-\,\dfrac{1}{L_{B}^{2}}\Ra\,
\quad &r_{*}\,\leq\,r\,\leq\,R_{-}\quad\text{(boundary core, strong AdS)}\,.
\end{cases}
\eeq
ensures continuity of $\chi$ at $r\,=\,r_{*}$ fixed as
\beq\label{BBA8}
\chi_{B}(r_{*})\,=\,1\,+\,\frac{r_{*}^{2}}{L_{I}^{2}}\,=\,\chi_{I}(r_{*})\,.
\eeq
Important, that the thermal excitations $\zeta,\,\epsilon$ discussed in Appendix~\ref{AppenB}, are
defined around the second minimum $\phi\,=\,\phi_{0}$ only, hence
their natural background is the boundary core AdS patch
$\chi_{B}(r)$.
The energy momentum tensor we apply for the boundary core system as well is a sum of two parts:
\beq\label{BBA9}
T_{\mu\nu}\,=\,\langle T_{\mu\nu}\rangle_{\beta}^{\rm th}\,
+\,g_{\mu\nu}\,V_{0}^{B}\;
\eeq
the first term here provides the thermal excitations in the region and presented in Appendix~\ref{AppenB}, see \eq{AppenB28}, the last
$g_{\mu\nu}\,V_{0}^{B}$ term carries the constant negative
background of the boundary core, see \eq{AA10}. 

  Further we use the results of Appendix~\ref{AppenB}, we obtained there that the thermal excitations of the core
live in boundary core AdS patch, have the locally measured temperature defined as
\beq\label{BBA10}
T_{\rm loc}(r)\,=\,\frac{T_{\infty}}{\sqrt{\chi_{B}(r)}}\,=\,
\frac{T_{\infty}}{\sqrt{1+r^{2}/L_{B}^{2}}}\,,
\qquad r_{*}\,\leq\,r\,\leq\,R_{-}\,,
\eeq
and the following energy density of the free modes:
\beq\label{BBA11}
\rho^{(a)}_{\rm th}(r)\,=\,
\int\!\frac{d^{3}k}{(2\pi)^{3}}\,
\frac{\om^{(a)}_{k}}{\exp\!\Le \om^{(a)}_{k}/T_{\rm loc}(r)\Ra-1}\,.
\qquad
\om^{(a)}_{k}\,=\,\sqrt{k^{2}+m_{a}^{2}}\,,
\eeq
The expression recover the  black body answer for $\zeta$ (massless) mode
\beq\label{BBA12}
\rho^{(\zeta)}_{\rm th}(r)\,=\,\frac{\pi^{2}}{30}\,T_{\rm loc}^{4}(r)\,=\,
\frac{\pi^{2}}{30}\,\frac{T_{\infty}^{4}}{(1+r^{2}/L_{B}^{2})^{2}}\,,
\quad
p^{\zeta}_{\rm th}\,=\,\frac{1}{3}\,\rho^{(\zeta)}_{\rm th}\,,
\eeq
see \eq{AA35},
whereas for the massive $\epsilon$, when $T_{\rm loc}\,\ll\,m_{\epsilon}$, the \eq{AA36} expression is reproduced:
\beq\label{BBA13}
\rho^{\epsilon}_{\rm th}(r)\,\simeq\,
m_{\epsilon}^{3/2}\,T_{\rm loc}^{5/2}(r)\,
e^{-m_{\epsilon}/T_{\rm loc}(r)}\,,
\quad
p^{(\epsilon)}_{\rm th}\,\simeq\,\rho^{(\epsilon)}_{\rm th}\,
\frac{T_{\rm loc}}{m_{\epsilon}}\,\ll\,\rho^{(\epsilon)}_{\rm th}\,.
\eeq
Thus we see, that the results, after the redefinition of the local temperature value, reproduce the previous ones.

 We obtained therefore that additionally to the negative background energy density, we have now a positive contribution of the thermal excitations obtaining for the full energy density
\beq\label{BBA14}
\rho_{\rm tot}(r)\,=\,V_{0}^{B}(r)\,+\,\rho^{\zeta}_{\rm th}(r)\,+\,\rho^{\epsilon}_{\rm th}(r)\,.
\eeq
Correspondingly, in the Einstein equation we will have the following mass function:
\beq\label{BBA15}
\chi(r)\,=\,1\,-\,\frac{8\pi}{m_{p}^{2}\,r}\,\int_{0}^{r}\,r'^{2}\,
\rho_{\rm tot}(r')\,dr'\,.
\eeq
Linearizing around the boundary--core AdS background we have an additional part we add to the mass function:
\beq\label{BBA16}
\de\chi(r)\,=\,-\,\frac{8\pi}{m_{p}^{2}\,r}\,\int_{r_{*}}^{r}\,r'^{2}\,
\Le \rho^{(\zeta)}_{\rm th}(r')+\rho^{(\epsilon)}_{\rm th}(r')\Ra\,dr'\,<\,0\,.
\eeq
Since the thermal contribution is positive while the background is
negative, the magnitude of $|\rho_{\rm tot}|$ is reduced in comparison to the net background.
The effective curvature radius
\beq\label{Leff}
L_{\rm eff, B}^{2}(r)\,=\,\frac{3}{\kappa\,|\rho_{\rm tot}(r)|}\,>\,L_{B}^{2}
\eeq
is enlarged in turn in the boundary shell, i.e. the AdS phase
is locally weakened by the thermal load and the
metric in the region is pulled toward the Minkowski one.

\subsection{Interpolating energy--momentum tensor}\label{BB2}

 For the whole core system, additionally to the \eq{BBA9} EMT
\beq\label{BBB1}
T_{\mu\nu}^{\rm B}\,=\,T^{(\zeta)}_{\mu\nu}\,+\,T^{(\epsilon)}_{\mu\nu}\,
+\,g_{\mu\nu}\,V_{0}^{B}\,,
\eeq
we have to acount the EMT of the inner core
\beq\label{BBB2}
T_{\mu\nu}^{\rm I}\,=\,g_{\mu\nu}\,V_{0}^{I}(\phi_{cl})\,,
\eeq
see \eq{SBB4} above. The two expressions have a different form and behace differently, so we interpolate between them with the help of
a radial profile function $f(r)$ that smoothly increases from $0$
in the deep core to $1$ in the boundary core. The interpolation we define then determines a full EMT of the core:
\beq\label{BBB3}
T_{\mu\nu}(r)\,=\,g_{\mu\nu}\,V_{0}^{I}(r)\,+\,f(r)\,\Le g_{\mu\nu}\,\Le V_{0}^{B}(r)\,-\,V_{0}^{I}(r)\Ra\,+\,T^{(\zeta)}_{\mu\nu}\,+\,T^{(\epsilon)}_{\mu\nu}\Ra\,=\,
g_{\mu\nu}\,V_{0}(r)\,+\,f(r)\,\Le T^{(\zeta)}_{\mu\nu}\,+\,T^{(\epsilon)}_{\mu\nu}\Ra\,
\eeq
with
\beq\label{BB4}
f(r)\,=\,\frac{1}{2}\!\left[1\,+\,\tanh\!\Le \frac{r-r_{*}}{\De}\Ra\right]\,,
\quad
r_{*}\,\lesssim\,R_{-}\,,\quad\De\,\ll\,r_{*}\,
\eeq
The two parameters $r_{*}$ and $\De$ in the expression control the radial location and width of the
field--space crossover from $\phi_{cl}\,\gg\,M$ (deep core) to the
$\phi\,\sim\,\phi_{0}$ (boundary core); physically they are
determined by the classical field profile $\phi(r)$ minimizing the
full action. We emphasize that \eq{BBB3} is a phenomenological modeling ansatz, the $f(r)$ is a smooth switch function and, since the classical profile $\phi(r)$ is not solved explicitly here, the width $\De$ is to be identified with the domain-wall thickness $l_{0}$ of the transition layer introduced in Section~\ref{SecAA1}. The expression which interpolates the values of the energy density and the pressure in these two regions of the core are the following therefore:
\beqar
\rho_{\rm tot}(r)\,&=&\,V_{0}(r)\,+\,f(r)\,
\Le \rho^{(\zeta)}_{\rm th}(r)\,+\,\rho^{(\epsilon)}_{\rm th}(r)\Ra\,,
\label{BB5}\\
p_{\rm tot}(r)\,&=&\,-\,V_{0}(r)\,+\,f(r)\,
\Le p^{(\zeta)}_{\rm th}(r)\,+\,p^{(\epsilon)}_{\rm th}(r)\Ra\,.
\label{BB6}
\eeqar
In the answer, the energy and pressure of the background vacuum satisfy  $p_{\rm vac}\,+\,\rho_{\rm vac}\,=\,0$
with $w\,=\,-1$ correspondingly. The thermal excitations of the boundary AdS patch, in turn, provide different EOS for the massive and massless modes. For the massless mode we have
$p^{(\zeta)}_{\rm th}\,=\,\frac{1}{3}\,\rho^{(\zeta)}_{\rm th}$
with $w\,=\,1/3$ which is radiation EOS. For the massive mode we have 
$p^{(\epsilon)}_{\rm th}\,\ll\,\rho^{(\epsilon)}_{\rm th}$ with
$w\,\ll\,1$ which is EOS of a dilute non--relativistic (Boltzmann) gas. The total equation of
state is therefore a continuous interpolation from $w\,=\,-1$ at
$r\,\ll\,r_{*}$ to a mixed EOS with additional positive contributions added by the thermal excitations.

 Next we substitute energy density \eq{BB5} expression into the integrated Einstein equation for the mass function obtaining
\beq\label{BB7}
\chi(r)\,=\,1\,-\,\frac{8\pi}{m_{p}^{2}\,r}\,\int_{0}^{r}\,r'^{2}\,
\rho_{\rm tot}(r')\,dr'\,=\,\chi_{\rm B-I}(r)\,+\,\de\chi_{\rm th}(r)\,,
\eeq
where the background $\chi_{\rm B-I}(r)$ function is given by \eq{BBA7} expression.
The thermal correction in the expression are negative and given by
\beq\label{BB8}
\de\chi_{\rm th}(r)\,=\,-\,\frac{8\pi}{m_{p}^{2}\,r}\,\int_{0}^{R_{-}}r'^{2}\,f(r')\,
[\rho^{(\zeta)}_{\rm th}(r')+\rho^{(\epsilon)}_{\rm th}(r')]\,dr'\,<\,0\,.
\eeq
We see, that the thermal correction reduces the $\chi$ relative to the boundary AdS
background pushing it to Minkowski like geometry form.

 The next step toward the simplification of the overall picture we can make is an assumption, based on the introduced structure of the core, that the boundary core region stays in a cold regime, i.e. the EMT tensor there is dominated by the massless Goldstone modes
$\zeta$ in the Tolman redshifted ideal gas approximation. Namely we take that 
\beq\label{BB9}
T_{\infty}\,\ll\,m_{\epsilon}\,,
\eeq
obtaining that the massive radial modes $\epsilon$ are exponentially
suppressed and the boundary--core thermal stress is dominated by
the massless Goldstone $\zeta$. In this limit the source is
isotropic, $p_{r,\rm th}\,=\,p_{\perp,\rm th}\,\equiv\,p_{\rm th}$,
and the Einstein equations reduce to a single TOV like radial
problem. Then 
the energy density and pressure enter in the Einstein equation we use
\beqar\label{BB10}
\rho_{\zeta}(r)\,&=&\,\frac{\pi^{2}}{30}\,T_{\rm loc}^{4}(r)\,=\,
\frac{\pi^{2}\,T_{\infty}^{4}}{30\,\chi_{B}^{2}(r)}\,,\\
p_{\zeta}(r)\,&=&\,\frac{1}{3}\,\rho_{\zeta}(r)\,.
\label{BB11}
\eeqar
are the same as for the ideal gas massless modes
in the boundary shell.

\subsection{Mass and redshift functions of the boundary shell}\label{SecEin}\label{BB3}

 So, the picture we have now is the following.
In the deep core, when $0\,\leq\,r\,<\,r_{*}$, the energy density is provided by the
constant background $V_{0}^{(I)}\,\to\,0^{-}$ and the mass function solution is a nearly Minkowski AdS patch:
\beq\label{BB12}
\chi_{I}(r)\,=\,1\,+\,\frac{r^{2}}{L_{I}^{2}}\,\approx\,1\,,
\qquad \Phi_{I}(r)\,=\,0\,.
\eeq
The redshift function $\Phi$ vanishes because $\rho\,+\,p\,=\,0$ in this region
identically, the AdS correction
$r^{2}/L_{I}^{2}$ is very small because $L_{I}\,\to\,\infty$ as
$\phi_{cl}\,\to\,\infty$. In turn, in the boundary shell, when $r_{*}\,\leq\,r\,\leq\,R_{-}$, the energy source has
both a constant vacuum piece $V_{0}^{B}$ and a thermal piece. Moreover, there an interpolating \eq{BB7} expression for the mass function we need to use:
\beq\label{BB13}
\chi_{B}(r)\,=\,1\,+\,\frac{r^{2}}{L_{B}^{2}}\,+\,
\frac{r_{*}^{3}}{r}\!\left(\frac{1}{L_{I}^{2}}\,-\,\frac{1}{L_{B}^{2}}\right)\,-\,\frac{8\pi}{m_{p}^{2}\,r}\,\int_{r_{*}}^{r}r'^{2}\,\rho_{\zeta}(r')\,dr'\,,
\eeq
we simplified here the thermal contribution integral in comparison to the \eq{BB8} preserving only integration over the size of the patch, see \eq{AppenB11}, and taking there $f(r')\,=\,1$.
Next, accounting that $L_{I}^{2}\,\gg\,L_{B}^{2}$ and correspondingly $1/L_{I}^{2}\ll 1/L_{B}^{2}$, we can write
\beq\label{BB14}
\chi_{B}(r)\,\simeq\,1\,+\,\frac{r^{2}}{L_{B}^{2}}\,-\,
\frac{r_{*}^{3}}{r\,L_{B}^{2}}\,
-\,\frac{8\pi}{m_{p}^{2}\,r}\,\int_{r_{*}}^{r}r'^{2}\,\rho_{\zeta}(r')\,dr'\,.
\eeq
At $r\,=\,r_{*}$ the thermal integral vanishes and we obtain $\chi_{B}(r_{*})\,=\,1\,\approx\,\chi_{I}(r_{*})$,
i.e. in this approximation the $\chi$ function is continuous through the regions boundary as required.

 Now, substituting \eq{BB10} into \eq{BB14} we obtain a non-linear
integral equation for $\chi_{B}(r)$:
\beq\label{BB14001}
\chi_{B}(r)\,=\,1\,+\,\frac{r^{2}}{L_{B}^{2}}\,-\,
\frac{r_{*}^{3}}{r\,L_{B}^{2}}\,
-\,\frac{4\,\pi^{3}\,T_{\infty}^{4}}{15\,m_{p}^{2}\,r}\,
\int_{r_{*}}^{r}\frac{r'^{2}}{\chi_{B}^{2}(r')}\,dr'\,.
\eeq
The following perturbative scheme of the integral calculation then can be established then. We assume that $\rho_{\zeta}\,\ll\,|V_{0}^{B}|$, i.e. we consider a quasi-equilibrium regime
of the core evolution where the thermal corrections are small
compared with the AdS background. Therefore, there is a close form LO result for the integral we obtain:
\beq\label{BB15}
\mathcal{I}_{\rm th}(r)\,\equiv\,\int_{r_{*}}^{r}\frac{r'^{2}}{(1+r'^{2}/L_{B}^{2})^{2}}\,dr'
\,=\,\frac{L_{B}^{3}}{2}\!\left[\arctan\!\Le\frac{r'}{L_{B}}\Ra
\,-\,\frac{r'/L_{B}}{1+r'^{2}/L_{B}^{2}}\right]_{r_{*}}^{r}\,.
\eeq
providing following answer for the mass function:
\beq\label{BB16}
\chi_{B}(r)\,=\,1\,+\,\frac{r^{2}}{L_{B}^{2}}\,-\,
\frac{r_{*}^{3}}{r\,L_{B}^{2}}\,
-\,\frac{2\,\pi^{3}\,T_{\infty}^{4}\,L_{B}^{3}}{15\,m_{p}^{2}\,r}\,
\bigl[\arctan(x)-x/(1+x^{2})\bigr]_{x_{*}}^{x}\;
\eeq
with $x\,=\,r/L_{B}$, $x_{*}\,=\,r_{*}/L_{B}$. At $x\,\gg\,1$ we obtain
\beq\label{BB17}
\chi_{B}(r)\,\approx\,1\,+\,\frac{r^{2}}{L_{B}^{2}}\,-\,
\frac{r_{*}^{3}}{r\,L_{B}^{2}}\,
-\,\frac{\pi^{4}\,T_{\infty}^{4}\,L_{B}^{3}}{15\,m_{p}^{2}\,r}\,,
\eeq
the last term is assumed to be small in comparison to the second one of course.

 An another approximation we can use in the calculation of the \eq{BB14} integral it is an approximation of the geometrically thin boundary core region, i.e. the
$R_{-}\,-\,r_{*}\,\ll\,r_{*}$ approximation. In this case we take
$\chi_{B}\,\equiv\,\chi_{B}(r_{*})\,\approx\,
1\,+\,r_{*}^{2}/L_{B}^{2}$ 
and the thermal integral becomes elementary:
\beq\label{BB18}
\mathcal{I}_{\rm th}^{\rm thin}(r)\,=\,
\frac{r^{3}-r_{*}^{3}}{3\,\chi_{B}^{2}}\,.
\eeq
providing
\beq\label{BB19}
\chi_{B}^{\rm thin}(r)\,=\,1\,+\,\frac{r^{2}}{L_{B}^{2}}\,-\,
\frac{r_{*}^{3}}{r\,L_{B}^{2}}\,
-\,\frac{4\,\pi^{3}\,T_{\infty}^{4}\,(r^{3}-r_{*}^{3})}{45\,m_{p}^{2}\,r\,\chi_{B}^{2}(r_{*})}\,.
\eeq
This form makes manifest that the thermal contribution behaves as a
positive--energy "shell of radiation'' enclosed between $r_{*}$ and
$r$, generating a Schwarzschild--like $1/r$ correction with positive
mass (subtracting from $\chi$). The thermal correction to the negative mass contained in
the boundary shell is
\beq\label{BB20}
\De M_{\rm th}\,=\,\frac{4\pi}{3}\,(R_{-}^{3}-r_{*}^{3})\,
\frac{\pi^{2}\,T_{\infty}^{4}}{30\,\chi_{B}^{2}}\,>\,0\,
\eeq
then.

 Considering the $\Phi(r)$ redshift function, we notice that in the deep core region $\Phi_{I}(r)\,=\,0$ because of the $w\,=\,-1$ equation of
state there. In the boundary shell, due to the thermal excitations,  we have that $\rho_{\zeta}\,+\,p_{\zeta}\,=\,
(4/3)\,\rho_{\zeta}\,>\,0$ that provides a non-trivial form of the $\Phi_{B}(r)$ through the
\beq\label{BB21}
\frac{d\Phi}{dr}\,=\,\frac{\kappa\,r\,(\rho_{\zeta}+p_{\zeta})}{\chi(r)}\,=\,
\frac{4\,\kappa\,r\,\rho_{\zeta}(r)}{3\,\chi(r)}\,=\,
\frac{2\,\pi^{2}\,\kappa\,T_{\infty}^{4}\,r}{45\,\chi^{3}(r)}\,
\eeq
equation. Again, taking to leading order approximation
$\chi\,=\,1\,+\,r^{2}/L_{B}^{2}$, we obtain:
\beq\label{BB22}
\Phi_{B}(r)\,=\,\frac{2\,\pi^{2}\,\kappa\,T_{\infty}^{4}}{45}\,
\int_{r_{*}}^{r}\frac{r'\,dr'}{(1+r'^{2}/L_{B}^{2})^{3}}\,=\,
\frac{\pi^{2}\,\kappa\,T_{\infty}^{4}\,L_{B}^{2}}{90}\,
\!\left[\frac{1}{(1+r_{*}^{2}/L_{B}^{2})^{2}}-\frac{1}{(1+r^{2}/L_{B}^{2})^{2}}\right]\,,\,\qquad\,r_{*}\,\leq\,r\,\leq\,R_{-}\,.
\eeq
This is everywhere positive and monotonically increasing in $r$,
so $g_{tt}(r)\,=\,e^{\Phi(r)}\,\chi(r)$ acquires an additional
positive redshift factor through the boundary shell. The
corresponding locally measured temperature at the wall is
\beq\label{BB23}
T_{\rm loc}(R_{-})\,=\,\frac{T_{\infty}}{\sqrt{\chi_{B}(R_{-})\,e^{\Phi_{B}(R_{-})}}}\,
\eeq
and it matches the local temperature at $r\,=\,r_{*}$.


\section{GR description of the crust}\label{SecCC}

\subsection{The crust metric set-up}\label{SecCC1}

 The crust layer in the proposed model is located in between an inner edge at $r\,=\,R_{+}$ and the Schwarzschild horizon as an outer
edge of the crust at $r\,=\,R_{S}\,=\,2\,G\,M_{\rm ADM}$, its mass is the $M_{\rm ADM}$ of the black hole. The layer constitutes the first quasi-stable minimum of the field at $\phi\,=\,0$;
expanding the field above the minimum, an approximate form of the potential is
\beq\label{CC1}
V(\phi)\,=\,\frac{1}{2}\,m^{2}\,\epsilon^{2}\,-\,
\frac{1}{2}\,\frac{m^{2}}{M^{2}}\,\epsilon^{4}\,+\,\mathcal{O}(\epsilon^{6})\,,
\qquad
\phi\,=\,\epsilon\,,\,\,\,\epsilon\,\sim\,0\,.
\eeq
see \eq{AA5}.
The important property of the field in the crust is that the potential there is identically zero 
\beq\label{CC2}
V\,\equiv\,V(0)\,=\,0\,
\eeq
and hence there is no constant background present. The quasi-particles in the crust are
massive positive energy thermal excitations $\epsilon$, the \eq{AA1} Lagrangian of the free quasi-particles expanded around
$\phi\,=\,0$ has a standard form therefore:
\beq\label{CC3}
\el\,=\,\frac{1}{2}\,\sqrt{-g}\,g^{\mu\nu}\,
\D_{\mu}\epsilon\,\D_{\nu}\epsilon\,-\,\frac{1}{2}\,\sqrt{-g}\,m^{2}\,\epsilon^{2}_{}\,.
\eeq
The $m$ in the expression is a bare quasi-particle mass and it is different from the effective masses of the corresponding excitation mode in the core. 
The energy--momentum tensor follows from \eq{SAA1} is the standard one:
\beq\label{CC4}
T_{\mu\nu}\,=\,
\D_{\mu}\epsilon\,\D_{\nu}\epsilon\,-\,\frac{1}{2}\,g_{\mu\nu}\!\left[
g^{\al\beta}\,\D_{\al}\epsilon\,\D_{\beta}\epsilon\,-\,m^{2}\,\epsilon^{2}\right]\,
\eeq
with the trace 
\beq\label{CC5}
T\,=\,g^{\mu\nu}\,T_{\mu\nu}\,=\,
-\,g^{\al\beta}\,\D_{\al}\epsilon\,\D_{\beta}\epsilon\,+\,2\,m^{2}\,\epsilon^{2}\,.
\eeq
The introduced field we consider is stationary and spatially isotropic in a static spherically symmetric background
\beq\label{CC6}
ds^{2}\,=\,F(r)\,dt^{2}\,-\,\chi^{-1}(r)\,dr^{2}\,-\,r^{2}\,d\Omega^{2}\,,
\eeq
further, throughout the Section, it is defined that
\beq\label{CC7}
F(r)\,\equiv\,g_{00}(r)\,=\,e^{\Phi(r)}\,\chi(r)\,,
\qquad
\sqrt{F}\,=\,e^{\Phi/2}\,\sqrt{\chi}\,,
\eeq
as a shorthand for the $00$ component of the metric. We see that the radial and temporal metric components differ by the
redshift factor $e^{\Phi}$ whenever $\rho\,+\,p\,\neq\,0$. 
Here the null projection vectors are defined as $\tau^{\mu}\,=\,(F^{-1/2},0,0,0)$ and $\tau_{\mu}\,=\,(F^{1/2},0,0,0)$, see \eq{SC45} and \eq{SC47}.
Using results of Appendix~\ref{AppenC}, we see that for the free quantum field in a WKB or local Rindler type basis
\beq\label{CC8}
\hat\epsilon(x)\,=\,\sum_{n}\Le f_{n}\,e^{-i\omega_{n}t/\sqrt{F}}\,\hat a_{n}\,
+\,f_{n}^{*}\,e^{i\omega_{n}t/\sqrt{F}}\,\hat a_{n}^{\dagger}\Ra\,,
\eeq
with $\omega_{n}$ as the locally measured frequency $\omega_{loc}$ from Appendix~\ref{AppenC}.  So, using Appendix~\ref{AppenC} results, we see then that the canonical relativistic ideal gas expressions for the energy density and pressure components have the following form:
\beqar\label{CC9}
\rho_{\rm crust}(r)\,&=&\,\rho\,=\,\int\!\frac{d^{3}k}{(2\pi)^{3}}\,
\frac{\om_{k}}{e^{\om_{k}/T_{\rm loc}(r)}-1}\,,\\
p_{\rm crust}(r)\,&=&\,p\,=\,\frac{1}{3}\,\int\!\frac{d^{3}k}{(2\pi)^{3}}\,
\frac{k^{2}/\om_{k}}{e^{\om_{k}/T_{\rm loc}(r)}-1}\,,
\label{CC9001}
\eeqar
here, for the sake of simplicity, we denote $k_{loc}$ as $k$ and $\omega_{k}$ is defined through the local dispersion relation in terms of
$k_{loc}$, see \eq{C24}.
The problem we solve then is a calculation of the static, spherically symmetric metric written in terms of the problem's parameters, the task is performed in Appendix~\ref{AppenD}.

 For the mass function which determines the form of the $\chi$ radial function we have:
\beq\label{CC10}
\chi(r)\,=\,1\,-\,\frac{2\,G\,m(r)}{r}\,,\qquad
\frac{dm}{dr}\,=\,4\pi\,r^{2}\,\rho(r)\,,
\eeq
see discussion in Appendix~\ref{AppenD} about the form of the function.
The equation of hydrodynamic equilibrium, the TOV equation, has the following form in turn: 
\beq\label{CC11}
\frac{1}{F}\,\frac{dF}{dr}\,=\,2\,G\,\frac{\Le m(r)\,+\,4\pi\,r^{3}\,p(r)\Ra}{\chi(r)\,r^{2}}\,,
\eeq
equivalently it can be written as
\beq\label{CC12}
\frac{dp}{dr}\,=\,-\,\frac{1}{2}\,
\Le\rho\,+\,p\Ra\,\frac{1}{F}\,\frac{dF}{dr}\,,
\eeq
see \eq{D15} and \eq{D17}.
The local temperature we use for the distribution function, see Appendix~\ref{AppenC}, has the standard form as well:
\beq\label{CC13}
T_{\rm loc}(r)\,=\,\frac{T_{\infty}}{\sqrt{F(r)}}\,.
\eeq
We note also, that the energy density and pressure expressions, \eq{CC9}-\eq{CC10}, we can rewrite in dimensionless form, more suitable for the approximation schemes, namely
in terms of $y(r)\,\equiv\,m/T_{\rm loc}(r)$ obtaining for the energy density 
\beq\label{CC14}
\rho(r)\,=\,\frac{m^{4}}{2\pi^{2}}\,\Le y^{-4}\,I_{\rho}(y)\Ra\,,\,\,\,
I_{\rho}(y)\,\equiv\,\int_{0}^{\infty}\!\frac{x^{2}\,\sqrt{x^{2}+y^{2}}}{e^{\sqrt{x^{2}+y^{2}}}-1}\,dx\,.
\eeq
and for the the pressure correspondingly
\beq\label{CC15}
p(r)\,=\,\frac{m^{4}}{6\pi^{2}}\,\Le y^{-4}\,I_{p}(y)\Ra\,,\,\,\,
I_{p}(y)\,\equiv\,\int_{0}^{\infty}\!\frac{x^{4}/\sqrt{x^{2}+y^{2}}}{e^{\sqrt{x^{2}+y^{2}}}-1}\,dx\,.
\eeq
It must not be confused that the $m$ in the expressions is a mass of the thermal excitations defined in \eq{CC3} Lagrangian.
As usual, because of the universality of the excitation behavior above some vacuum state, there are two different limits for the
$I_{\rho}(y)$, $I_{p}(y)$ integrals. They are the following. The first one is when $y\,\ll\,1$, i.e. hot regime aka radiation regime:
\beq\label{CC16}
\rho^{\rm hot}\,=\,\frac{\pi^{2}}{30}\,T_{\rm loc}^{4}\,;\qquad
p^{\rm hot}\,=\,\frac{1}{3}\,\rho^{\rm hot}\,;\qquad
w_{\rm hot}\,=\,\frac{1}{3}\,,
\eeq
see \eq{AA35} answer.
The second limit is a cold regime when $y\,\gg\,1$, it is a non-relativistic or  dust-like state:
\beq\label{CC17}
\rho^{\rm cold}\,\simeq\,\Le \frac{m\,T_{\rm loc}}{2\pi}\Ra^{3/2}\,e^{-m/T_{\rm loc}}\,m\,;\qquad
p^{\rm cold}\,\simeq\,\Le \frac{m\,T_{\rm loc}}{2\pi}\Ra^{3/2}\,e^{-m/T_{\rm loc}}\,T_{\rm loc}\,;
\qquad w_{\rm cold}\,=\,T_{\rm loc}/m\,\ll\,1\,,
\eeq
see \eq{AA36} expression. The crossover between the two limits is at $T_{\rm loc}\,=\,m$ and it is smooth.

  The system of equations we discuss, can be reduced then to a coupled pair of first order
ODEs for the two metric functions. Namely, we use the Tolman law \eq{CC13} and write the $\rho$ and $p$ entirely in terms of the
$F\,=\,e^{\Phi}\chi$ redshift function. Then the two metric functions obey
\beqar\label{CC18}
\frac{d\chi}{dr}\,&=&\,\frac{1\,-\,\chi}{r}\,-\,\kappa\,r\,\rho(F)\,,
\\
\frac{d\Phi}{dr}\,&=&\,\kappa\,\frac{r\,}{\chi}\,\Le\rho(F)\,+\,p(F)\Ra\,;
\label{CC1801}
\eeqar
the first expression here is the differentiated Misner--Sharp relation, the second one is the redshift equation \eq{D12}. 
They are coupled only through $F\,=\,e^{\Phi}\chi$ in the
source terms. Equivalently, using \eq{D13}, the pair can be rewritten as the single integro-differential equation for $F$ alone:
\beq\label{CC19}
\frac{dF}{dr}\,=\,\frac{\kappa}{r^{2}}\,e^{\Phi(r)}\,\Le\int_{R_{+}}^{r}r'^{2}\,\rho(F(r'))\,dr'\,+\,r^{3}\,p(F(r))\Ra\,,
\eeq
here we took $C\,=\,0$, see Appendix ~\ref{AppenD}.
The boundary conditions defined at $r\,=\,R_{+}$ and $r\,=R_{S}$ provide in turn:
\beqar\label{CC20}
\Phi(R_{+})\,&=&\,\Phi_{+}\,,\qquad\chi(R_{+})\,=\,1\,,
\\
\Phi(R_{S})\,&=&\,\Phi_{S}\,,\qquad F(R_{S})\,=\,e^{\Phi(R_{S})}\,\chi(R_{S})\,=\,e^{\Phi_{S}}\,\Le 1\,-\,\frac{2\,G\,M_{\rm ADM}}{R_{S}}\Ra\,.
\label{CC2001}
\eeqar
i.e.\ at the inner edge we fix some value of the redshift function, which we match with the transition layer further; the fixing of
the outer edge form we discuss further.

  In order to discuss an approximate form of the solution of the crust metric we, first of all, define the following small parameter of the problem:
\beq\label{CC21}
\eta\,\equiv\,\frac{l_{+}}{R_{+}}\,=\,\frac{R_{S}\,-\,R_{+}}{R_{+}}\ll\,1\,,
\eeq
i.e.we assume that the width of the crust is small in comparison to the characteristic radii of the BH and core. Then, we parametrize the radial distance inside the shell as
\beq\label{CC22}
r(\eta)\,=\,R_{+}\,(1\,+\,\eta\,s)\,,\,\,\,s\,\in\,[0,1]\,;\qquad r(0)\,=\,R_{+}\,,\,\,\, r(1)\,=\,R_{S}\,
\eeq
and expand any quantity of interest with respect to the $\eta$. The boundary conditions we choose in the form that 
the outer edge of the crust at $r\,=\,R_{S}\,=\,2\,G\,M_{\rm ADM}$ will remain a
regular surface with the local temperature stays finite there:
\beq\label{CC23}
T_{\rm loc}(R_{S})\,=\,\frac{T_{\infty}}{\sqrt{F_{S}}}\,=\,T_{\infty}\,\frac{e^{-\Phi_{S}/2}}{\sqrt{\chi(R_{S})}}\,,
\eeq
see discussion further.The Misner-Sharp mass function of the
crust solution, in turn, starts from zero at the inner edge
\beq\label{CC24}
m(R_{+})\,=\,C\,=\,0\,,
\eeq
see Eq.~\eq{D10} in Appendix~\ref{AppenD}, so that the crust's gravitating mass in the model is supplied  by its integrated
thermal density:
\beq\label{CC25}
m(r)\,=\,4\pi\!\int_{R_{+}}^{r}\!r'^{2}\,\rho_{\rm crust}(r')\,dr'\,,
\qquad
m(R_{+})\,=\,0\,,\quad m(R_{S})\,=\,M_{\rm ADM}\,.
\eeq
So, now we can use a perturbative theory with respect to the $R_{+}\,\eta\,s$ parameter of \eq{CC22} in order to calculate the quantity of interest which depend on regularized 
\eq{CC23} local temperature.

\subsection{Regularization of the outer edge temperature divergence and different temperature regimes for the crust}\label{SecCC2}

 Now we choose the \eq{CC2001} outer edge of the crust boundary condition defining
\beq\label{CC26}
e^{-\Phi_{S}/2}\,\chi^{-1/2}(R_{S})\,=\,B\,
\eeq
with
\beq\label{CC26001}
T_{\rm loc}(R_{S})\,=\,T_{\infty}\,B\,
\eeq
and 
\beq\label{CC26002}
F(R_{S})\,=\,B^{-2}\,.
\eeq
The condition provides a finite and constant local temperature
to the edge, the constant $B$ is dimensionless and equal to the edge temperature measured in units of the asymptotic Tolman
temperature. So, at finite $B$ we achieve finiteness of the quantities at $r\,=\,R_{S}$ and can calculate the metric's parameters over the whole radial interval of the crust.
Next we expand the $F$ function of \eq{CC19} with respect to our parameter, obtaining to leading order
\beq\label{CC27}
F(s)\,=\,F_{0}\,+\,\eta\,F_{1}(s)\,+\,\mathcal{O}(\eta^{2})\,,
\qquad
F_{0}\,=\,F(R_{+})\,=\,e^{\Phi_{+}}\chi(R_{+})\,=\,e^{\Phi_{+}}\,,
\eeq
with
\beq\label{CC28}
m(r)\,=\,4\pi R_{+}^{3}\,\bar\rho_{\rm crust}\,\eta \,s\,+\,\mathcal{O}(\eta^{2})
\eeq
as the mass at given $r$. Then the \eq{CC19} can be written as a perturbative series differential equation, to leading order it reads
\beq\label{CC29}
\frac{dF_{1}}{ds}\,=\,\kappa\,R_{+}^{2}\,F_{0}\!\Le
\bar{\rho}_{\rm crust}\,s\,+\,p_{\rm crust}(F_{0})\Ra\,,
\eeq
where $\bar{\rho}_{\rm crust}$ is the radius--averaged density, equal
to the local $\rho_{\rm crust}(F_{0})$ to leading order. Therefore, next, integrating from $s\,=\,0$ to $s\,=\,1$, we obtain:
\beq\label{CC30}
F(R_{S})\,=\,F(R_{+})\,+\,\frac{\kappa}{2}\,R_{+}^{2}\,F(R_{+})\,\Le\bar{\rho}_{\rm crust}\,+\,2\,p_{\rm crust}(F(R_{+}))\Ra\,.
\eeq
To the same precision order, the integrated mass relation \eq{CC28} provides the following constraint at $r\,=\,R_{S}$, i.e. at $s\,=\,1$ in the case:
\beq\label{CC31}
4\pi\,R_{+}^{2}\,\bar{\rho}_{\rm crust}\,l_{+}\,=\,M_{\rm ADM}\,,
\eeq
see \eq{CC21} definition.
The thermal EOS \eq{CC16}--\eq{CC17} expresses $\rho_{\rm crust}$ and
$p_{\rm crust}$ as functions of $T_{\rm loc}\,=\,T_{\infty}/\sqrt{F}$.
The local temperature at $r\,=\,R_{S}$ is finite now
\beq\label{CC32}
T_{\rm loc}\,=\,\frac{T_{\infty}}{\sqrt{F(R_{S})}}\,=\,T_{\infty}\,B\,,
\eeq
so we can discuss next the two temperature regimes introduced in \eq{CC16}-\eq{CC17}.

 The first regime is a hot one when $T_{\rm loc}\,>\,m$, see \eq{CC16}. In this case we have:
\beq\label{CC33}
\frac{M_{\rm ADM}}{4\pi\,R_{+}^{2}\,l_{+}}\,\approx\,
\frac{\pi^{2}}{30}\,T_{\rm loc}^{4}\,
=\,\frac{\pi^{2}}{30}\,T_{\infty}^{4}\,B^{4}\,,
\eeq
obtaining
\beq\label{CC34}
l_{+}\,\simeq\,\frac{15}{2\pi^{3}}\,
\frac{M_{\rm ADM}}{R_{S}^{2}\,T_{\infty}^{4}\,B^{4}}\,
=\,\frac{15}{2\pi^{3}}\,\frac{F(R_{S})^{2}\,M_{\rm ADM}}{R_{S}^{2}\,T_{\infty}^{4}}\,,
\eeq
where the second form uses $B^{-2}\,=\,F(R_{S})$, see \eq{CC26002}. The $l_{+}$
is finite and
generically macroscopic in the case of the finite $B$. If instead we take $l_{+}\,\sim\,\ell_{P}$, i.e. minimally possible width, we obtain
\beq\label{CC35}
B^{4}\,\simeq\,\frac{15}{2\pi^{3}}\,
\frac{M_{\rm ADM}}{R_{S}^{2}\,T_{\infty}^{4}\,\ell_{P}}\,,
\eeq
i.e. the expression determines the redshift function at the outer edge for the crust of the Planck width.
In the case of the cold regime, see \eq{CC17} when $T_{\rm loc}(R)\,<\,m$, the same mass expression provides:
\beq\label{CC36}
l_{+}\,\simeq\,\sqrt{\frac{\pi}{2}}\,e^{m/(T_{\infty}B)}\,\frac{M_{\rm ADM}}{R_{+}^{2}\,m^{5/2}\,\Le T_{\infty}\,B \Ra^{3/2}}\,
\eeq
which again for finite $B$ gives finite $l_{+}$. The $l_{+}\,\sim\,\ell_{P}$ condition provides in turn:
\beq\label{CC37}
B^{3/2}\,e^{-\,m/(T_{\infty}B)}\,\simeq\,\sqrt{\frac{\pi}{2}}\,\frac{M_{\rm ADM}}{R_{+}^{2}\,m^{5/2}\,T_{\infty}^{3/2}\,\ell_{P}}\,.
\eeq
As mentioned above, the crossover between the two regimes happens at $T_{\rm loc}\,=\,m$ approximately.

 The next condition we need to check is a WKB approximation validity we used in Appendix~\ref{AppenC}.
The WKB/geometric optics averaging there requires that the dominant thermal wavelength inside the shell satisfies
\beq\label{CC38}
l_{+}\,\gtrsim\,\lambda_{T}(R_{S})\,\sim\,T_{\rm loc}^{-1}(R_{S})\,
=\,\frac{1}{T_{\infty}\,B}\,.
\eeq
We see then that the WKB/geometric optics approximation used requires a finite value of $B$; when $B\,\rightarrow\,\infty$ the whole sum over the $l$ must be performed explicitly, with a corresponding redefinition of the thermally averaged energy density and pressure in which $T_{\rm loc}$ has been used.

 Therefore, we need to estimate the value of the $B$ appears in brick-wall regime with $l_{+}\,\sim\,\ell_{P}$. The simple estimate of the value of the $B$ we can perform as following then. Consider the Planck units defined as $G\,=\,c\,=\,\hbar\,=\,k_{B}\,=\,1$, $\ell_{P}\,=\,1$, masses in Planck masses $M_{P}$. Next, assume that naturally 
\beq\label{CC3801}
T_{\infty}\,=\,T_{H}\,=\,\frac{1}{8\pi G M_{\rm ADM}}\,=\,\frac{1}{4\pi R_{S}}\,
\eeq
and using \eq{CC35} for the hot regime, we obtain consequently:
\beq\label{CC3802}
B^{4}\,=\,\frac{15}{2\pi^{3}}\,\frac{M_{\rm ADM}}{R_{S}^{2}\,T_{\infty}^{4}\,\ell_{P}}\,
=\,7680\,\pi\,M_{\rm ADM}^{3}\,,
\eeq
so that
\beq\label{CC3803}
B\,\simeq\,(7680\,\pi)^{1/4}\,M_{\rm ADM}^{3/4}\,
\approx\,12.5\,\Le\frac{M_{\rm ADM}}{M_{P}}\Ra^{3/4}\,.
\eeq
The result we obtain is that the $B$ constant grows as $M_{\rm ADM}^{3/4}$ and is large for any
astrophysical mass:
\begin{center}
\begin{table}
\begin{tabular}{lll}
\toprule
$M_{\rm ADM}$ &\,\,\, $R_{S}\,=\,2M$ &\,\,\,\,\, $B$ (hot) \\
\midrule
$1\,M_{P}$            &\,\,\, $2\,\ell_{P}$             &\,\,\, $\approx 1.2\times10^{1}$ \\
$10^{3}\,M_{P}$       &\,\,\, $2\times10^{3}\,\ell_{P}$ &\,\,\, $\approx 2.2\times10^{3}$ \\
$10^{6}\,M_{P}$       &\,\,\, $2\times10^{6}\,\ell_{P}$ &\,\,\, $\approx 3.9\times10^{5}$ \\
$10^{20}\,M_{P}$      &\,\,\, $2\times10^{20}\,\ell_{P}$&\,\,\, $\approx 1.2\times10^{16}$ \\
$1\,M_{\odot}\,(1.18\times10^{38}M_{P})$ &\,\,\, --- &\,\,\, $\approx 4.5\times10^{29}$ \\
Sgr~A$^{*}\,(\sim4\times10^{6}M_{\odot})$ &\,\,\, --- &\,\,\, $\approx 3.9\times10^{34}$ \\
\bottomrule
\end{tabular}
\label{CC38030}
\end{table}
\end{center}
The conclusion is that $B\,\to\,\infty$ with the growth of the BH mass $M_{\rm ADM}$; this means that the local edge temperature greatly exceeds the
asymptotic one, the result is expected of course. Storing the full mass $M_{\rm ADM}$ in a shell of Planckian thickness demands a very large
local energy density. So, as mentioned above, this Planck width of the crust is not compatible with the WKB approximation we use; this width demands
another way of constructing the crust's EOS, at least. In the cold regime, the solution can be explored only numerically and, as it seems, for
practically possible $M_{\rm ADM}$ masses there is no window for such a $l_{+}\,\sim\,\ell_{P}$ crust width to exist.

\subsection{Consequences of the finite $T_{\rm loc}$ temperature }\label{SecCC3}

 The finite temperature regulating boundary condition \eq{CC26} imposed at the
edge $r\,=\,R_{S}$ is equivalent to the statement that
\beq\label{CC39}
\Phi(r)\,=\,-\,\ln\!\Le B^{2}\,\chi(r)\Ra\,
=\,-\,2\ln B\,-\,\ln\chi(r)\,,
\qquad r\,\to\,R_{S}^{-}\,.
\eeq
This is the form of $\Phi$ in the vicinity of $R_{S}$ we imposed; as $\chi\,\to\,0$,
the redshift function diverges logarithmically, $\Phi\,\sim\,-\ln\chi\,\to\,+\infty$, while the observable $F\,=\,e^{\Phi}\chi\,=\,B^{-2}$ and the local temperature
$T_{\rm loc}\,=\,T_{\infty}B$ remain finite.

 The first and simplest consequence of this form of the redshift function is the following.
Differentiating \eq{CC39} and combining with the \eq{CC1801} expression we obtain:
\beq\label{CC40}
\frac{d\Phi}{dr}\,=\,-\,\frac{\chi'(r)}{\chi(r)}\,=\,\kappa\,\frac{r}{\chi(r)}\,\Le\rho(r)\,+\,p(r)\Ra\,,
\eeq
here as usual the prime denotes $d/dr$. So, at $r\,\to\,R_{S}$ we obtain:
\beq\label{CC41}
\rho(r)\,+\,p(r)\,=\,-\,\frac{\chi'(r)}{\kappa\,r}\,,
\eeq
we see that the divergence of the redshift function $\Phi$ does not affect the physical source $\rho+p$, which stays finite.
Next, let us approximate
\beq\label{CC42}
\chi(r)\,\simeq\,\varkappa\,(R_{S}-r)\,,\qquad\varkappa\,\equiv\,-\,\chi'(R_{S})\,>\,0\,,
\eeq
with $\varkappa$ as the finite magnitude of the slope at the horizon, and it provides in turn:
\beq\label{CC43}
\rho(R_{S})\,+\,p(R_{S})\,=\,\frac{\varkappa}{\kappa\,R_{S}}\,=\,\mbox{finite}\,,\qquad \kappa\,\equiv\,8\pi G\,=\,\frac{8\pi}{m_{p}^{2}}\,.
\eeq
Further simplification comes from the observation that at $r\,=\,R_{S}$ where $\chi\,\to\,0$  we can write
\beq\label{CC44}
\chi'(R_{S})\,=\,\frac{1}{R_{S}}\,-\,\kappa\,R_{S}\,\rho(R_{S})\,=\,-\,\varkappa\,,
\eeq
obtaining consequently
\beq\label{CC45}
\rho(R_{S})\,=\,\frac{1\,+\,\varkappa R_{S}}{\kappa\,R_{S}^{2}}\,.
\eeq
Therefore, from \eq{CC43}, the pressure is fixed purely by the geometry in the vicinity of the crust:
\beq\label{CC46}
p(R_{S})\,=\,\Le\rho+p\Ra\,-\,\rho\,=\,-\,\frac{1}{\kappa\,R_{S}^{2}}\,,
\eeq
which is negative and independent of $\varkappa$. The corresponding equation of state at the edge therefore is
\beq\label{CC47}
w(R_{S})\,\equiv\,\frac{p(R_{S})}{\rho(R_{S})}\,=\,-\,\frac{1}{1\,+\,\varkappa} R_{S}\,.
\eeq
As before, we can as well consider different temperature regimes for the crust in relation to the \eq{CC43} expression.
In the \eq{CC16} hot regime, we have:
\beq\label{CC50}
\rho_{\rm crust}+p_{\rm crust}\,=\,\frac{2\pi^{2}}{45}\,T_{\infty}^{4}B^{4}\,,
\qquad
\varkappa\,=\,\frac{16\pi^{3}}{45}\,G\,R_{S}\,T_{\infty}^{4}B^{4}\,,
\eeq
so that the dimensionless slope is
\beq\label{CC51}
\varkappa\, R_{S}\,=\,\frac{16\pi^{3}}{45}\,G\,R_{S}^{2}\,T_{\infty}^{4}B^{4}\,=\,\frac{4}{3}\,\frac{R_{S}}{l_{+}}\,,
\eeq
see \eq{CC34} expression. The corresponding EOS is then
\beq\label{CC52}
w(R_{S})\,=\,-\,\frac{3\,l_{+}}{3\,l_{+}\,+\,4\,R_{S}}\,\simeq\,-\,\frac{3}{4}\,\frac{l_{+}}{R_{S}}\,\rightarrow\,0^{-},
\eeq
i.e. it is of a dust-like type. When the crust excitations are non-relativistic, we have a cold regime, and the thermal source acquires the Boltzmann form:
\beq\label{CC53}
\rho_{\rm crust}\,\simeq\,m\,n\,,\qquad  p_{\rm crust}\,\simeq\,n\,T_{\rm loc}\,\ll\,\rho_{\rm crust}\,,\qquad
n\,=\,\Le\frac{m\,T_{\rm loc}}{2\pi}\Ra^{3/2}e^{-m/T_{\rm loc}}\,.
\eeq
Therefore, the source combination is dominated by the rest mass density:
\beq\label{CC54}
\rho_{\rm crust}\,+\,p_{\rm crust}\,\simeq\,m\,n\,
=\,m\,\Le\frac{m\,T_{\infty}B}{2\pi}\Ra^{3/2}e^{-m/(T_{\infty}B)}\,.
\eeq
In this case \eq{CC43} provides the following form of the dimensionless slope
\beq\label{CC55}
\varkappa\,R_{S}\,=\,2\sqrt{\frac{2}{\pi}}\,G\,R_{S}^{2}\,m^{5/2}(T_{\infty}B)^{3/2}\,e^{-m/(T_{\infty}B)}\,
\eeq
which drives
$\varkappa\, R_{S}\,\to\,0$ much faster than the power-law smallness of the
hot regime. Indeed, taking
\beq\label{CC56}
l_{+}\,=\,\sqrt{\pi/2}\,e^{m/T_{\rm loc}}M_{\rm ADM}/(R_{+}^{2}m^{5/2}T_{\rm loc}^{3/2})\,,
\eeq
see \eq{CC36}, we obtain
\beq\label{CC57}
\varkappa_{\rm cold}\,l_{+}\,=\,\frac{2\,G M_{\rm ADM}\,R_{S}}{R_{+}^{2}}\,\frac{m\,+\,T_{\rm loc}}{m}\,.
\eeq
In the cold regime when $T_{\rm loc}\,\ll\,m$, hence $(m+T_{\rm loc})/m\to1$ and $R_{+}\,\simeq\,R_{S}$, $R_{S}\,=\,2GM_{\rm ADM}$,
it gives therefore:
\beq\label{CC58}
\varkappa_{\rm cold}\,=\,\frac{1}{l_{+}}\,,
\qquad
\varkappa_{\rm cold}\,R_{S}\,=\,\frac{R_{S}}{l_{+}}\,.
\eeq
The two regimes share the same structure $\varkappa\, R_{S}\,=\,c\,R_{S}/l_{+}$, with
$c\,=\,4/3$ (hot) and $c\,=\,1$ (cold), i.e. we have:
\beq\label{CC59}
\varkappa\,=\,\frac{c}{l_{+}}\,,\qquad \varkappa\, R_{S}\,=\,c\,\frac{R_{S}}{l_{+}}\,,
\qquad
c\,=\,\tfrac{4}{3}\;(\text{hot}),\quad c\,=\,1\;(\text{cold}).
\eeq
A thinner crust forces a steeper interior slope; the same gravitating mass packed
into a smaller width makes the mass factor $\chi$ drop faster towards the horizon. The EOS we explore then are the following:
\beq\label{CC60}
w_{\rm hot}(R_{S})\,=\,-\,\frac{3\,l_{+}}{4R_{S}\,+\,3\,l_{+}}\,,
\qquad
w_{\rm cold}(R_{S})\,=\,-\,\frac{l_{+}}{R_{S}\,+\,l_{+}}\,,
\eeq
with the limits
\beqar\label{CC60001}
l_{+}\,\ll\,R_{S} &:& \varkappa\, R_{S}\,\gg\,1,\quad w(R_{S})\,\to\,0^{-}
\quad(\text{dust});
\\
l_{+}\,=\,c\,R_{S} &:& \varkappa\, R_{S}\,=\,1,\quad w(R_{S})\,=\,-\tfrac12
\quad(\text{in-between the dust and dS});
\label{CC60002}\\
l_{+}\,\gg\,R_{S} &:& \varkappa\, R_{S}\,\to\,0,\quad w(R_{S})\,\to\,-1
\quad(\text{de~Sitter}).
\label{CC60003}
\eeqar
The obtained result is simple, for any physically admissible thin crust when $l_{+}\,\ll\,R_{S}$, the
thermally sourced slope exceeds the Schwarzschild slope by the factor
$c\,R_{S}/l_{+}\,\gg\,1$, and the corresponding edge EOS type is a dust with
$w(R_{S})\,\to\,0^{-}$.

\subsection{Matching solutions from two sides of a BH horizon}\label{SecCC4}

   Due to the proposed construction of the BH, in which the redshift function in the vicinity of the horizon is non-zero from inside but vanishes from outside,
we can treat $r=R_{S}\,=\,2GM_{\rm ADM}$ as a junction surface regulated by
a proper small distance parameter $\delta r$. There are an inner, noted as $-$, crust region and an outer, noted as $+$, Schwarzschild vacuum region,
which we match together with the data in the vicinity of the horizon.
The notations must not be confused with the similar notations introduced for the crust-core system of the interior, of course. So, in this Section, we discuss
the matching of the corresponding BH quantities evaluated at $R_{S}\mp\delta r$, with the limit
$\delta r\,\to\,0$ taken at the end.

 The introduced boundary conditions \eq{CC2001} and \eq{CC26} fix the redshift function:
\beq\label{CC61}
\chi_{-}(R_{S}\,-\,\delta r)\,\simeq\,\chi_{-}(R_{S})\,+\,\varkappa\,\delta r\,=\,\varkappa\,\delta r \,,\qquad
F_{-}\,=\,e^{\Phi}\chi\,=\,\frac{1}{B^{2}}\,=\,\mbox{const}\,,\qquad
\frac{F_{-}'}{F_{-}}\,=\,0\,,
\eeq
where $\varkappa\,=\,-\chi'(R_{S})\,>\,0$ is the interior slope of the mass
factor, the $F_{-}$ function is flat and finite because of the implemented boundary condition.
On the Schwarzschild vacuum side of the construction, where $\Phi_{+}\,=\,0$, the vacuum mass function $m_{+}\,=\,M_{\rm ADM}$ and we have:
\beq\label{CC62}
\chi_{+}(R_{S}\,+\,\delta r)\,=\,1\,-\,\frac{2GM_{\rm ADM}}{R_{S}\,+\,\delta r}\,\simeq\,\frac{\delta r}{R_{S}}\,,\qquad
F_{+}\,=\,\chi_{+}\,,\qquad
\frac{F_{+}'}{F_{+}}\,=\,\frac{\chi_{+}'}{\chi_{+}}\,=\,\frac{1}{\delta r}\,.
\eeq
The exterior $F_{+}\,\to\,0$ as $\delta r\,\to\,0$, i.e. the outer side is a regular Schwarzschild horizon.

 Now we see that the mass function $\chi$ itself is continuous, but its slope and the redshift function are not:
\beqar\label{CC63}
{}[\chi]\,&=&\,\chi_{+}\,-\,\chi_{-}\,\to\,0\,,\qquad [\chi']\,=\,-\frac{1}{R_{S}}\,+\,\varkappa\,=\,\varkappa\,-\,\frac{1}{R_{S}}\,\simeq\,\frac{1}{l_{+}}\,-\,
\frac{1}{R_{S}}\,\approx\,\frac{1}{l_{+}}\,;
\\
{}[\Phi]\,&=&\,\Phi_{+}\,-\,\Phi_{-}\,=\,0\,-\,\Le -\ln(B^{2}\,\varkappa\,\delta r)\Ra\,=\,\ln\!\Le B^{2}\,\varkappa\,\delta r\Ra\,\xrightarrow{\delta r\to 0}\,-\infty\,;
\label{CC64}
\\
{}[F]\,&=&\,F_{+}\,-\,F_{-}\,=\,\frac{\delta r}{R_{S}}\,-\,\frac{1}{B^{2}}\,\xrightarrow{\delta r\to 0}\,-\frac{1}{B^{2}}\,,
\label{CC65}
\eeqar
see \eq{CC51}.
Thus we obtain that the metric has a discontinuity across the horizon.
The surface $r\,=\,R_{S}$ is therefore not an ordinary continuous $F$ junction but a membrane-like surface which separates a finite-temperature interior from the
vacuum exterior region. There is the logarithmic jump $[\Phi]\,\to\,-\infty$, which is the
infinite relative redshift between the two states, regulated by a finite $\delta r$.

 Next we consider the matching conditions between the two sides of the horizon. The horizon in this case is a time-like hypersurface of constant areal radius
defined by 
\beq\label{CC65001}
\mathcal{S}:\quad \Psi(x)\,\equiv\,r-R_{S}\,=\,0\,,\qquad
R_{S}\,=\,2GM_{\rm ADM}\,
\eeq
with induced coordinates $y^{a}=(t,\theta,\phi)$ on the hypersurface. The
normal is directed along the gradient of the defining function,
\beq\label{CC66}
n_{\mu}\,\propto\,\D_{\mu}\Psi\,=\,(\,0,\,1,\,0,\,0\,)\,,
\eeq
so it has only the radial component and the surface contains the
$\D_{t},\D_{\theta},\D_{\phi}$ directions with the radial direction
orthogonal to all of them. Normalizing with $g_{rr}=-\chi^{-1}$ we have:
\beq\label{CC67}
n_{\mu}n^{\mu}\,=\,g^{rr}n_{r}^{2}\,=\,-\chi\,n_{r}^{2}\,\equiv\,\varepsilon_{n}\,,
\eeq
We choose an outward orientation of the normal vector $n^{r}>0$ to the horizon hypersurface $\mathcal{S}$, so it has the following form:
\beq\label{CC68}
n_{\mu}\,=\,\Le\,0,\,-\chi^{-1/2},\,0,\,0\,\Ra\,,\qquad
n^{\mu}\,=\,g^{\mu\nu}n_{\nu}\,=\,\Le\,0,\,+\sqrt{\chi},\,0,\,0\,\Ra\,,\qquad
\varepsilon_{n}\,=\,n_{\mu}n^{\mu}\,=\,-1\,.
\eeq
The $\mathcal{S}$ is a time-like wall, the induced metric is
\beq\label{CC69}
h_{ab}\,=\,g_{ab}\,-\,\varepsilon_{n}\,n_{a}n_{b}\,=\,\mathrm{diag}\Le\,F,\,-r^{2},\,-r^{2}\sin^{2}\theta\,\Ra\Big|_{r=R_{S}}\,,
\eeq
with the $r$ row and $r$ column removed.
The extrinsic curvature, second fundamental form, is defined then as
\beq\label{CC70}
K_{ab}\,=\,h_{a}{}^{\mu}h_{b}{}^{\nu}\nabla_{\mu}n_{\nu}\,,
\eeq
with the $+$ sign convention, i.e. with outward normal, sphere convex outward with $K^{\theta}{}_{\theta}>0$. In coordinates adapted to
$\mathcal{S}$ this reduces, for tangential indices $a,b\in\{t,\theta,\phi\}$
and the single nonzero $n_{r}$, to
\beq\label{CC71}
K_{ab}\,=\,\nabla_{a}n_{b}\,=\,\D_{a}n_{b}-\Gamma^{c}{}_{ab}\,n_{c}\,
=\,-\,\Gamma^{r}{}_{ab}\,n_{r}\,.
\eeq
Evaluating \eq{CC71} gives the covariant components
\beq\label{CC72}
K_{00}\,=\,\frac{\sqrt{\chi}}{2}\,F'\,,\qquad
K_{\theta\theta}\,=\,-\,r\sqrt{\chi}\,,\qquad
K_{\phi\phi}\,=\,-\,r\sqrt{\chi}\,\sin^{2}\theta\,,
\eeq
and, raising with $h^{ab}$, the mixed components and trace, we obtain:
\beq\label{CC73}
K^{0}_{0}\,=\,\sqrt{\chi}\,\frac{F'}{2F}\,,\qquad
K^{\theta}_{\theta}\,=\,K^{\phi}_{\phi}\,=\,\frac{\sqrt{\chi}}{r}\,,\qquad
K\,=\,K^{0}_{0}\,+\,2\,K^{\theta}_{\theta}\,=\,\sqrt{\chi}\Le\frac{F'}{2F}+\frac{2}{r}\Ra\,,
\eeq
where $F'=dF/dr$. Every component carries the overall factor
$\sqrt{\chi}$, which vanishes at the horizon on both faces.

 On the exterior, Schwarzschild side of the horizon, the functions are given by the \eq{CC62} expressions. We have then:
\beqar
K^{0}_{0}\big|_{+}\,&=&\,\sqrt{\chi_{+}}\,\frac{F_{+}'}{2F_{+}}\,
\simeq\,\frac{1}{2}\sqrt{\frac{\delta r}{R_{S}}}\Le\frac{1}{\delta r}-\frac{1}{R_{S}}\Ra\,,
\label{CC74}
\\
K^{\theta}_{\theta}\big|_{+}\,=\,K^{\phi}_{\phi}\big|_{+}\,&=&\,
\frac{\sqrt{\chi_{+}}}{R_{S}}\,\simeq\,\frac{1}{R_{S}}\sqrt{\frac{\delta r}{R_{S}}}\,.
\label{CC75}
\eeqar
On the interior, crust, side of the horizon the functions in turn are written in \eq{CC61}, and we obtain:
\beqar
K^{0}_{0}\big|_{-}\,&=&\,\sqrt{\chi_{-}}\,\frac{F_{-}'}{2F_{-}}\,=\,0\,,
\label{CC76}
\\
K^{\theta}_{\theta}\big|_{-}\,=\,K^{\phi}_{\phi}\big|_{-}\,&=&\,
\frac{\sqrt{\chi_{-}}}{R_{S}}\,\simeq\,\frac{\sqrt{\varkappa\,\delta r}}{R_{S}}\,.
\label{CC77}
\eeqar
The interior $K^{0}_{0}$ vanishes identically because the crust has a
flat redshift block ($F_{-}=\mathrm{const}$), which makes the tension driven purely by the exterior.

 So now we can determine the jumps and the Israel relations between the introduced quantities. 
Defining $[X]\equiv X_{+}-X_{-}$, we have:
\beq\label{CC78}
[K^{0}_{0}]\,=\,\sqrt{\chi_{+}}\,\frac{F_{+}'}{2F_{+}}\,,\qquad
[K^{\theta}_{\theta}]\,=\,[K^{\phi}_{\phi}]\,
=\,\frac{1}{R_{S}}\Le\sqrt{\chi_{+}}\,-\,\sqrt{\chi_{-}}\Ra\,,
\eeq
and the surface stress follows from
\beq\label{CC79}
S^{a}_{b}\,=\,-\,\frac{1}{\kappa}\,\Le[K^{a}_{b}]\,-\,\de^{a}_{b}\,[K]\Ra\,,
\qquad
S^{a}_{b}\,=\,\mathrm{diag}\Le\si_{H},\,-\Theta_{H},\,-\Theta_{H}\Ra\,,
\eeq
providing the following answers
\beqar
\si_{H}\,&=&\,\frac{2}{\kappa R_{S}}\,\Le\sqrt{\chi_{+}}\,-\,\sqrt{\chi_{-}}\Ra\,,
\label{CC80}
\\
\Theta_{H}\,&=&\,-\,\frac{1}{\kappa}\,\Bigg[\sqrt{\chi_{+}}\,\frac{F_{+}'}{2F_{+}}\,+\,\frac{1}{R_{S}}\Le\sqrt{\chi_{+}}-\sqrt{\chi_{-}}\Ra\Bigg]\,.
\label{CC81}
\eeqar
We see therefore that the surface energy density has the following form in our construction:
\beq\label{CC82}
\si_{H}\,=\,-\,\frac{2}{\kappa\,R_{S}}\,\Le\sqrt{\frac{\delta r}{R_{S}}}\,-\,\sqrt{\varkappa\,\delta r}\Ra\,
=\,\frac{2\sqrt{\delta r}}{\kappa\,R_{S}^{3/2}}\,\Le\sqrt{\varkappa\, R_{S}}\,-\,1\Ra\,\xrightarrow{\delta r\to 0}\,0\,.
\eeq
The result we obtained is that the surface energy density vanishes as $\delta r\,\to\,0$. The membrane carries no localized
rest energy in the limit, and this is consistent with the mass factor $\chi$ being continuous across $R_{S}$.

 Surface tension, in turn, is dominated by the exterior term:
\beq\label{CC83}
\Theta_{H}\,=\,-\,\frac{1}{\kappa}\,\Bigg[\sqrt{\frac{\delta r}{R_{S}}}\,\frac{1}{2\delta r}
\,+\,\frac{1}{R_{S}}\Le\sqrt{\frac{\delta r}{R_{S}}}\,-\,\sqrt{\varkappa\delta r}\Ra\Bigg]\,=\,
-\,\frac{1}{2\kappa\,\sqrt{\delta r\,R_{S}}}\,+\,\mathcal{O}(\sqrt{\delta r})\,.
\eeq
We see that it diverges as $\delta r^{-1/2}$ because of the
exterior factor $\sqrt{\chi_{+}}\,F_{+}'/(2F_{+})\,=\,\tfrac12\sqrt{\chi_{+}}/\delta r$,
i.e.\ because of the redshifted surface gravity of the Schwarzschild horizon on the outer edge. Again we notice that the interior contributes nothing to the divergence because there $F$ is flat. The divergence of $\Theta_{H}$ is a coordinate/redshift artifact and can be removed
by the introduction of the proper tension $\Theta_{H}^{\rm proper}\,=\,\Theta_{H}\,\sqrt{F_{+}}$ measured by a
static observer on the surface:
\beq\label{CC84}
\Theta_{H}^{\rm proper}\,=\,\Theta_{H}\,\sqrt{F_{+}}\,
=\,\Theta_{H}\,\sqrt{\frac{\delta r}{R_{S}}}\,\xrightarrow{\delta r\to 0}\,-\,\frac{1}{2\kappa\,R_{S}}\,.
\eeq
The scale of the proper tension is set by the Schwarzschild surface
gravity then. With $F_{+}\,=\,\chi_{+}\,=\,1-R_{S}/r$ we have
\beq\label{CC85}
\kappa_{\rm sg}\,=\,\tfrac12\,|F_{+}'(R_{S})|\,=\,\frac{1}{2R_{S}}\,,\qquad
T_{H}\,=\,\frac{\kappa_{\rm sg}}{2\pi}\,=\,\frac{1}{4\pi R_{S}}\,,
\eeq
so that
\beq\label{CC86}
\Theta_{H}^{\rm proper}\,=\,-\,\frac{1}{2\kappa\,R_{S}}\,=\,-\,\frac{\kappa_{\rm sg}}{\kappa}\,=\,-\,\frac{T_{H}}{4G}\,,
\eeq
the result is that the membrane tension is fixed by the horizon temperature, and that the regulated
surface is a kind of tension membrane whose proper tension is equal to one quarter of the
Hawking temperature in Planck units.

 The whole horizon construction therefore has a simple structure. From outside at $r\,=\,R_{S}$ there is a regular event horizon with
$F_{+}\,\to\,0$, surface gravity $\kappa_{\rm sg}\,=\,1/2R_{S}$ and 
Hawking temperature $T_{H}\,=\,1/4\pi R_{S}$. The horizon, in turn, is dressed by a
massless membrane with finite proper tension that mediates between the interior aka crust and the vacuum exterior.


\section{Transition layer description}\label{SecDD}

 In this Section we discuss the transition layer, we can name also as a wall, placed between the core and crust at
\beq\label{DD1}
R_{-}\,\leq\,r\,\leq\,R_{+}\,,\qquad l_{0}\,\equiv\,R_{+}-R_{-}\,
\eeq
region. 
Throughout the whole interior the metric we use is defined in the static spherically symmetric form:
\beq\label{DD2}
ds^{2}\,=\,F(r)\,dt^{2}\,-\,\chi^{-1}(r)\,dr^{2}\,-\,r^{2}\,d\Omega^{2}\,,
\qquad
F(r)\,\equiv\,e^{\Phi(r)}\,\chi(r)\,
\eeq
with different values of the introduced functions of course, in particular for the transition layer we write
\beq\label{DD3}
ds^{2}\,=\,F_{w}(r)\,dt^{2}\,-\,\chi_{w}^{-1}(r)\,dr^{2}\,-\,r^{2}\,d\Omega^{2}\,,
\qquad
F_{w}(r)\,\equiv\,e^{\Phi_{w}(r)}\,\chi_{w}(r)\,,
\eeq
with the Misner-Sharp mass function $m_{w}(r)$ defined by
\beq\label{DD4}
\chi_{w}(r)\,=\,1\,-\,\frac{2\,G\,m_{w}(r)}{r}\,.
\eeq
In general, the wall is the region located at vicinity of the potential's sign change from positive, $V>0$ crust side,
to negative, $V<0$, core side, value. Or, roughly defining, this is a part of the potential which connects it's maximum with the second minimum.
We denote further the negative overall mass of the core as $m_{-}<0$, then it is assumed that the wall consists $|m_{-}|>0$ mass plus an additional negative mass
provided by the negative part of the potential included in the layer. The overall mass of the layer then is positive,
at $r\,=\,R_{+}$ a zero gravity hypersurface is present by construction therefore, see further.
The Einstein equations we have for the layer are the same as in Appendix~\ref{AppenD}, introducing isotropic fluid like EMT
$T^{\mu}_{\nu}=\diag(\rho_{w},-p_{w},-p_{w},-p_{w})$, they are
\beqar
\frac{dm_{w}}{dr}\,&=&\,4\pi\,r^{2}\,\rho_{w}(r)\,,
\label{DD5}
\\
\frac{F_{w}'}{F_{w}}\,&=&\,\frac{d\Phi_{w}}{dr}\,+\,\frac{\chi_{w}'}{\chi_{w}}\,
=\,2\,G\,\frac{m_{w}(r)\,+\,4\pi\,r^{3}\,p_{w}(r)}{r^{2}\,\chi_{w}(r)}\,,
\label{DD6}
\\
\frac{dp_{w}}{dr}\,&=&\,-\,\Le\rho_{w}+p_{w}\Ra\,\frac{F_{w}'}{2F_{w}}\,,
\label{DD7}
\eeqar
where \eq{DD6} is the redshift equation, identical to \eq{D14} of
Appendix~D, and \eq{DD7} is the TOV balance.

 There are two following interfaces we have then in the interior:
\beq\label{DD8}
\Si_{-}\,:\,r=R_{-}\quad(\text{boundary core}\,\to\,\text{wall})\,,
\qquad
\Si_{+}\,:\,r=R_{+}\quad(\text{wall}\,\to\,\text{crust})\,.
\eeq
Correspondingly, the task is invented for the determination of the two following sets of wall's data.
The first  concerns the junction conditions we have, namely it is about the continuity of the induced
metric and the Israel jump of the extrinsic curvature at $\Si_{-}$ and
$\Si_{+}$, which determine the surface energy densities and tensions which carries by the
wall. The second is about the possible form of the metric inside the wall, i.e. about the
$F_{w}(r),\,\chi_{w}(r)$ defined in $[R_{-},R_{+}]$ interval. The metric, of course, must be determined from the Einstein equations
sourced by the wall's EMT we define.

  The interior solutions we match are known and they are the following. The first one is form the core boundary side, $r\leq R_{-}$, it is given by \eq{BB14}:
\beq\label{DD9}
\chi_{B}(r)\,=\,1+\frac{r^{2}}{L_{B}^{2}}
+\frac{r_{*}^{3}}{r}\!\Le\frac{1}{L_{I}^{2}}-\frac{1}{L_{B}^{2}}\Ra\,-\,\frac{8\pi}{m_{p}^{2}\,r}\!\int_{r_{*}}^{r}\!r'^{2}\rho_{\zeta}\,dr'\,,
\eeq
the $\Phi_{B}(r)$ is defined by \eq{BB21}-\eq{BB22} expressions. At the inner interface we have therefore:
\beq\label{DD10}
\chi_{B}(R_{-})\,=\,1+\frac{R_{-}^{2}}{L_{B}^{2}}+\dots\,>\,1\,,
\qquad
F_{B}(R_{-})\,=\,\chi_{B}(R_{-})\,e^{\Phi_{B}(R_{-})}\,.
\eeq
At the crust side at $r\geq R_{+}$, taking the \eq{D10} with $C\,=\,0$ from Appendix~D, we have: 
\beq\label{DD11}
m(r)\,=\,4\pi\!\int_{R_{+}}^{r}r'^{2}\rho_{\rm crust}\,dr'\,.
\eeq
At $r\,=\,R_{+}$ edge, the crust mass function starts from zero therefore:
\beq\label{DD12}
m_{\rm crust}(R_{+})\,=\,0\,,\qquad
\chi_{\rm crust}(R_{+})\,=\,1\,,\qquad
F_{\rm crust}(R_{+})\,=\,e^{\Phi_{+}}\,=\,1
\eeq
with $\Phi_{+}\equiv\Phi(R_{+})\,=\,0$ is the inner-edge redshift equal to zero because the crust
base carries zero enclosed mass. At the inner enclosed surface the signed Misner--Sharp mass  is negative in turn:
\beq\label{DD13}
m_{\rm core}(R_{-})\,=\,\frac{R_{-}}{2G}\Le1-\chi_{B}(R_{-})\Ra\,<\,0\,.
\eeq
Important that this is only a part of the negative core, i.e it is the mass enclosed at
$R_{-}$, not the whole core mass $-|m_{-}|$. The whole core mass is the
separate quantity $m_{-}<0$ of Appendix~D, residing at $r<R_{-}$ in the inner
and boundary core.
The introduced wall straddles the potential barrier, so the wall density $\rho_{w}(r)$ changes sign at an interior radius $r_{0}\in(R_{-},R_{+})$ where
\beq\label{DD14}
\rho_{w}(r_{0})\,=\,0\,,\qquad
\rho_{w}(r)\,<\,0\;\;(R_{-}<r<r_{0})\,,\qquad
\rho_{w}(r)\,>\,0\;\;(r_{0}<r<R_{+})\,,
\eeq
the inner sub-shell carrying the captured negative core tail and the outer
sub-shell the ordinary (positive-energy) barrier matter. The wall mass integral
therefore splits into two contributions of opposite sign,
\beq\label{DD15}
\Delta m_{w}\,\equiv\,4\pi\!\int_{R_{-}}^{R_{+}}\!r^{2}\rho_{w}\,dr\,
=\,4\pi\!\int_{r_{0}}^{R_{+}}\!r^{2}\rho_{w}\,dr\,+\,4\pi\!\int_{R_{-}}^{r_{0}}\!r^{2}\rho_{w}\,dr\,=\,|m_{-}|\,-\,m_{0}\,>\,0\,;
\qquad |m_{-}|\,>\,0\,,\;m_{0}\,>\,0\,,
\eeq
which defines $|m_{-}|$ as the positive sub-shell integral on
$[r_{0},R_{+}]$ alone and $-m_{0}$ as the negative sub-shell integral on
$[R_{-},r_{0}]$. 

  The two interfaces we introduced are time-like hypersurfaces of constant areal radius:
\beq\label{DD16}
\Si_{\pm}:\quad \Psi_{\pm}(x)\equiv r-R_{\pm}\,=\,0\,,
\eeq
with induced coordinates $y^{a}=(t,\theta,\ph)$. The unit normal directed
toward increasing $r$ (outward), normalised with $g^{rr}=-\chi$, is
\beq\label{DD17}
n_{\mu}\,=\,\Le0,\,-\chi^{-1/2},\,0,\,0\Ra\,,\qquad
n^{\mu}\,=\,\Le0,\,+\sqrt{\chi},\,0,\,0\Ra\,,\qquad
n_{\mu}n^{\mu}\,=\,-1\,,
\eeq
identical in structure to \eq{CC68}. The induced metric is
\beq\label{DD18}
h_{ab}=\diag(F,-r^{2},-r^{2}\sin^{2}\theta)|_{r=R_{\pm}}\,,
\eeq
and the extrinsic curvature $K_{ab}=h_{a}^{\mu}h_{b}{}^{\nu}\nabla_{\mu}n_{\nu}$, as above in \eq{CC71}, has the following nonzero mixed components
\beq\label{DD20}
K^{0}_{0}\,=\,\sqrt{\chi}\,\frac{F'}{2F}\,,\qquad
K^{\theta}_{\theta}\,=\,K^{\ph}_{\ph}\,=\,\frac{\sqrt{\chi}}{r}\,,\qquad
K\,=\,\sqrt{\chi}\Le\frac{F'}{2F}+\frac{2}{r}\Ra\,,
\eeq
in agreement with \eq{CC73}. The jump notation we use is $[X]\equiv X_{\rm out}-X_{\rm in}$ and they are the same as defined above. The corresponding surface
stress tensor is also the standard one:
\beq\label{DD21}
S^{a}_{b}\,=\,-\,\frac{1}{8\pi G}\Le[K^{a}_{b}]-\de^{a}_{b}\,[K]\Ra\,,
\qquad
S^{a}_{b}\,=\,\diag\Le\si,\,-\Theta,\,-\Theta\Ra\,.
\eeq
It's projection onto the diagonal blocks gives the surface energy density and tension 
\beqar
\si\,&=&\,S^{t}_{t}\,=\,\frac{1}{4\pi G}\,[K^{\theta}_{\theta}]\,
=\,\frac{1}{4\pi G\,R}\Le\sqrt{\chi_{\rm out}}\,-\,\sqrt{\chi_{\rm in}}\Ra\,,
\label{DD22}
\\
\Theta\,&=&\,-\,S^{\theta}_{\theta}\,=\,-\,\frac{1}{8\pi G}\Le[K^{t}_{t}]\,+\,[K^{\theta}_{\theta}]\Ra\,
=\,-\,\frac{1}{8\pi G}\Bigg[\Big[\sqrt{\chi}\,\frac{F'}{2F}\Big]\,+\,\frac{1}{R}\Le\sqrt{\chi_{\rm out}}-\sqrt{\chi_{\rm in}}\Ra\Bigg]\,,
\label{DD23}
\eeqar
evaluated at particular $r=R$ for the interface in question.

\subsection{Inner and outer interfaces: $\Si_{-}$ and $\Si_{+}$}\label{SecDD1}

 The request of the continuity of the induced metric at $r=R_{-}$ provides
\beq\label{DD24}
\chi_{w}(R_{-})\,=\,\chi_{B}(R_{-})\,=\,1\,+\,\frac{R_{-}^{2}}{L_{B}^{2}}\,+\,\dots\,,
\qquad
F_{w}(R_{-})\,=\,F_{B}(R_{-})\,=\,\chi_{B}(R_{-})\,e^{\Phi_{B}(R_{-})}\,,
\eeq
which through \eq{DD4} fixes the negative inner wall mass \eq{DD13},
$m_{w}(R_{-})\,=\,m_{\rm core}(R_{-})<0$.
If functions $\chi$ and $F$ join the boundary-core solution smoothly, i.e. without  thin shell, then $[\chi']\,=\,[F']\,=\,0$ as well and the 
conjunction is stress-free, $\si_{-}\,=\,\Theta_{-}\,=\,0$ and the transition is carried entirely by the smooth bulk
metric of the wall we discuss further. In the thin wall idealisation, with $\chi$ continuous but the slopes jumping, \eq{DD22}-\eq{DD23} give
\beqar
\si_{-}\,&=&\,\frac{1}{4\pi G\,R_{-}}\Le\sqrt{\chi_{w}(R_{-})}-\sqrt{\chi_{B}(R_{-})}\Ra\,=\,0\,,
\label{M2}
\\
\Theta_{-}\,&=&\,-\,\frac{1}{8\pi G}\Bigg[
\sqrt{\chi_{w}(R_{-})}\,\frac{F_{w}'(R_{-})}{2F_{w}(R_{-})}\,
-\,\sqrt{\chi_{B}(R_{-})}\,\frac{F_{B}'(R_{-})}{2F_{B}(R_{-})}\Bigg]\,,
\label{DD25}
\eeqar
where $\si_{-}=0$ because $\chi$ is continuous at $\Si_{-}$. The inner conjunction thus carries no surface rest energy and any localized stress is a pure tension fixed by
the mismatch of redshift gradients $F'/F$ between the boundary-core and wall
solutions. In this case, the $\Si_{-}$ is a domain-wall type surface and not a mass shell.
The second interface continuity, when $r=R_{+}$ and the crust data is given by \eq{DD12}, fixes in turn:
\beq\label{DD26}
\chi_{w}(R_{+})\,=\,\chi_{\rm cr}(R_{+})\,=\,1\,,
\qquad
F_{w}(R_{+})\,=\,F_{\rm cr}(R_{+})\,=\,1\,,
\eeq
so through \eq{DD4} the condition $\chi_{w}(R_{+})=1$ enforces
\beq\label{DD27}
m_{w}(R_{+})\,=\,0\,,
\eeq
consistent with the crust constant $C=0$ introduced in Appendix~\ref{AppenD}. 
Since $\chi$ is continuous and equal to unity on both sides, \eq{DD22} gives a
vanishing surface energy density,
\beq\label{DD28}
\si_{+}\,=\,\frac{1}{4\pi G\,R_{+}}\Le\sqrt{\chi_{\rm cr}(R_{+})}-\sqrt{\chi_{w}(R_{+})}\Ra\,=\,0\,,
\eeq
so $\Si_{+}$ is a massless conjuction as well. It's tension follows from the jump of
$K^{0}_{0}$, the wall side contributes nothing by null gravity,
$K^{0}_{0}|_{w}=\sqrt{\chi}\,F_{w}'/(2F_{w})|_{R_{+}}=0$ and the crust side provides
\beq\label{DD29}
K^{0}_{0}\big|_{\rm cr}\,=\,\sqrt{\chi}\,\frac{F_{\rm cr}'}{2F_{\rm cr}}\bigg|_{R_{+}}
\,=\,4\pi G\,R_{+}\,p_{\rm crust}(R_{+})\,.
\eeq
Now, because
\beq \label{DD30}
[K^{0}_{0}]\,=\,4\pi G R_{+}\,p_{\rm crust}(R_{+})\,,\qquad 
[K^{\theta}_{\theta}]=0\,;
\eeq
the Israel tension carried by the outer conjunction is
\beq\label{DD31}
\Theta_{+}\,=\,-\,\frac{1}{8\pi G}\,[K^{0}_{0}]\,
=\,-\,\frac{1}{2}\,R_{+}\,p_{\rm crust}(R_{+})\,.
\eeq
If it is assumed that the inner-edge pressure is small, then $\Theta_{+}$ is finite and of order
$R_{+}\,|p_{\rm crust}(R_{+})|$. Like the horizon membrane of previous Section,
$\Si_{+}$ is a massless tension surface, i.e. $\si_{+}=0$, $\Theta_{+}\neq0$ and it is the
junction-level realization of the null-gravity hypersurface, see further.

\subsection{Metric of the transition layer}\label{SecDD2}

 Now we construct the explicit static spherically symmetric metric which describes obtained above results.
First of all, we introduce the layer parametrization through
\beq\label{DD35}
s\,\equiv\,\frac{r-R_{-}}{l_{0}}\,\in\,[0,1]\,,\qquad
r(s)\,=\,R_{-}+l_{0}\,s\,,\qquad l_{0}=R_{+}-R_{-}\,,\qquad r(0)\,=\,R_{-}\,,\qquad r(1)\,=\,R_{+}\,.
\eeq
Inside $[R_{-},R_{+}]$ the scalar climbs and descends the potential barrier,
$V(\phi_{w})$ running from $0^{+}$ at the crust minimum through the maximum
$V_{\rm top}>0$ toward the negative core background. The wall source is the
gradient stress of the static$\phi=\phi_{w}(r)$ profile: 
\beqar
\rho_{w}\,&=&\,\tfrac12\,\chi_{w}\,(\phi_{w}')^{2}\,+\,V(\phi_{w})\,,
\label{DD36}
\\
p_{r,w}\,&=&\,\tfrac12\,\chi_{w}\,(\phi_{w}')^{2}\,-\,V(\phi_{w})\,,
\label{DD37}
\\
p_{\perp,w}\,&=&\,-\tfrac12\,\chi_{w}\,(\phi_{w}')^{2}\,-\,V(\phi_{w})\,,
\label{DD38}
\eeqar
the standard anisotropic thin-wall stress with $\rho_{w}+p_{r,w}=
\chi_{w}(\phi_{w}')^{2}\geq0$ driving the redshift, and effective gravitational
source $\rho_{w}+p_{r,w}+2p_{\perp,w}=-2V(\phi_{w})$ that changes sign at the
null-gravity point.

  The mass function we discuss then must satisfy the boundary values
$m_{w}(R_{-})=m_{\rm core}(R_{-})=-(|m_{-}|-m_{0})$ and $m_{w}(R_{+})=0$. It's  deepest value $-|m_{-}|$, in turn, is an attaining
at the density sign change radius $r_{0}$, i.e. where $\rho_{w}=0$ and hence $m_{w}'=0$. It also must have $m_{w}'(R_{\pm})=0$ so,
definining  $s_{0}\equiv(r_{0}-R_{-})/l_{0}$, we can write the following representative profile for the function:
\beq\label{DD39}
m_{w}(s)\,=\,
\begin{cases}
-\,(|m_{-}|-m_{0})\,-\,m_{0}\,h\!\Le\dfrac{s}{s_{0}}\Ra\,, & 0\leq s\leq s_{0}\,,\\[8pt]
-\,|m_{-}|\,+\,|m_{-}|\,h\!\Le\dfrac{s-s_{0}}{1-s_{0}}\Ra\,, & s_{0}\leq s\leq 1\,,
\end{cases}
\qquad
h(u)\,\equiv\,3u^{2}-2u^{3}\,,
\eeq
which descends from $m_{w}(0)=-(|m_{-}|-m_{0})$ to the minimum
$m_{w}(s_{0})=-|m_{-}|$ and rises back to $m_{w}(1)=0$, with $m_{w}'$ vanishing
at $s=0,s_{0},1$. The corresponding density
\beq\label{DD40}
\rho_{w}(r)\,=\,\frac{m_{w}'(r)}{4\pi r^{2}}\,
=\,\begin{cases}
\displaystyle -\,\frac{6\,m_{0}}{4\pi r^{2}\,l_{0}\,s_{0}}\,
\frac{s}{s_{0}}\Le1-\frac{s}{s_{0}}\Ra\,<\,0\,, & 0<s<s_{0}\,,\\[10pt]
\displaystyle +\,\frac{6\,|m_{-}|}{4\pi r^{2}\,l_{0}\,(1-s_{0})}\,
\frac{s-s_{0}}{1-s_{0}}\Le1-\frac{s-s_{0}}{1-s_{0}}\Ra\,>\,0\,, & s_{0}<s<1\,,
\end{cases}
\eeq
changes sign at $r_{0}$ as required by \eq{DD14}. The radial metric function is the following then:
\beq\label{DD41}
\chi_{w}(r)\,=\,1\,-\,\frac{2G\,m_{w}(r)}{r}\,,
\eeq
with $m_{w}(r)$ from \eq{DD39}, it interpolates 
\beq\label{DD42}
\chi_{w}(R_{-})\,=\,1\,-\,\frac{2G\,m_{\rm core}(R_{-})}{R_{-}}\,=\,\chi_{B}(R_{-})\,>\,1\,,
\qquad
\chi_{w}(R_{+})\,=\,1\,=\,\chi_{\rm cr}(R_{+})\,,
\eeq
and reaching its maximum $\chi_{w}(r_{0})=1+2G|m_{-}|/r_{0}$ at the sign-change
radius.

 The pressure in the wall could be fixed by requiring the Tolman--Komar active mass to interpolate
to zero at the null-gravity edge. With
\beq\label{DD43}
\mathcal{M}_{\rm Komar}(r)\,\equiv\,m_{w}(r)\,+\,4\pi r^{3}\,p_{w}(r)\,,
\qquad
\mathcal{M}_{\rm Komar}(R_{+})\,=\,0\,,
\eeq
a convenient closed representative with the correct endpoints behavior is the following:
\beq\label{DD44}
\mathcal{M}_{\rm Komar}(r)\,=\,\mathcal{M}_{\rm Komar}(R_{-})\,\Le1-s\Ra^{2}\Le1+2s\Ra\,,
\qquad
p_{w}(r)\,=\,\frac{\mathcal{M}_{\rm Komar}(r)-m_{w}(r)}{4\pi r^{3}}\,,
\eeq
giving $p_{w}(R_{+})=0$ automatically by \eq{DD27} and \eq{DD35}. The redshift function can be determined by integrating of \eq{DD6}, it
has the following form:
\beq\label{DD45}
\Phi_{w}(r)\,=\,\Phi_{B}(R_{-})\,+\,2G\!\int_{R_{-}}^{r}\!
\frac{\mathcal{M}_{\rm Komar}(r')}{r'^{2}\,\chi_{w}(r')}\,dr'\,-\,
\ln\!\frac{\chi_{w}(r)}{\chi_{w}(R_{-})}\,,
\qquad
F_{w}(r)\,=\,e^{\Phi_{w}(r)}\,\chi_{w}(r)\,.
\eeq
We see that the integrand vanishes at $r=R_{+}$ where $s\,=\,1$, see \eq{DD44} definition above. The inner value
$\Phi_{w}(R_{-})=\Phi_{B}(R_{-})$ secures continuity of $F$ at $\Si_{-}$, while
$F_{w}(R_{+})=1$ matches the crust value \eq{DD12} with $\Phi(R_{+})=0$.

\subsection{Hypersurface of zero gravity}\label{SecDD3}

 The hypersurface of zero gravity, we introduced above, is the time-like hypersurface
$\mathcal{Z}:\,r=R_{+}$ on which the proper acceleration vanishes. Namely, by construction it is placed at $r=R_{+}$ and a static observer
on it feels no proper acceleration. It means the following, with 
\beq\label{DD46}
\tau^{\mu}=(F^{-1/2},0,0,0)
\eeq 
the only
acceleration component is 
\beq\label{DD47}
a^{r}=\tfrac12\,\chi\,F'/F\,,\qquad |a|\,\equiv\,\sqrt{-g_{rr}\,(a^{r})^{2}}\,=\,\frac{\sqrt{\chi}\,|F'|}{2F}\,,
\eeq
it definesthe proper magnitude of the force per unit mass needed to hold the observer
static. So the no-acceleration condition we have is
\beq\label{DD48}
a^{r}(R_{+})\,=\,0\quad\rightarrow\quad F_{w}'(R_{+})\,=\,0\,.
\eeq
Using \eq{DD27} condition in the  redshift equation \eq{DD6}, we obtain:
\beq\label{DD49}
\frac{F_{w}'(R_{+})}{F_{w}(R_{+})}\,
=\,2\,G\,\frac{m_{w}(R_{+})+4\pi R_{+}^{3}\,p_{w}(R_{+})}{R_{+}^{2}}\,=\,0\,
=\,8\pi G\,R_{+}\,p_{w}(R_{+})\,\quad\rightarrow\quad p_{w}(R_{+})\,=\,0\,,
\eeq
and, therefore, correspondingly
\beq\label{DD50}
\mathcal{M}_{\rm Komar}(R_{+})\,\equiv\,m_{w}(R_{+})+4\pi R_{+}^{3}\,p_{w}(R_{+})\,=\,0\,.
\eeq
So, the edge is the locus where attraction from the positive-mass layer of the upper part of the wall  and
repulsion from the whole negative-energy core mass  are canceled.  This is also the "factory"  surface of
maximal density anticipated in Section~\ref{SecAA1}.

  From the outer interface analysis of Subsection~\ref{SecDD1}-Subsection{DD2}, the surface $\Si_{+}$ coinciding with massless conjunction $\si_{+}=0$ \eq{DD28},
because $\chi$ is continuous and equal to unity on both sides. The tension on it 
is fixed by the jump of $K^{0}_{0}$
\beq\label{DD51}
\Theta_{+}\,=\,-\,\frac{1}{8\pi G}\,[K^{t}{}_{t}]\,
=\,-\,\frac{1}{2}\,R_{+}\,p_{\rm crust}(R_{+})\,,
\eeq
see \eq{DD31}. We can define the hypersurface of zero gravity as a special membrane therefore, it carries no surface energy but only a finite tension
$\Theta_{+}$ sourced entirely by the crust pressure at the inner edge. So the three characteristic properties of the hypersurface, which are 
\eq{DD48}, \eq{DD50} and that $\si_{+}=0$ with $\Theta_{+}$ finite, are
equivalent and define the hypersurface from the bulk, the source and the junction
points of view respectively.

 The behavior of the scalar particles at vicinity of the $\mathcal{Z}$ zero gravity hypersurface is clarified in Appendix ~\ref{AppenE}.
In terms ofthe quasi-particles, the result depends on the sign of the effective mass which is negative for the quasi-particles we discuss and positive for the
normal state matter. What we obtained is that there are stable oscillations in the system and runaway behaviour, see \eq{E16}. The physical picture then is that 
the positive matter is trapped in the redshift well at $\mathcal{Z}$ and accumulates there, making the $\mathcal{Z}$  as surface of maximal regular matter
density. A negative energy quasi-particles, which populate core, in turn are expelled from the $\mathcal{Z}$. They are repelled
inward in direction toward the core if displaced inward and they are repelled outward in direction toward the horizon $R_{S}$ if they are displaced outward. We see that the surface sorts the two sorts of particles. Important that the outward flow of quasi-particles directed toward the
horizon, belong to the crust layer then and therefore they fate is set by the crust dynamics and interactions with the ordinary matter in the crust, not by
$\mathcal{Z}$ itself. Also we underline that 
these results are statements about the static background, in a dynamical regime the $k_{\rm eff}$  drifts together with
$\mathcal{M}_{\rm Komar}'(R_{+})$ and the trap/expulsion rates which track the growth or
evaporation of the hole.

 In the proposed picture the $\mathcal{Z}$ is a stable attractor for positive mass matter, the ordinary matter accumulates against it from both
sides. Therefore, the regular matter density $\rho(r)$ has a maximum at $\mathcal{Z}$ and then decreases on either side of the $\mathcal{Z}$:
\beq\label{DD52}
\rho(R_{+})\,=\,\rho_{\max}\,,\qquad
\rho'(R_{+})\,=\,0\,,\qquad
\rho''(R_{+})\,<\,0\,.
\eeq
Namely, the density gradient we have points toward $\mathcal{Z}$ from both sides, so we consider the $\mathcal{Z}$ as the surface of the maximal
regular-matter density with maximum $\rho_{\max}\equiv\rho_{\rm cr}(R_{+})$ attained on $\mathcal{Z}$ itself. 
As stated above, we assumed that the quasi-particle creation requires the local density to exceed some critical value $\rho_{c}$, i.e. the introduced "factory"
begins operate when 
\beq\label{DD53}
\rho_{\rm cr}(r)\,\geq\,\rho_{c}\,.
\eeq
Given the monotonic profile \eq{DD52}, the condition \eq{DD53} supposed to take place in a thin
shell adjacent to $\mathcal{Z}$ and centered on the density maximum:
\beq\label{DD54}
\mathcal{S}_{c}:\quad R_{+}-\de_{\rm in}\,\leq\,r\,\leq\,R_{+}+\de_{\rm out}\,,
\qquad
\rho(R_{+}-\de_{\rm in})\,=\,\rho(R_{+}+\de_{\rm out})\,=\,\rho_{c}\,.
\eeq
Here the inner edge $R_{+}\,-\,\de_{\rm in}$ is in the wall positive sub-shell with
$r_{0}<R_{+}\,-\,\de_{\rm in}$ and the outer edge $R_{+}+\de_{\rm out}$ is in the crust.
Expanding about the maximum, $\rho(r)\simeq\rho_{\max}\,-\,\tfrac12|\rho''(R_{+})|(r-R_{+})^{2}$,
the threshold $\rho\,=\,\rho_{c}$ gives to leading order the followinh equal half-widths of this thin layer:
\beq\label{DD55}
\de_{\rm in}\,\simeq\,\de_{\rm out}\,\equiv\,\de\,
=\,\sqrt{\frac{2\,(\rho_{\max}-\rho_{c})}{|\rho''(R_{+})|}}\,,
\eeq
positive whenever $\rho_{\max}>\rho_{c}$, i.e. whenever the density at $\mathcal{Z}$ is supercritical. 
We notice, that beyond the leading order, the wall side and crust side curvatures
differ, so $\de_{\rm in}\neq\de_{\rm out}$ in general.

 Consider now the quasi-particle creation provided by \eq{E14}-\eq{E15} expressions.
Introducing a positive displacement
\beq\label{DD56}
x_{q}=r_{q}\,-\,R_{+}\,>\,0\,
\eeq
which describe the quasi-particles move outward, i.e. in direction to the crust, we can write the solution of the equation of motion for them 
at vicinity of the $\mathcal{Z}$ as 
\beq\label{DD57}
x(\tau)\,\simeq\,x_{0}\,\cosh(\Om\,\tau)\,.
\eeq
As mentioned, it describes a quasi-particle moves in the direction of the decreasing density. So, tt reaches the outer edge of
the creation shell, $x=\de_{\rm out}$, after the following proper time approximately:
\beq\label{DD58}
\tau_{\rm dis}\,=\,\frac{1}{\Om}\,\mathrm{arccosh}\!\Le\frac{\de_{\rm out}}{x_{q}}\Ra\,
\xrightarrow[x_{q}\ll\de_{\rm out}]\,\frac{1}{\Om}\,\ln\!\frac{2\de_{\rm out}}{x_{q}}\,.
\eeq
In this construction, the local density is decreasing to $\rho_{c}$  at $r\,=\,R_{+}+\de_{\rm out}$ and beyond it we have 
$\rho\,<\,\rho_{c}$, so the creation condition \eq{DD53} fails and the quasi-particle can
no longer be sustained, it dissociates inside the crust at
\beq\label{DD59}
r\,=\,R_{+}+\de_{\rm out}\,;\quad \rho\,\sim\,\rho_{c}\,
\eeq
reheating it by the process.
Of course, there are also quasi-particles which move toward the core and condense in it changing the overall core's depth.

 Finally we notice in this Section, that the location $r\,=\,R_{+}$ and the \eq{DD50} balance equation were obtained in the case of the static,
spherically symmetric background constructed in Sections~\ref{SecBB}--\ref{SecDD}.
In the evolving black hole the surface is only approximately null-gravity, with 
infalling fluxes present, there is a shift of $R_{\pm}$. Then also the crust pressure $p_{\rm crust}(R_{+})$
drifts with the temperature regime of Section~\ref{SecCC2}, and $\mathcal{Z}$ follows to the moving balance point. In the model, the growth $\dot M\,>\,0$ and evaporation $\dot M\,<\,0$ regimes correspond to the inward/outward imbalance of fluxes across
$\mathcal{Z}$ which we discussed in Section~\ref{SecA}.


\section{Results and conclusion}\label{SecEE}

 In the article we generalize the ideas proposed in \cite{BondBH}, which concern the black hole interior structure, to the case of the relativistic scalar field in the framework of general relativity.
The problem is very complex, for different reasons; therefore, first of all, we begin with a short summary of the main conceptual ideas and proposals behind the presented calculations, the technical details of these ideas we presented mainly in Section~\ref{SecAA}. There we did not consider a GR background, discussing instead the framework in the flat space-time in order to clarify its main properties and ideas.

 The first and most important idea of the proposed construction is a proposal that, talking about a BH, it is constructive to discuss a physical object with finite physical characteristics\footnote{See for example different views on the BH in \cite{Alt1,Alt2,Alt3,Alt4,Alt5,Alt6}.}. In general, we can discuss the quantum gravity properties of a black hole, its reinterpretation as an exotic object, but it is always useful to strive to model a black hole as a physical object without singularities and/or infinite parameters characterizing its properties. Of course, any reformulation of the BH interior behavior must then be performed in terms of known finite physics\footnote{There are possible quantum gravity effects which can affect the BH interior properties, but we assume that a possible account of these effects will as well not make the related physics infinite.}. Therefore, the first proposition of \cite{BondBH} was to avoid the
notion of an infinitely dense state of matter, introducing instead a notion of a new matter state of negative energy density created through the channels of a macroscopic gravitational collapse and
microscopic attraction between the matter's constituents in the no-escape conditions of the BH interior. The discussions concerning the possibility of this matter state can be found in 
\cite{Zeld1,Zeld2} for example, but of course it yet remains speculative, we do not know precisely about a possible existence and microscopic description of such states. 
Nevertheless, from a physical point of view, nothing special is proposed.
The introduced matter state reproduces an AdS EOS\footnote{In relativistic covariant QFT, a negative energy density of the state inevitably leads to a positive pressure value, i.e. to the AdS vacuum state if we talk about a vacuum.} and, preventing the infinite-density paradox, it also cancels the possible singularity scenario for the BH evolution by introducing a negative energy density
core in the center of the BH interior\footnote{In \cite{Int3}, a negative pressure interior was introduced for the same purposes. In relativistic QFT negative constant pressure means positive constant energy density, i.e. it is a dS interior type which as well helps to avoid a singularity.}. Moreover, if we consider the BH in general and its interior as a patch of space-time where we need to pack the largest amount of energy in the minimum possible volume,
the negative energy density is the solution; the negativity, through matter rebinding, enlarges the amount of energy we can put inside.
Thus, the proposed inversion of the sign of the energy density turns the entire problem of the black hole interior into a complex but solvable task, completely formulated in terms of the corresponding equation of state, thermodynamics and general relativity. The proposed idea, in turn, raises additional questions. The main ones are the following: what is the structure of the interior then, and what is the matter content of the proposed structure. Namely, the idea of the negative energy core immediately dictates a many-layer structure of the whole interior, the BH is an object of a positive ADM mass and we need a place in the interior for the regular mass location. Then, the simplest construction we can discuss is a three-layer BH interior which consists of the negative energy core, the crust layer which provides the external ADM mass, and the transition layer which provides a transition between the two previous ones, see Fig.~\ref{Interior}. 
In this framework, the black hole appears not as a ``black box'' object with unknown interior but as a physical system whose properties can be defined and calculated from first principles. The further clarification of the BH structure then requires a definition of what kind of matter populates the interior.

 In \cite{BondBH} we considered two types of quasi-particles in the interior instead of a full microscopic matter description of it. For the crust we proposed that for the clarification of the regular matter
behavior there it is enough to consider scalar quasi-particles above some regular vacuum state; in general we assume that already in the crust the main interactions between these constituents
are attractive. In turn, for the core, we consider scalar composite quasi-particles with only binding potential energy and zero kinetic energy present, 
which provide a negative energy density vacuum in the core. This special state has very special thermodynamical properties, including a negative temperature parameter, which describe a negative energy density value and its dynamics; a detailed description of these quasi-particle thermodynamics 
can be found in \cite{BondBH}. Yet, the two-type picture of the quasi-particles introduced in the previous paper was not fully satisfactory. First of all, the framework there was not relativistic and was not defined on a GR background; secondly, it operated with a flat--band EOS, see \cite{FB,FB1,FB2},
which is a highly unusual EOS in relativistic physics; thirdly, the negative pressure appearing there was a result of a regularization of the phase space integrals for the particles with no kinetic energy.
Because of these reasons, in the present manuscript we performed the task differently. Namely, in the absence of a fundamental microscopic physics of the crust and core, we introduced a relativistic scalar 
field with a non-linear self-interaction potential whose properties depend on the density of the field, see Fig.~\ref{Poten}. At lower densities, corresponding to the crust, the potential describes a normal zero-valued vacuum state and then massive thermal excitations on top of it; these excitations undergo attractive interactions in the crust and the whole system is quasi-stable therefore. When the density is growing, the corresponding part of the scalar field potential describes a negatively valued potential well with thermal fluctuations about this second minimum present; in this case the excitations are of two types, they are massive and massless. This region of the potential we defined as a boundary core region, it is a boundary which provides a heat interaction with the normal matter crust. When the density is growing further we face a deep core region, there the scalar field has no kinetic energy but negative and close to zero potential energy, this is a genuine AdS deep core vacuum region. 
This field density change correlates with the change of radial coordinate, the field density grows from normal matter in the vicinity of the Schwarzschild radius in the crust to the center of the AdS inner
vacuum core at $r\,\sim\,0$ distances.
The description of all these layers of the interior, as well as the regular matter behavior in the crust and in the core, is achieved by only one QFT scalar field Lagrangian then. This construction drastically simplifies the technical calculations in the task and does not obscure the essence of the problem we consider.

 The proposed construction of the interior was analyzed from the point of view of classical thermodynamics in \cite{BondBH} and in the present article. The reason for that is simple and two-fold.
The first reason for the classical thermodynamic introduction is that in the paper we consider static solutions, making only a snapshot of the system at some given moment of time. It is not quite informative in the sense that, in general, the BH interior is a dynamical system which changes with time and the changes can be essential. Then classical thermodynamics can be extremely useful in the clarification of the evolution of the system depending on the system's instantaneous states. In this sense, the thermodynamic laws replace the description of the dynamics of the system's evolution. The second reason is that simple thermodynamics perfectly clarifies the snapshot properties of the complex system consisting of different parts. Namely, we describe a two-temperature thermodynamic system for the interior. The negative potential of the core is described by the negative temperature parameter coupled to the negative potential energy, and the positive thermal excitations in the core and crust are described in terms of a regular temperature; the properties of such a two-temperature system are unusual and classical thermodynamics clarifies the behavior of such a system.

 The main thermodynamic property of the system constructed this way is that it is unstable with respect to heat/matter inward fluxes or heat/matter outward leaks. Reformulating, we can say that this interior system, not being fully isolated, is unstable from the thermodynamical point of view. The instability appears in the interior in two opposite directions. When the inward flow contribution is the dominant one, it leads to a ``charging'' of the core, i.e. the core grows both in radial size and in the absolute value of the negative potential well\footnote{It is interesting to note that this behavior has some similarity with the steady-state Universe scenario of \cite{Ho1,Ho2,Ho3,Ho4,Ho5}.}. The core in this mechanism plays the role of a heat reservoir which absorbs the inward heat/matter flows, whereas the crust's purpose is to deliver the incoming heat/matter to the special hypersurface where the transition to the negative energy density takes place, the hypersurface we named a ``factory'' before, see above and further. When, in turn, the leading heat flow is directed outward, by the Hawking radiation process, the core, aka heat reservoir, begins to ``discharge'', feeding the matter of the crust and preserving a quasi-static flow of the evaporation. This direction of the thermodynamic process evolution is underlined by two 
factors. The first one is the general direction of the AdS space-time evolution, see for example \cite{NEM}, and the second is the heat flow direction between the negative and positive temperature sub-systems, which is always directed from the negative to the positive one. Therefore, without the ``charging'' but with only Hawking radiation caused by crust processes, the core will release heat through the boundary to the crust, preserving the quasi-stationarity of the outgoing radiation. In this sense, the whole interior works as a kind of heat machine which heats the vacuum exterior of the BH at the cost of the energy stored initially in the core.

 It is crucial that in this framework the interior is not a vacuum anymore but consists of layered space-time regions with different EOS and corresponding EMT. Indeed, the next step performed is a consideration of the layers' matter as sources of the corresponding gravity equations, separately for each one, with the transition and matching between the layers introduced. The first region we considered then is the deep core region of Section~\ref{SecAB}. The GR solution there is the simplest one, the potential provides a negative constant value of energy density, in the static approximation, and the gravity solution of the whole region is an AdS patch given by
the \eq{SBB8} expression. As is seen from \eq{SBB9}, the radius of this AdS space-time is large because of its $1/\varepsilon$ behavior, therefore
the patch, aka the central region of the BH, is close to the flat Minkowski space-time form. Of course, as we underlined already, this is a static picture. In general the region changes dynamically in both processes of growing or evaporation of the BH. Yet, we assume that whereas the boundary of this inner core is changing, see 
the \eq{AA57}-\eq{AA59} discussion, there is still an inner AdS core present from some not-very-initial stages of collapse until some not-very-final BH evaporation states.

 Going radially, the next layer we have is a boundary core region which connects the core with the crust and conducts the heat between the regions, through the transition layer of course. This region is not an AdS patch any more. More precisely, it is an AdS region plus a positive contribution from the thermal excitations in it, see 
\eq{BBA9} and \eq{BBA14}. It leads to the softening of the AdS regime, the thermal excitations push the AdS radius to a larger value and the whole region locally 
to a flatter one. Of course, because we now have two regions of the core with different properties, we need to interpolate between them, it is done in Subsection~\ref{BB2}. In general, the metric there, because of the matter content, has a non-zero redshift function, see \eq{BB21}-\eq{BB22}, and this is an important result we obtained. The positive-mass thermal excitations provide a non-vacuum form of the Schwarzschild solution inside the black hole regions.

 The outermost layer we introduced is the crust. First of all, because the core has an overall negative energy density, the crust carries the $M_{\rm ADM}$ mass of the BH. The negative mass of the core is then compensated by the positive and equal mass of the transition layer placed between the crust and core. Therefore, 
in the proposed construction, all the $M_{\rm ADM}$ mass is concentrated in the crust, whose width is supposed to be much smaller than the radial size of the core. The crust then is a matter-like state with positive and massive thermal excitations on top of a zero-valued vacuum. This matter presence in the crust leads to non-trivial physical consequences in the model. First of all, the metric inside the crust remains regular, the usual sign flip of the radial and temporal metric components does not occur inside the crust, it takes place only at $r\,=\,R_{S}$, where the interior crust solution is matched to the exterior Schwarzschild vacuum. There, until $r\,=\,R_{S}$, the metric is regular and defined by the usual form with mass function and
non-trivial redshift functions through the corresponding EOS, see the Subsection~\ref{SecCC1} discussion. Yet the $\chi$ function at $r\,=\,R_{s}$ is divergent, and so is the
corresponding local temperature appearing in the thermal EOS. Therefore, we introduce a new, dimensionless parameter $B$, aka the redshift regularization parameter, through the \eq{CC26} definition. It regularizes the local temperature of the crust at $r\,=\,R_{S}$ and determines the value of the redshift function there. This parameter relates the crust's width and other characteristics of the BH at different temperature regimes, see \eq{CC34} and \eq{CC36}. When the crust's width approaches the Planck value this parameter becomes infinite, see \eq{CC3803} and the discussion after the equation. This regularization parameter appears not only in the definition of the local temperature and redshift function. Using the \eq{CC39} expression, we obtain then an EOS for the crust in the vicinity of its outer boundary at $r\,\to\,R_{S}$. So it allows
us to clarify the behavior of the crust's matter there and, see \eq{CC60001}-\eq{CC60003}, in the thin crust approximation it is simply a dust EOS. What is also very important is that we now have two different Schwarzschild solutions on the two sides of the horizon, which must be matched through it, see \eq{CC63}-\eq{CC65} jumps of the corresponding functions.
The result of that is a discontinuity of the metric across the horizon, and $r\,=\,R_{S}$ is a membrane-like surface which separates the matter of the crust from the vacuum of the exterior. This surface has no energy density, the $\chi$ is continuous across it, but it has the surface tension given by the \eq{CC84} expression,
valid for a static observer.

 The important part of the whole construction is the transition layer between the crust and core discussed in Section~\ref{SecDD}. The appearance of this wall is required by the presence of the different types of matter in the crust and core we have, it must interpolate between them. Then its description is a standard one 
and similar to the other Schwarzschild-type solutions we discussed in the paper. The difference is only that now we have two matching hypersurfaces, for the wall-core and crust-wall systems, and we need a metric which interpolates between these two edges of the wall. The task is an algebraic one, the metric is presented in 
Subsection~\ref{SecDD2}. The unusual property of the introduced transition layer is a zero-gravity hypersurface located as the boundary between the crust and this wall. 
Because of the mass bookkeeping we implemented for the whole interior, there is a special hypersurface on which the gravity is zero, namely, the total mass enclosed by it, which is the negative core mass together with the positive part of the wall's mass interior to it, vanishes there. This special hypersurface is proposed to be the ``factory'' we introduced and discussed in 
Sections~\ref{SecA}--\ref{SecAA}. The main property of this hypersurface is that it attracts the positive-mass normal matter and repels the effective negative-mass particles. The maximum matter density is achieved there therefore, and this is the hypersurface where initially the negative-mass quasi-particles, in the terminology of 
\cite{BondBH}, are produced and from which they afterwards condense into the core. We analyze the behavior of the matter in the vicinity of the hypersurface in Appendix~\ref{AppenE}; the appearance of such a ``factory'' is a non-trivial consequence of the proposed model of the interior.

 Let us now note some important issues that we did not address in the manuscript, partly due to the already large length of the article and partly because we need to discuss these topics separately. The first interesting and important issue that we mentioned but did not discuss in the article is the issue of a possible microscopic description of negative energy density states. Whereas in quantum mechanics such virtual states are normal, in the paper we consider a classical negative energy state which arises because of a very special combination of matter density, temperature, microscopic attraction channel and self-gravitating collapse process. At some physical intuitive level, such states are understandable; they represent a kind of elementary molecule of a new state of matter, when the attraction, aka binding potential between the constituent parts of this molecule, is so strong and the kinetic energy is so small that a negative energy density is effectively obtained.
However, it is important to test the possibility of creating such and other exotic states of matter using the rules of microscopic physics, adapted, of course, for the special macroscopic conditions that we discuss in the article. Therefore, the issues related to these states of matter form an important direction of future investigation of black hole physics.

 Another important problem, which we only touched on in Appendix~\ref{AppenA}, is the problem of the description of the non-stationarity of the BH interior.
In general, we have formalisms intended to describe non-equilibrium thermodynamics and condensed matter physics, see for example \cite{Zub,Zub1,Zub2}. Nevertheless, the non-stationary description of the interior is a different and very complex task. As we demonstrated in Appendix~\ref{AppenA}, even a simple modification of the stationary Lagrangian leads to a time-dependent mixed EOS and correspondingly to a time-dependent metric. Taking into account that we have to bookkeep not one region but many layers of the non-equilibrium interior, the task becomes even more complicated. This investigation direction could be very important for the definition of the mechanics
and thermodynamics of the BH interior as well. We have to mention, nevertheless, that the problem may be investigated by quasi-equilibrium methods and approaches which describe the properties of all sub-systems of the interior and their characteristic times. Indeed, the simplest analogous example we have is a moving car. The car is built from many parts and in general has an extremely complicated structure, but in order to describe its motion we need to know only a few important things. We must account for the fact that the car must be charged during some period of time, by fuel or electricity; that the discharge takes some period of time, generally much larger than the charging; we have to know the simplest principles of the car's engine, and we need to know the kinetic principles of the car's motion. The analogy with the BH as a heat engine is then straightforward. We introduce the core, aka heat reservoir; it takes some period of time to enlarge it, this is the analogue of charging; we have the Hawking radiation which leads to the BH evaporation which we can consider as a discharge, it usually takes a very long period of time; basing on the thermodynamics of the interior we can assume how the heat is transferred between the BH layers, aka the engine work principles. After that we can obtain the characteristics of the BH evolution which in turn can be verified or falsified by observations. This direction of investigation is an interesting one, it allows one, from first principles concerning the interior, to describe also the external properties of the BH.

 In this sense, the proposed form of the structure of the crust, for example, allows to discuss an origin of the BH temperature and entropy from the very first principles. Let's consider a very short example of calculations can be done in the proposed framework. We assume, for simplicity, that the crust's density is homogeneous
\beq\label{Conc1}
\rho\,\equiv\,\rho_{\rm crust}\,=\,
\frac{M_{\rm ADM}}{V_{\rm crust}}\,,\qquad
V_{\rm crust}\,=\,\frac{4\pi}{3}\Le R_{S}^{3}-R_{+}^{3}\Ra\,
\eeq
and denote the crust width as before by
\beq\label{Conc2}
\ell_{0}\,\equiv\,R_{S}\,-\,R_{+}\,,\qquad R_{+}\,=\,R_{S}-\ell_{0}\,.
\eeq
Next we consider a thin film sitting on the top of the crust with mass equal approximately to
\beq\label{Conc3}
\delta M\,=\,4\pi R_{S}^{2}\,\rho\,\ell_{f}\,
=\,\frac{3\,M_{\rm ADM}\,R_{S}^{2}}{R_{S}^{3}-R_{+}^{3}}\,\ell_{f}\,,
\eeq
here $\ell_{f}$ is a widt of the film.
The amount of mass which is interior to the film, to leading order, in this the thin-film limit is the full crust mass:
\beq\label{Conc4}
M_{\rm below}\,\simeq\,m(R_{S})\,=\,M_{\rm ADM}\,.
\eeq
For sake of simplicity, consider the Newtonian gravitational potential energy of the film in the field of the mass beneath, it is
\beq\label{Conc5}
E_{\rm gr}\,=\,-\,\frac{G\,M_{\rm below}\,\delta M}{R_{S}}\,
=\,-\,\frac{3\,G\,M_{\rm ADM}^{2}\,R_{S}\,}{R_{S}^{3}\,-\,R_{+}^{3}}\,\ell_{f}\,.
\eeq
Expanding the denominator of \eq{Conc5}
\beq\label{Conc6}
R_{S}^{3}\,-\,R_{+}^{3}\,=\,3R_{S}^{2}\ell_{0}\Le
1\,-\,\frac{\ell_{0}}{R_{S}}\,+\,\frac{1}{3}\frac{\ell_{0}^{2}}{R_{S}^{2}}\Ra\,,
\eeq
we obtain for the gravitational surface energy:
\beq\label{Conc7}
\sigma_{gr}\,=\,\frac{\abs{E_{\rm gr}}}{4\pi R_{S}^{2}}\,=\,\frac{1}{8\pi R_{S}^{2}}\,M_{\rm ADM}\,\Le 1\,-\,\frac{\ell_{0}}{R_{S}}\,+\,\frac{1}{3}\frac{\ell_{0}^{2}}{R_{S}^{2}} \Ra^{-1}\,
\frac{\ell_{f}}{\ell_{0}}\,=\,\frac{1}{16\pi R_{S}}\,\Le 1\,-\,\frac{\ell_{0}}{R_{S}}\,+\,\frac{1}{3}\frac{\ell_{0}^{2}}{R_{S}^{2}} \Ra^{-1}\,\frac{\ell_{f}}{G\,\ell_{0}}\,,
\eeq
here the 
\beq\label{Conc8}
\frac{G M_{\rm ADM}}{R_{S}}\,=\,\frac{M_{\rm ADM}}{2}
\eeq
identity was used.
Considering this energy as a thermal one with a Hawking temperature
\beq\label{Conc9}
\sigma_{gr}\,=\,n_{f}\,T_{H}
\eeq
we, assuming that for the introduced film
\beq\label{Conc10}
n_{f}\,=\,\frac{1}{4\,G}\,\frac{\ell_{f}}{\ell_{0}}\,,
\eeq
obtain immediately
\beq\label{Conc11}
T_{H}(\ell_{0})\,=\,\frac{1}{4\pi R_{S}}\Le 1\,+\,\frac{\ell_{0}}{R_{S}}\,+\,\frac{2}{3}\frac{\ell_{0}^{2}}{R_{S}^{2}}+\dots\Ra\,.
\eeq
Of course, the calculation we made is a pretty rough estimate of what we can consider as the Hawking temperature. It is important, nevertheless, the this value is finite because of
the Subsection~\ref{SecCC2} regularization of redshift function introduced that allows to the corresponding temperature to stay finite. The further exploration of the 
BH properties basing on the proposed matter states of the interior we postpone for the additional publications.

 Another direct consequence of the proposed three-layer BH structure is that the radius of the black hole of the same mass can, in general, be different from 
the "normal" BH.
Indeed, because the black hole mass in the model is placed fully in the crust and there is an additional negative energy density core, it is not really required that the radius of this BH, 
which we denote as $R_{3L}$, will be the same as $R_{S}\,=\,R_{reg}$, where by $R_{reg}$ we denote the regular value of the Schwarzschild radius. The question is under which conditions the material radius $R_{3L}$ of the crust coincides with $R_{reg}$ at fixed $M_{\rm ADM}$. Therefore, we have to understand whether there is a window for the size of these three-layer black holes. Introduce then the geometric defect parameter and the core fraction as follows:
\beq\label{Conc12}
\delta \,\equiv\, \frac{\Rreg}{\RtL}\,, \qquad u \,\equiv\, \frac{\Rcore}{\RtL}\in[0,1], \qquad \frac{\RtL\,-\,\Rcore}{\RtL}\,=\,1\,-\,u\, .
\eeq
Using the critical density expression for the regular BH
\beq\label{Conc13}
\rhocr \,=\, \frac{3\Madm}{4\pi R_{S}^{3}} 
\eeq
we have then
\beq\label{Conc14}
\frac{\rhocrust}{\rhocr}\,=\, \frac{R_{S}^{3}}{\RtL^{3}-\Rcore^{3}}\,=\,\frac{\Rreg^{3}}{\RtL^{3}\left(1-u^{3}\right)}\,=\,\frac{\delta^{3}}{1-u^{3}}\,. 
\eeq
Solving the expression for the $\delta$ provides the result in terms of the two densities only:
\beq\label{Conc15}
\delta \,=\, \frac{\Rreg}{\RtL}\,=\,\left[\frac{\rhocrust}{\rhocr}\left(1-u^{3}\right)\right]^{1/3}\,,\qquad u=\frac{\Rcore}{\RtL}\,.
\eeq
The density ratio $\rhocrust/\rhocr$ is assumed to run from $1$, at the limit the crust is dilute as the
regular BH, up to $\rhoP/\rhocr$ which denotes a Planck-density crust.
We see therefore, that the three-layer BH we discuss, reproduces the regular radius, i.e. $\delta=1$ ration, along the one-parameter locus
\beq\label{Conc16}
\rhocrust\left(1-u^{3}\right)\,=\,\rhocr\,.
\eeq
In particular, for an extremelly thin core configurations, $u$the match is exact when
$\rhocrust=\rhocr$: the three-layer object degenerates into the regular black
hole. Above the locus ($\delta>1$) the object is over-compact, $\RtL<R_{S}$;
below it ($\delta<1$) it is ``puffy'', $\RtL>R_{S}$. The whole family collapses
onto the single combination $\rhocrust(1-u^{3})$---crust density times crust
volume fraction---so two very different three-layer configurations share the
same defect whenever this product agrees.
\begin{figure}[h]
\centering
\includegraphics[width=\textwidth]{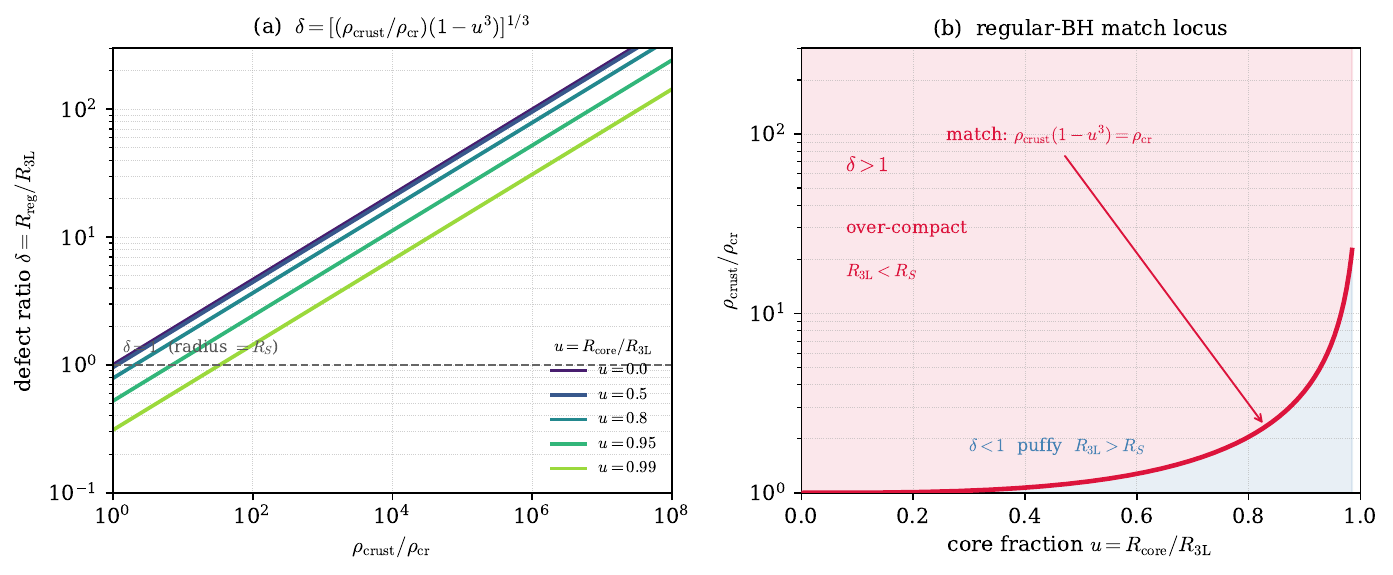}
\caption{\textbf{(a)} Defect ratio $\delta=\Rreg/\RtL$ versus the density contrast $\rhocrust/\rhocr$, for several
core fractions $u=\Rcore/\RtL$; the cube-root scaling and the $\delta=1$
reference (dashed) are shown. \textbf{(b)} The regular-BH match
locus in the $(u,\ \rhocrust/\rhocr)$ plane, separating the
over-compact region ($\delta>1$, $\RtL<R_{S}$) from the defective/puffy region
($\delta<1$, $\RtL>R_{S}$).}
\label{fig:defect}
\end{figure}
We see that there is a set of BHs with $\RtL=\Rreg=R_{S}$ exists. Them, when the core is large, the
crust must become denser to hold all of $\Madm$ in the shrinking shell volume.
The limits we have are transparent therefore.
When  $u=0$, i.e. no core structure, the $\rhocrust=\rhocr$  and the object is the regular
black hole. Whereas we consider small $u$ with $\rhocrust\simeq\rhocr\,(1+u^{3})$ , there is a small rise of the density.
With $u\to 1$, i.e. with core fills almost everything, we have $\rhocrust\to\infty$ with
vanishingly thin crust needs arbitrarily high density.
Because $\rhocrust$ cannot exceed the Planck density $\rhoP$, the family exists
only up to a maximal core fraction where the crust saturates that bound,
\beq\label{Conc17}
\frac{\rhocr}{1\,-\,u_{\max}^{3}}\,=\,\rhoP\qquad\rightarrow\qquad u_{\max}\,=\,\left(1\,-\,\frac{\rhocr}{\rhoP}\right)^{1/3}.
\eeq
Since $\rhocr=3/(32\pi G^{3}\Madm^{2})\propto\Madm^{-2}$, it is far below $\rhoP$ for any astrophysical mass. The $u_{\max}$ then lies extremely close to unity, i.e. the core
can occupy essentially the whole object while still maintaining $\delta=1$, until the crust would require super-Planckian density.
The conclusion is simple then, the three-layer BH can looks as regular BH with $R_{3L}\,=\,R_{S}\,=\,R_{reg}$ for any given $M_{ADM}$. 

 The next natural question to ask, therefore, is about the possible observation signatures of the proposed complex interior structure of a BH.
This is a subject of separated research, of course, but we want here to indicate on the possible directions to approach the problem.
We note, that in the description of the interior we have a parameter which can appear in corrections to the leading order answers of some calculable quantities. 
Consider Subsection~\ref{SecCC2} and the dimensionless quantity introduced there:
\beq\label{Conc18}
\ep\,\equiv\,\frac{1}{B^{2}}\,=\,F(\RS)\,=\,\frac{\Tinf^{2}}{\Tloc^{2}}\,\ll\,1\,
\eeq
is the natural expansion parameter, we have $\ep\,\to\,0$ when $B\,\to\,\infty$ is the Schwarzschild limit for the crust in which the crust redshift becomes infinite and the finite-redshift surface becomes a true horizon, see discussion in Subsection~\ref{SecCC2}. Since $B\to\infty$ with growing $\Madm$, the $\ep$ is small for large astrophysical masses and every observable is expected to take the form
\beq\label{Conc19}
\mathcal{O}_{\rm 3L}=\mathcal{O}_{\rm Schw}(\Madm)\,
\bigl[\,1+c_{1}\,\ep^{p}+\dots\bigr],
\eeq
with coefficients built from the crust/membrane data introduced in Section~\ref{SecCC}, i.e. from  $B$, $\varkappa$, the surface tension $\Theta_{H}$, the 
edge EOS $w(R_{S})$) and etc., see there. This \eq{Conc19} disperancy between the value of the possible observables
enters through a boundary condition at $\RS$, where the membrane replacing the perfectly
absorbing horizon, rather than through the exterior bulk. The possible effects we can mention are the following therefore.
\begin{enumerate}
\item 
The horizon absorption / tidal heating effects. A classical horizon absorbs
low-frequency radiation with a universal coefficient; a partially reflecting
membrane absorbs slightly less. In an inspiral this changes the tidal-heating
rate and hence the orbital energy balance, imprinting on the gravitational-wave
phase, see \cite{HawkingHartle1972,Hartle1973,PoissonSasaki1995,Datta2020}.
\item 
A quasinormal-mode (QNM) frequency shift. The ringdown modes solve
the perturbation equation on the Schwarzschild exterior with a boundary
condition at $R_S$. A reflecting membrane shifts the complex frequencies,
$\omega_{\ell n}\,=\,\omega_{\ell n}^{\rm Schw}\,+\,\delta\omega$. Unlike a horizonless
object, which yields echoes, a  membrane correction's could give a
small complex shift, see \cite{ReggeWheeler1957,Zerilli1970,KokkotasSchmidt1999,CardosoFranzinPani2016}.
\item 
A modification of the tidal Love number. A regular black hole has vanishing tidal
Love number $k_{2}\,=\,0$ exactly. Surface structure yields a nonzero $k_{2}$, which is the leading nonzero effect in the proposed construction.
The effect can be imprinted at high post-Newtonian order on the inspiral
waveform, see \cite{Hinderer2008,BinningtonPoisson2009,Charalambous2021}.
\item 
Modification of thermal emission spectrum / greybody spectrum. An appearance of the finite crust
temperature $\Tloc=\Tinf B$ and the membrane modify the source boundary condition and hence the emission
spectrum, see \cite{Hawking1975,Page1976}. Have to mention nevertheless, that this effect is formally present, but negligible for
astrophysical masses.
\end{enumerate}
Yet, not all these effects can be fixed by an far observer, namely an effect is accessible at infinity only if it imprints on radiation that
reaches infinity. For example, a direct emission from the surface is not observable, i.e. anything emitted outward from a surface at $\RS$ is infinitely
redshifted in the Schwarzschild limit and suppressed; then it cannot be a clean signal for a far observer. Therefore, possibly important effects appear as 
a signals how the BH responds to an external perturbation, in this case a presence of the assumed membrane can change the response of the three layer BH in comparison to regular BH.
There are three effects then that, perhaps, could be observed because the membrane's effect.
If a black hole has an inspiral companion, or, in other words, a non-trivial inspiral matter around, then the modified tidal heating and Love number act back on it, and the companion's inspiral gravitational waves, generated in the exterior, propagates to infinity and carries the imprint of the black hole's structure in it's phase.
Similarly, the QNM shift modifies modes of the exterior space-time radiation that is outgoing to infinity by definition and which, in principle, can be obreved and analyzed. We notice again, this subject is a very important and quite complicated, it deserves an additional exploration when we talk about the possible physical effects and consequences of the introduced model of the interior.

 Finally, we underline, that the in the article we discuss a possible internal structure of a black hole proposed to clarify theoretical issues related to black hole description in GR. In our framework, instead operate by infinite quantities, we proposed to treat the interior and whole black hole as a kind of physical object  
with finite characteristics and finite properties of the matter inside. The main consequences of the construction then are possibilities to calculate from the first principles the classical characteristics of the BH, such as temperature, entropy and clarify the possible mechanisms of he Hawking radiations basing on the introduced form of the interior. We did not discuss these issues in the present article, mainly because the article's length, but we hope that the presented formalism will be useful for the for the further development and exploration of these and other research subjects.


\section*{Acknowledgments}

 The author expresses his gratitude to R. Singh for helpful discussions concern the topic of this article.


\section*{Declaration of use of generative Anthropic Claude Opus 4.8 AI engine}

 During the preparation of the manuscript, the author used the Anthropic Claude Opus 4.8 AI software engine to check symbolic computations, 
to check grammar, improve text style and formatting, and create explanatory graphics.
After using this tool, the author reviewed and edited the content as needed and bears full responsibility for the content of the published article.


\appendix
\newpage
\section{On heat transfer by an external current }\label{AppenA}
\renewcommand{\theequation}{A.\arabic{equation}}
\setcounter{equation}{0}

 In this Appendix we present a simplest case of a system with external current included which may serve for an investigation
of non-stationary solutions of the model. 
Namely, we consider a scalar field interacting with the external current provided by the first term from \eq{AA16}. The total Lagrangian we have then is
the following one:
\beq\label{AppenA1}
\el_{tot}\,=\,\el_{0}\,+\,\el_{J}\,,\qquad
\el_{J}\,=\,\sqrt{-g}\;g^{\mu\nu}\,\phi_{i}\,D_{\nu}J_{\mu}^{i}\,,
\eeq
with $\el_{0}$ as the \eq{AA1} Lagrangian and $J_{\mu}^{i}$ as the external current describes a simplest heat
transfer rates into and out of particular regions of the
configuration. Throughout we take $D_{\nu}$ to be the gauge-covariant
derivative,  the internal index $i$ marks a number of scalar fields and it is related to an internal rotation symmetry of the field components.
Further, we treat the different components of the field on the same ground\footnote{It could be  not the case when we consider some other regions of the interior.}, 
so in the expressions without the sum over $i$ added, instead the summation a proper overall coefficient appears which is for the two components field
is equal to $2$ simply.
In our construction we further simplify the problem restricting the current to be a purely radial and time dependent only:
\beq\label{AppenA2}
J_{\mu}^{i}\,=\,\Le 0\,,\,J_{r}^{i}(t)\,,\,0\,,\,0\Ra\,,
\eeq
The background metric is taken to be spherically symmetric, in the same
form used in the main text, see \eq{SC40}-\eq{SC41}.
Using the standard definition
\beq\label{AppenA3}
T_{\mu\nu}\,=\,\frac{2}{\sqrt{-g}}\,\frac{\delta \el}{\delta g^{\mu\nu}}\,,
\eeq
one obtains the following additional to the \eq{SDD4001}-\eq{SDD4004} components of the tensor:
\beq\label{AppenA4}
T^{J}_{\mu\nu}\,=\,2\,\phi_{i}\,D_{(\mu}J_{\nu)^{i}}\,-\,g_{\mu\nu}\,\phi_{i}\,D_{\alpha}J^{i\alpha}\,,
\eeq
where $D_{(\mu}J_{\nu)}^{i}\,=\,\tfrac{1}{2}(D_{\mu}J_{\nu}^{i}+D_{\nu}J_{\mu}^{i})$
is the symmetrized covariant derivative.
The trace of this new part of the tensor, in turn, is 
\beq\label{AppenA5}
{\cal Q}(t,r)\,\equiv\,g^{\mu \nu}\,T^{J}_{\mu\nu}\,=\,
-2\,\phi_{i}\,g^{\alpha \beta}\,D_{\alpha}J^{i}_{\beta}\,=\,2\,\phi_{i}\,g^{\alpha\beta}\,\Ga^{\rho}_{\alpha\beta}\,J_{\rho}^{i}
\,=\,2\,\phi_{i}\,g^{\alpha\beta}\,\Ga^{r}_{\alpha\beta}\,J_{r}^{i}(t)\,=\,
4\,\phi\,g^{\alpha\beta}\,\Ga^{r}_{\alpha\beta}\,J_{r}(t)\,
\eeq
For the metric \eq{SC38}, the relevant Christoffel symbols we use are
\beqar\label{AppenA6}
\Ga^{r}_{00}\,&=&\,\tfrac{1}{2}\,e^{\Phi}\,\chi\,\Le \chi\,\D_{r}\Phi\,+\,\chi'\Ra\,,\\
\Ga^{r}_{rr}\,&=&\,-\,\frac{\chi'}{2\,\chi}\,,
\label{AppenA6001}
\\
\Ga^{r}_{\theta\theta}\,&=&\,-\,r\,\chi\,,\,\,\,\Ga^{r}_{\varphi\varphi}\,=\,-\,r\,\chi\,\sin^{2}\theta\,,
\label{AppenA6002}
\\
\Ga^{r}_{0r}\,&=&\,-\,\frac{\dot{\chi}}{2\,\chi}\,,
\label{AppenA6003}
\eeqar
inserting them into \eq{AppenA5} we have
\beq\label{AppenA7}
{\cal Q}(t,r)\,=\,2\,\phi_{i}\,J_{r}^{i}(t)\,
\Le e^{-\Phi}\chi^{-1}\,\Ga^{r}_{00}\,-\,\chi\,\Ga^{r}_{rr}\,-\,\frac{2}{r^{2}}\,\Ga^{r}_{\theta\theta}\Ra\,
\eeq
and finally
\beqar\label{AppenA8}
{\cal Q}(t,r)\,&=&\,2\,\phi_{i}\,J_{r}^{i}(t)\,
\left[\frac{\chi\,\D_{r}\Phi\,+\,\chi'}{2}\,+\,\frac{\chi'}{2}\,+\,\frac{2\,\chi}{r}\right]
\,=\,2\,\phi_{i}\,J_{r}^{i}(t)\,
\left[\frac{\chi\,\D_{r}\Phi}{2}\,+\,\chi'\,+\,\frac{2\,\chi}{r}\right]\,;
\\
T^{J}_{\mu\nu}\,&=&\,2\,\phi_{i}\,D_{(\mu}J_{\nu)^{i}}\,+\,\frac{1}{2}\,g_{\mu\nu}\,{\cal Q}\,.
\label{AppenA8001}
\eeqar
The corresponding new additional components of the energy--momentum tensor become then
\beqar\label{AppenA9}
T^{J}_{00}\,&=&\,2 \ph_{i}\,D_{0} J^{i}_{0}\,+\,\frac{1}{2}\,g_{00}\,{\cal Q}\,=\,2\,\phi_{i}J^{i}_{r}\,\frac{\chi^{2}\,e^{\Phi}}{r}\Le 1\,-\,\frac{1}{4}\,r\,\D_{r}\Phi \Ra\,;
\\
T^{J}_{0r}\,&=&\,\phi_{i}\,D_{0}J_{r}^{i}\,=\,\phi_{i}\,\Le \dot{J}_{r}^{i}(t)\,+\,\frac{\dot{\chi}}{\chi}\,J_{r}^{i}\Ra\,;
\label{AppenA9001}
\\
T^{J}_{rr}\,&=&\,2 \ph_{i}\,D_{r} J^{i}_{r}\,+\,\frac{1}{2}\,g_{rr}\,{\cal Q}\,=\,-\,2\,\frac{\phi_{i}J^{i}_{r}}{r}\Le 1\,+\,\frac{1}{4}\,r\,\D_{r}\Phi \Ra\,;
\label{AppenA9002}
\\
T^{J}_{\theta\theta}\,&=&\,2 \ph_{i}\,D_{\theta} J^{i}_{\theta}\,+\,\frac{1}{2}\,g_{\theta \theta}\,{\cal Q}\,=\,-\,r^{2}\,\phi_{i}J^{i}_{r}\,
\Le \frac{\chi\,\D_{r}\Phi}{2}\,+\,\chi'\Ra\,;
\label{AppenA9003}
\\
T^{J}_{\varphi\varphi}\,&=&\,\sin^{2}\theta\,T^{J}_{\theta\theta}\,,
\label{AppenA9004}
\eeqar
where the dot denotes $d/dt$.
Using \eq{SC4} definitions of the energy density and pressure we obtain now the following answers.
For the energy density:
\beq\label{AppenA11}
\rho_{J}\,=\,\frac{1}{2}{\cal Q}(t,r)-\,\phi_{i}J^{i}_{r}\,\Le \chi\,\D_{r}\Phi\,+\,\chi' \Ra\,,
\eeq
and for the pressure correspondingly
\beq\label{AppenA12}
p_{J}\,=\,-\,\frac{1}{3}\,g^{ij}\,T_{ij}\,=\,-\,\frac{1}{2}\,{\cal Q}(t,r)\,-\,\frac{2}{3}\,\phi_{k}\,g^{ij}\,D_{i}J_{j}^{k}\,=
\,-\,\frac{1}{2}\,{\cal Q}(t,r)\,+\,\frac{1}{3}\,\phi_{i}J_{r}^{i}\,\Le \chi'\,+\frac{4\chi}{r}\Ra\,.
\eeq
The additional part  of the energy-momentum tensor contributions changes an effective equation
of state which already not really of the AdS/dS type with $w\,=\,-1$.

 Namely, consider the full contributions to the energy density and pressure, we have:
\beq\label{AppenA120001}
\rho_{\rm tot}\,=\,\rho\,+\,\rho_{J}\,
=\,\rho\,+\,\frac{1}{2}\,{\cal Q}\,+\,\De\rho_{J}\,,
\eeq
and
\beq\label{AppenA120002}
p_{\rm tot}\,=\,p\,+\,p_{J}\,
=\,p\,-\,\frac{1}{2}\,{\cal Q}\,+\,\De p_{J}\,.
\eeq
Here we defined a leading pieces of the quantities which concists also the
$\pm\,{\cal Q}/2$ contributions and the corrections:
\beq\label{AppenA120003}
\rho_{J}\,=\,\rho_{J}^{(0)}\,+\,\De\rho_{J}\,,\qquad
p_{J}\,=\,p_{J}^{(0)}\,+\,\De p_{J}\,,
\eeq
with
\beq\label{AppenA120004}
\rho_{J}^{(0)}\,=\,+\,\frac{1}{2}\,{\cal Q}\,,\qquad
p_{J}^{(0)}\,=\,-\,\frac{1}{2}\,{\cal Q}\,,
\eeq
and
\beq\label{AppenA120005}
\De\rho_{J}\,=\,-\,\ph_{i}J^{i}_{r}\,\Le\chi\,\D_{r}\Phi\,+\,\chi'\Ra\,,
\qquad
\De p_{J}\,=\,+\,\frac{1}{3}\,\ph_{i}J^{i}_{r}\,
\Le\chi'\,+\,\frac{4\,\chi}{r}\Ra\,.
\eeq
At the leading level $\rho_{J}^{(0)}\,+\,p_{J}^{(0)}\,=\,0$, so the
current sector alone is of the AdS/dS--type with $w\,=\,-1$. The
$\Phi'$-- and $\chi'$--dependent corrections break this property, we
treat them as small corrections which changing relatively slow preserving a quasi-static nature of the processes we discuss. 

  The effective equation of state ratio is then
\beq\label{AppenA120006}
w_{\rm tot}\,\equiv\,\frac{p_{\rm tot}}{\rho_{\rm tot}}\,=\,
\frac{\,p\,-\,\dfrac{1}{2}\,{\cal Q}\,+\,\De p_{J}\,}
{\,\rho\,+\,\dfrac{1}{2}\,{\cal Q}\,+\,\De\rho_{J}\,}\,.
\eeq
In the perturbative scheme we assume we obtain therefore
\beq\label{AppenA120007}
w_{\rm tot}\,\approx\,
\frac{p\,-\,\dfrac{1}{2}\,{\cal Q}}{\rho\,+\,\dfrac{1}{2}\,{\cal Q}}\,+\,
\frac{\De p_{J}}{\rho\,+\,\dfrac{1}{2}\,{\cal Q}}\,-\,
\frac{\Le p\,-\,\dfrac{1}{2}\,{\cal Q}\Ra\,\De\rho_{J}}
{\Le\rho\,+\,\dfrac{1}{2}\,{\cal Q}\Ra^{2}}\,,
\eeq
displaying the zeroth order ratio plus the two first order corrections. Now, taking $p\,=\,-\rho$ as a leading order EOS, we obtain in turn: 
\beq\label{AppenA120008}
w_{\rm tot}\,\approx\,-\,1\,+\,
\frac{\De p_{J}\,+\,\De\rho_{J}}{\rho\,+\,\dfrac{1}{2}\,{\cal Q}}\,,
\eeq
that provides after the \eq{AppenA120005} expressions inserting
\beq\label{AppenA120009}
w_{\rm tot}\,\approx\,-\,1\,+\,
\frac{\ph_{i}J^{i}_{r}\,\Le\,\dfrac{4\,\chi}{3\,r}\,-\,\dfrac{2\,\chi'}{3}\,-\,\chi\,\D_{r}\Phi\Ra}
{\rho\,+\,\ph_{i}J^{i}_{r}\,\Le\,\dfrac{\chi\,\D_{r}\Phi}{2}\,+\,\chi'\,+\,\dfrac{2\,\chi}{r}\Ra}\,.
\eeq
There are the following regimes of the effective EOS we have.

 The first regime is when $|{\cal Q}|\gg|\rho|,|p|$, i.e. the scalar perfect fluid
contributions are subleading in this case. Dropping $\rho$  relative to
${\cal Q}/2$ in \eq{AppenA120009}, we obtain then the leading and first subleading terms:
\beq\label{AppenA120010}
w_{\rm tot}\,\approx\,-1\,+\,
\frac{\De p_{J}\,+\,\De\rho_{J}}{{\cal Q}/2}\,=\,
-1\,+\,
\frac{\dfrac{4\,\chi}{3\,r}\,-\,\dfrac{2\,\chi'}{3}\,-\,\chi\,\D_{r}\Phi}
{\dfrac{\chi\,\D_{r}\Phi}{2}\,+\,\chi'\,+\,\dfrac{2\,\chi}{r}}\,
\equiv\,-\,1\,+\,\De w(r,t)\,.
\eeq
An additional condition of the applicability of that particular scheme is 
\beq\label{AppenA12001001}
\left|\frac{\dfrac{4\,\chi}{3\,r}\,-\,\dfrac{2\,\chi'}{3}\,-\,\chi\,\D_{r}\Phi}
{\dfrac{\chi\,\D_{r}\Phi}{2}\,+\,\chi'\,+\,\dfrac{2\,\chi}{r}}\right|\,<\,1\,.
\eeq
The correction here is purely geometrical one, there no external current in the expression and deviation from the AdS EOS is fixed entirely by the local metric profile.
We can estimate the corrections using the  static AdS core profile as first order input for the calculations. Taking
\beq\label{AppenA120011}
\chi(r)\,=\,1\,+\,r^{2}/L^{2}\,,\qquad\,\Phi\,=\,0
\eeq
with
\beq\label{AppenA120012}
\chi'\,=\,\frac{2\,r}{L^{2}}\,,\qquad
\D_{r}\Phi\,=\,0\,.
\eeq
we obtain:
\beq\label{AppenA120013}
\De w\,=\,\frac{4/(3\,r)}{2/r\,+\,4\,r/L^{2}}\,
=\,\frac{2\,L^{2}}{3\,(L^{2}\,+\,2\,r^{2})}\,.
\eeq
The answer interpolates between
\beq\label{AppenA12001301}
\De w(r\to 0)\,=\,\frac{2}{3}\,,\qquad
\De w(r\gg L)\,=\,\frac{L^{2}}{3\,r^{2}}\,\to\,0\,
\eeq
regions, so $w_{\rm tot}$ moves from $-1/3$ in the inner, locally-flat region toward $-1$ at
$r\gtrsim L$ where the AdS curvature dominates the metric. Namely, we see that taking
$\chi\,=\,1$, $\chi'\,=\,0$, $\D_{r}\Phi\,=\,0$ we obtain
\beq\label{AppenA120014}
\De w_{\rm flat}\,=\,\frac{4/(3\,r)}{2/r}\,=\,\frac{2}{3}\,,
\qquad w_{\rm tot}\,=\,-\,\frac{1}{3}\,,
\eeq
i.e. the EOS reduces to  a radial-tension ( or string network like source), consistent with the anisotropy
$p_{r}\neq p_{\perp}$ of the induced energy-momentum tensor.

 The second regime is a scalar vacuum dominated one when $|\rho|\,\gg\,{\cal Q}/2$.
The current--carried term in the denominator is dropped and the
correction scales as $\ph_{i}J^{i}_{r}/\rho$,
\beq\label{AppenA120015}
w_{\rm tot}\,\to\,-\,1\,+\,
\frac{\ph_{i}J^{i}_{r}}{\rho}\,
\Le\dfrac{4\,\chi}{3\,r}\,-\,\dfrac{2\,\chi'}{3}\,-\,\chi\,\D_{r}\Phi\Ra\,.
\eeq
Here is assumed that the scalar AdS vacuum dominates and preserves the $w$ near $-1$, the current
correction is parametrically suppressed in this approximation, it fits the assumption of quasi-stationarity of the evolving processes.

 A third, distinct regime occurs when the denominator
$\rho\,+\,\frac{1}{2}{\cal Q}$ approaches zero, i.e. when the current
contribution nearly compensates the scalar AdS density:
\beq\label{AppenA120016}
\rho\,+\,\frac{1}{2}\,{\cal Q}\,\to\,0\,.
\eeq
For the core with negative scalar density $\rho\,=\,-|\rho|$ this is
\beq\label{AppenA120017}
\frac{1}{2}\,{\cal Q}\,\to\,|\rho|\,.
\eeq
The perturbative EOS equation then develops a small denominator and the
linearized expansion we use is breaking, i.e. subleading terms become comparable to the leading $-\,1$. In our framework, 
this regime corresponds to the current ${\cal Q}$ being tuned
to compensate the negative AdS core density. The effective vacuum
density approaches zero, the geometry approaches flatness and physically, as it seems, it corresponds to an final processes of the core's evaporation.

 Next we solve the $G_{\mu\nu}\,=\,\kappa\,T_{\mu\nu}$ Einstein equations.
As in \eq{SC38}, we use the following metric:
\beq\label{AppenA19}
ds^{2}\,=\,e^{\Phi(t,r)}\,\chi(t,r)\,dt^{2}\,-\,\chi^{-1}(t,r)\,dr^{2}\,-\,r^{2}\,d\Omega^{2}\,,
\qquad
\chi(t,r)\,=\,1\,+\,\frac{2\,\kappa\,|\mu(t,r)|}{r}\,,
\eeq
see also \eq{SBB8}.
We define for the further use $f(0,r)\equiv e^{\Phi}\chi$ for the $00$ component, so we have $g_{00}=f$, $g_{rr}=-\chi^{-1}$,
$g_{\theta\theta}=-r^{2}$, $g_{\varphi\varphi}=-r^{2}\sin^{2}\theta$. As usual, primes denote
$\D_{r}$, overdots $\D_{t}$.
For the \eq{AppenA19} metric a direct computation gives, with $\chi=\chi(t,r)$, $\Phi=\Phi(t,r)$, the following independent components of the Einstein equations:
\beqar
G_{0}{}^{0}\,&=&\,-\,\frac{1}{r^{2}}\,\Le \chi\,+\,r\,\chi'\,-\,1\Ra\,,
\label{AppenA20}
\\
G_{r}{}^{r}\,&=&\,-\,\frac{1}{r^{2}}\,\Le \chi\,+\,r\,\chi\,\Phi'\,+\,r\chi'\,-\,1\Ra\,,
\label{AppenA21}
\\
G_{0r}\,&=&\,-\,\frac{\dot{\chi}}{r\,\chi}\,,
\label{AppenA22}
\eeqar
The mixed component of the Einstein tensor is the only place where $\dot{\chi}$ and correspondingly
$\dot{\mu}$ enter.

 Now, using $\chi = 1 + 2\kappa|\mu|/r$ in the $G_{0}{}^{0}$ component, we obtain the
regular Misner--Sharp form of the equation:
\beq\label{AppenA22001}
\chi\,+\,r\,\chi'\,-\,1\,=\,2\,\kappa\,\D_{r}|\mu(t,r)|\quad\rightarrow\quad
G_{0}{}^{0}\,=\,-\,\frac{2\,\kappa}{r^{2}}\,\D_{r}|\mu(t,r)|\,.
\eeq
So, for the mass function $\mu(t,r)\,=\,-\,|\mu(t,r)|$ and with the use of the \eq{AppenA8001} form of the energy-momentum tensor, we obtain:
\beq\label{AppenA23}
\D_{r}\mu(t,r)\,=\,\frac{1}{2}\,r^{2}\,\Le \rho\,+\,\frac{1}{2}\,{\cal Q}\,+\,\De\rho_{J}\Ra\,;
\eeq
the expression can be written as well in the following form:
\beq\label{AppenA24}
\mu(t,r)\,=\,\frac{1}{2}\,\int_{0}^{r}\,r'^{2}\,
\Le \rho\,+\,\frac{1}{2}\,{\cal Q}\,+\,\De\rho_{J}\,\Ra\,dr'\,+\,\mu(t,0)\,.
\eeq
The non-static current contributes through ${\cal Q}(t,r)$ even when the
underlying scalar profile is static. Also, the expression fixes the sign of the \eq{AppenA1} current and definition of it's direction. The plus sign of the current term in the \eq{AppenA1}
and correspondingly the plus sign of the ${\cal Q}(t,r)$ all together leads to decrease of the absolute value of the $\mu(t,r)$, i.e. in the case of the core it corresponds to the outward
flux of the negative energy-density quasi-particles and increase of the negative energy density toward zero.

 For the $(rr)$ equations component we have in turn:
\beq\label{AppenA25}
-\,\frac{1}{r^{2}}\,\Le \chi\,+\,r\,\chi\,\D_{r}\Phi\,+\,r\chi'\,-\,1\Ra\,=\,\kappa\,g^{rr}\,T_{rr}\,=\,-\,\kappa\,p\,+
2\,\kappa\,\phi_{i}J^{i}_{r}\,\frac{\chi}{r}\Le 1\,+\,\frac{1}{4}\,r\,\D_{r}\Phi \Ra\,\,.
\eeq
Then, 
isolating the expression for $\Phi'$, we obtain:
\beqar
\D_{r}\Phi(t,r)\,&=&\,\frac{\kappa}{\chi}\,\frac{p\,r}{1\,+\,\kappa\,r\,\phi_{i}J^{i}_{r}/2}\,-\,
\frac{2\,\kappa\,r}{\chi}\,\frac{\Le \phi_{i}J^{i}_{r}\,\chi/r\,-\,\D_{r}\mu(t,r)/r^{2}\Ra}{1\,+\,\kappa\,r\,\phi_{i}J^{i}_{r}/2}\,=\,
\nn
\\
&=&
\frac{\kappa}{\chi}\,\frac{p\,r}{1\,+\,\kappa\,r\,\phi_{i}J^{i}_{r}/2}\,+\,
\frac{\kappa\,r}{\chi}\,\frac{\Le \rho\,+\,\frac{1}{2}\,{\cal Q}\,+\,\De\rho_{J}\,-\,2\,\phi_{i}J^{i}_{r}\,\chi/r\Ra}{1\,+\,\kappa\,r\,\phi_{i}J^{i}_{r}/2}\,=\,
\nn
\\
&=&
\frac{\kappa\,r}{\chi}\,\frac{p\,+\,\rho_{tot}}{1\,+\,\kappa\,r\,\phi_{i}J^{i}_{r}/2}\,-\,
2\,\kappa\,\frac{\phi_{i}J^{i}_{r}}{1\,+\,\kappa\,r\,\phi_{i}J^{i}_{r}/2}\,
\label{AppenA26}
\eeqar
here the \eq{AppenA22}-\eq{AppenA23} were used.


  Next we consider the $(0r)$ component of the equation, which provides a non-staticity of the energy flux.
\beq\label{AppenA28}
-\,\frac{\dot{\chi}(t,r)}{r\,\chi(t,r)}\,=\,\phi_{i}\,\Le \dot{J}_{r}^{i}(t)\,+\,\frac{\dot{\chi}}{\chi}\,J_{r}^{i}\Ra\,.
\eeq
Using $\chi = 1+2\kappa|\mu|/r$, we obtain an equation describes an evolution of
the Misner--Sharp mass:
\beq\label{AppenA29}
\D_{t}\mu(t,r)\,=\,\frac{1}{2}\,r^{2}\,\chi(t,r)\,
\phi_{i}\,\Le \dot{J}_{r}^{i}(t)\,+\,\frac{\dot{\chi}}{\chi}\,J_{r}^{i}\Ra\,.
\eeq
We see thus, that the radial energy flux provided by the current
evolves the mass function in time. 

 The $(\theta \theta)$ component of Einstein equation, in turn, has the following form in this particular case:
\beqar\label{AppenA30}
G_{\theta}{}^{\theta}\,=\,G_{\varphi}{}^{\varphi}\,&=&\,
-\,\frac{1}{2}\,\chi\,\Phi''\,
-\,\frac{1}{4}\,\chi\,(\Phi')^{2}\,
-\,\frac{3}{4}\,\Phi'\,\chi'\,
-\,\frac{1}{2}\,\chi''
\nn\\
&&\,
-\,\frac{\chi\,\Phi'}{2\,r}\,
-\,\frac{\chi'}{r}\,
+\,e^{-\Phi}\,\Biggl[\,
-\,\frac{\ddot\chi}{2\,\chi^{2}}\,
+\,\frac{\dot\Phi\,\dot\chi}{4\,\chi^{2}}\,
+\,\frac{(\dot\chi)^{2}}{\chi^{3}}\,\Biggr]\,.
\eeqar
The two different parts of the expression has a simple physical interpretation.
The first six terms, which represent a spatial part, depend only on radial
derivatives of $\chi$ and $\Phi$. They contain the second radial
derivatives $\Phi''$ and $\chi''$ which are absent in $G_{0}{}^{0}$
and $G_{r}{}^{r}$. The
$\theta\theta$ equation is then a second order radial that makes it different from first three equations.
In turn, the time derivative terms, which were are absent in the strictly
static case, in general must be small in the
quasi-stationary approximation we assume and use.
For the r.h.s. of the equation we need to define corresponding energy-momentum tensor component. It, as well, has two parts. The first one is 
provided by static EOS, see \eq{SDD4003} and the second one is provided by the additional, current induced, term of the tensor, see \eq{AppenA9003}.
So, the AdS static contribution then is
\beq\label{AppenA31}
T^{\theta}{}_{\theta}\,=\,g^{\theta\theta}\,T^{}_{\theta\theta}\,=\,-\,p\,,
\eeq
identical in form to the radial principal pressure
$T^{r}{}_{r}\,=\,-\,p$, as expected for an isotropic perfect fluid we have in the case of the static solution.
The current contribution contribution provided by $T^{J}_{\theta\theta}$ component, is not
equal to the radial principal current pressure in turn. Indeed, we have
\beq\label{AppenA32}
T^{J}_{\theta\theta}\,=\,-\,r^{2}\,\ph_{i}\,J^{i}_{r}\,
\Le\frac{1}{2}\,\chi\,\Phi'\,+\,\chi'\Ra\,,
\eeq
that provides
\beqar\label{AppenA32001}
T^{J\theta}{}_{\theta}\,&=&\,g^{\theta\theta}\,T^{J}_{\theta\theta}\,=\,
\Le-\,\frac{1}{r^{2}}\Ra\Le-\,r^{2}\,\ph_{i}J^{i}_{r}\Ra
\Le\frac{1}{2}\,\chi\,\Phi'\,+\,\chi'\Ra
\nn\\
&=&\,\ph_{i}\,J^{i}_{r}\,
\Le\frac{1}{2}\,\chi\,\Phi'\,+\,\chi'\Ra\,.
\eeqar
The answer is not equal to 
\beq\label{AppenA33}
T^{Jr}{}_{r}\,=\,(2\chi\ph_{i}J^{i}_{r}/r)(1+\frac{1}{4}r\Phi')\,;
\eeq
there is a difference
\beq\label{AppenA34}
T^{Jr}{}_{r}\,-\,T^{J,\theta}{}_{\theta}\,=\,
\ph_{i}\,J^{i}_{r}\,\Le\frac{2\,\chi}{r}\,-\,\chi'\Ra\,,
\eeq
i.e. in the non-static system with the current present, a radial–tangential anisotropy appears.
The tangential Einstein equation $G_{\theta}{}^{\theta}\,=\,\kappa\,T_{\theta}{}^{\theta}$ we want then reads explicitly as
\beqar\label{AppenA35}
&&\,-\,\frac{1}{2}\,\chi\,\Phi''\,
-\,\frac{1}{4}\,\chi\,(\Phi')^{2}\,
-\,\frac{3}{4}\,\Phi'\,\chi'\,
-\,\frac{1}{2}\,\chi''\,
-\,\frac{\chi\,\Phi'}{2\,r}\,
-\,\frac{\chi'}{r}\,
\nn\\
&&\,+\,e^{-\Phi}\,\Biggl[\,
-\,\frac{\ddot\chi}{2\,\chi^{2}}\,
+\,\frac{\dot\Phi\,\dot\chi}{4\,\chi^{2}}\,
+\,\frac{(\dot\chi)^{2}}{\chi^{3}}\,\Biggr]\,
=\,-\,\kappa\,p\,+\,\kappa\,\ph_{i}\,J^{i}_{r}\,
\Le\frac{1}{2}\,\chi\,\Phi'\,+\,\chi'\Ra\,.
\eeqar
The equation can be used thus in the two different but complimentary ways. The first one is that it can be resolved inside the system of Einstein equations as an independent equation for the 
$\D_{t}\Phi$ derivative, i.e. for the determination of the time dependence of the $\Phi$ function. In this case the energy-momentum tensor conservation 
conditions aka Bianchi identity, 
\beq\label{AppenA36}
\nabla_{\mu}\,G^{\mu}{}_{\nu}\,\equiv\,\nabla_{\mu}\,T^{\mu}{}_{\nu}\,=\,0\,,
\eeq
can be used as a verification equations, i.e. for the consistency check of the possible solution. 
The another way is the opposite, we can use the Bianchi identity for the determination of the solution and then the \eq{AppenA35} will be used for the consistency check. 

 In the present case, the identity has the following form:
\beq\label{AppenA37}
\nabla_{\mu}\,T^{\mu}{}_{\nu}\,=\,\nabla_{\mu}\,
\Le T^{\mu}{}_{\nu}\,+\,T^{J\mu}{}_{\nu}\Ra\,=\,0
\eeq
it as well consists two parts, the static one and the current induced one.
The static contribution has a standard form
\beq\label{AppenA38}
\nabla_{\mu}T^{\,\mu}{}_{t}\,=\,
\D_{t}\rho\,-\,\frac{(\rho+p)\,\dot\chi}{2\chi}\,,
\eeq
\beq\label{AppenA39}
\nabla_{\mu}T^{\,\mu}{}_{r}\,=\,
-\,\D_{r}p\,-\,\frac{(\rho+p)}{2}\Le\Phi'\,+\,\frac{\chi'}{\chi}\Ra\,.
\eeq
In turn, for the current induced part, a
direct computation provides the following answers.
For the time, $\nu=0$, component of the identity we have:
\beqar\label{AppenA40}
\nabla_{\mu}T^{J\,\mu}{}_{0}\,&=&\,
-\,\frac{\chi\,\ph\,J\,\D_{t}\D_{r}\Phi}{1}\cdot\tfrac{1}{2}
\,-\,\frac{\chi\,J\,\Phi'\,\dot\ph}{2}\,
\,-\,\frac{\ph\,J\,\Phi'\,\dot\chi}{2}\,
\,-\,\ph\,J\,\D_{t}\D_{r}\chi
\,-\,J\,\dot\chi\,\ph'
\nn\\[2pt]
&&\,-\,\chi\,\ph\,\dot J\,\Phi'
\,-\,\chi\,\dot J\,\ph'
\,-\,\ph\,\dot J\,\chi'
\,+\,\frac{2\,\chi\,J\,\dot\ph}{r}\,.
\eeqar
Correspondingly for the radial, $\nu=r$, component we obtain:
\beqar\label{AppenA41}
\nabla_{\mu}T^{J\,\mu}{}_{r}\,&=&\,
\frac{\chi\,\ph\,J\,(\Phi')^{2}}{2}\,
+\,\frac{\chi\,\ph\,J\,\Phi''}{2}\,
+\,\frac{\chi\,\Phi'\,\ph'\,J}{2}\,
+\,\ph\,J\,\Phi'\,\chi'\,
\nn\\[2pt]
&&\,+\,\frac{2\,\chi\,\ph'\,J}{r}\,
+\,\frac{2\,\chi\,\ph\,J}{r^{2}}\,
+\,e^{-\Phi}\,\Biggl[\,
\frac{\ph\,\ddot J}{\chi}\,
+\,\frac{\dot\ph\,\dot J}{\chi}\,
-\,\frac{\ph\,\dot\Phi\,\dot J}{2\,\chi}\,
\nn\\[2pt]
&&\,
+\,\frac{\ph\,J\,\ddot\chi}{\chi^{2}}\,
+\,\frac{\dot\chi\,J\,\dot\ph}{\chi^{2}}\,
-\,\frac{\ph\,J\,\dot\Phi\,\dot\chi}{2\,\chi^{2}}\,
-\,\frac{2\,\ph\,J\,(\dot\chi)^{2}}{\chi^{3}}\,\Biggr]\,.
\eeqar
So, combining the static and current contributions $\nabla_{\mu}T^{\mu}{}_{\nu}=0$, we write finally the following full answers.
The time component of the full identity, i.e. energy conservation law, has the following form:
\beq\label{AppenA42}
\D_{t}\rho\,-\,\frac{(\rho+p)\,\dot\chi}{2\chi}\,+\,
\Sigma^{(t)}_{J}\,=\,0\;
\eeq
with
\beqar\label{AppenA43}
\Sigma^{(t)}_{J}\,&=&\,
-\,\frac{\chi\,\ph\,J\,\D_{t}\Phi'}{2}\,
-\,\frac{\chi\,J\,\Phi'\,\dot\ph}{2}\,
-\,\frac{\ph\,J\,\Phi'\,\dot\chi}{2}\,
-\,\ph\,J\,\D_{t}\chi'
-\,J\,\dot\chi\,\ph'
\nn\\
&&\,-\,\chi\,\ph\,\dot J\,\Phi'
\,-\,\chi\,\dot J\,\ph'
\,-\,\ph\,\dot J\,\chi'
\,+\,\frac{2\,\chi\,J\,\dot\ph}{r}\,.
\eeqar
The radial momentum equation has the following form then:
\beq\label{AppenA44}
-\,\D_{r}p\,-\,\frac{(\rho+p)}{2}\Le\Phi'+\frac{\chi'}{\chi}\Ra
\,+\,\Sigma^{(r)}_{J}\,=\,0\
\eeq
with
\beqar\label{SigmaR}
\Sigma^{(r)}_{J}\,&=&\,
\frac{\chi\,\ph\,J\,(\Phi')^{2}}{2}\,
+\,\frac{\chi\,\ph\,J\,\Phi''}{2}\,
+\,\frac{\chi\,\Phi'\,\ph'\,J}{2}\,
+\,\ph\,J\,\Phi'\,\chi'\,
+\,\frac{2\,\chi\,\ph'\,J}{r}\,
+\,\frac{2\,\chi\,\ph\,J}{r^{2}}\,
\nn\\[2pt]
&&\,+\,e^{-\Phi}\,\Biggl[\,
\frac{\ph\,\ddot J}{\chi}\,
+\,\frac{\dot\ph\,\dot J}{\chi}\,
-\,\frac{\ph\,\dot\Phi\,\dot J}{2\,\chi}\,
+\,\frac{\ph\,J\,\ddot\chi}{\chi^{2}}\,
+\,\frac{\dot\chi\,J\,\dot\ph}{\chi^{2}}\,
-\,\frac{\ph\,J\,\dot\Phi\,\dot\chi}{2\,\chi^{2}}\,
-\,\frac{2\,\ph\,J\,(\dot\chi)^{2}}{\chi^{3}}\,\Biggr]\,.\qquad
\eeqar
In general, using these identities, we can investigate the corrections for energy density and pressure appear in the static AdS EOS background
because of the current, see \eq{AppenA120001}-\eq{AppenA120002}. As mentioned above, we underline again that the system of equations which describe the framework with the time-dependent radial current is complete now and can be used for the calculation and construction of the non-stationary metric of the deep core region introduced in the manuscript.

 We conclude then that in the Appendix we shortly discuss an extension of the static AdS like scalar vacuum model by introducing a radial energy-transfer (heat) current that drives the system out of equilibrium. The idea explored is that an initially vacuum-dominated AdS configuration can be treated as an open system exchanging energy with its surroundings through a prescribed radial current. This what is assumed to be an example of a non-static solution of the model proposed in the paper. namely, it is shown that the probe current we introduced, generates an additional stress-energy contribution which can quasi-stationary modify the effective energy density, pressure, and equation of state of the initial medium. As a consequence, for example, the Misner–Sharp mass becomes time dependent, and we face a dynamical geometry instead the originally static one. The resulting evolution can be interpreted as a gradual depletion or evaporation of the negative-energy AdS core. So, as we wrote in the Introduction, at the end of the day we would like to establish a 
connection between vacuum EOS and transport phenomena which deforms it. The radial current acts not merely as an ordinary matter flux but as a mechanism that continuously reshapes the vacuum state itself. In this picture, deviations from the pure $w\,=\,-1$ vacuum behavior emerge from the interaction between geometry and energy transport, while sufficiently strong currents may compensate the negative vacuum density and drive the system toward a nearly flat final state.
In general, this task of an exploration of the non-stationary behavior of the proposed system of an interior is beyond the scope of the present article, we postpone
this discussion and calculations for a separated publication.


\newpage
\section{Thermal excitations in the boundary core region}\label{AppenB}
\renewcommand{\theequation}{B.\arabic{equation}}
\setcounter{equation}{0}

 There are two thermal excitations we introduced in this AdS patch, these are $\zeta$, which is massless Goldstone mode, and
$\epsilon$, which is massive radial mode. The mass $m_{\epsilon}$ is defined above the
second minimum $\phi\,=\,\phi_{0}$ of the underlying potential in the boundary region of the core. The background metric at signature we use then is defined as 
\beq\label{AppenB1}
ds^{2}\,=\,\chi(r)\,dt^{2}\,-\,\chi^{-1}(r)\,dr^{2}\,-\,r^{2}d\Omega^{2}\,,
\qquad\,
\chi_{B}(r)\,=\,1\,+\,\frac{r^{2}}{L_{B}^{2}}\,=\,1\,+\,\frac{\kappa\,m^{2}M^{2}}{12}\,r^{2}\,,\qquad\,
\sqrt{-g}\,=\,r^{2}\sin\theta\,,
\eeq
with the AdS radius
\beq\label{AppenB2}
L_{B}^{2}\,=\,\frac{3}{\kappa\,|V_{0}|}\,=\,\frac{12}{\kappa\,m^{2}M^{2}}\,;
\eeq
here we also take $\Phi\,=\,0$ in comparison with \eq{BBA1} metric. The non-trivial redshift function appears because of the thermal excitations; therefore
it provides a next-to-leading-order contribution to the free field solutions of the Klein-Gordon equation we discuss further.
With the $a\,\in\,\{\zeta,\epsilon\}$ as indexes of the two modes, the
relevant quadratic actions are
\beq\label{AppenB3}
S^{a}\,=\,\int\!d^{4}x\,\sqrt{-g}\,\el^{a}\,,\qquad
\el^{a}\,=\,\frac{1}{2}\,g^{\mu\nu}\,\D_{\mu}\phi^{a}\,\D_{\nu}\phi^{a}\,
-\,\frac{1}{2}\,m_{a}^{2}\,(\phi^{a})^{2}\,,
\eeq
with $m_{\zeta}\,=\,0$ and $m_{\epsilon}\,\neq\,0$ correspondingly.

 Next we solve an equation of motion of the minimally coupled scalar field in the external gravitational field, the form of the equation is standard of course. With
\beq\label{AppenB4}
\Box\phi^{a}\,=\,\frac{1}{\sqrt{-g}}\,\D_{\mu}\!\Le\sqrt{-g}\,g^{\mu\nu}\,\D_{\nu}\phi^{a}\Ra\,,
\eeq
in the component form, the usual Klein--Gordon equation reads as
\beq\label{AppenB5}
-\Le \Box\,+\,m_{a}^{2}\Ra\,\phi^{a}\,=\,-\,\frac{1}{\chi}\,\D_{0}^{2}\phi^{a}\,+\,\frac{1}{r^{2}}\,\D_{r}\!\Le r^{2}\,\chi\,\D_{r}\phi^{a}\Ra\,
+\,\frac{1}{r^{2}}\,\nabla_{S^{2}}^{2}\phi^{a}\,-\,m_{a}^{2}\,\phi^{a}\,=\,0\,,
\eeq
here we denoted
\beq\label{AppenB6}
-\,\frac{1}{\sqrt{-g}}\!\left[\D_{\theta}(\sqrt{-g}g^{\theta\theta}\D_{\theta}\phi^{a})
+\D_{\ph}(\sqrt{-g}g^{\ph\ph}\D_{\ph}\phi^{a})\right]\,=\,\frac{1}{r^{2}}\,\nabla_{S^{2}}^{2}\phi^{a}\,.
\eeq
The separation of the variables in the equation could be performed with the help of the following ansatz then:
\beq\label{AppenB7}
u_{n\ell m}^{a}(t,r,\theta,\ph)\,=\,e^{-i\,\om_{n\ell}^{a}\,t}\,
\frac{R_{n\ell}^{a}(r)}{r}\,Y_{\ell m}(\theta,\ph)\,,
\eeq
where $\D_{t}^{2}u\,=\,-\,\om^{2}u$ and $Y_{\ell m}(\theta,\ph)$ are spherical functions of the $\ell$ order which satisfy
$\nabla_{S^{2}}^{2}Y_{\ell m}\,=\,-\,\ell(\ell+1)Y_{\ell m}$ identity.
Substituting the ansatz into \eq{AppenB5} and dividing the obtained expression by
$e^{-i\om t}Y_{\ell m}/r$, we obtain an equation for the radial part of the ansatz $R_{n\ell}^{a}(r)$:
\beq\label{AppenB8}
\frac{\om^{a 2}_{n\ell}}{\chi}\,\frac{R^{a}_{n\ell}}{r}\,+\,\frac{1}{r^{2}}\,\D_{r}\!\Le r^{2}\,\chi\,\D_{r}\!\frac{R^{a}_{n\ell}}{r}\Ra\,
-\,\frac{\ell(\ell+1)}{r^{2}}\,\frac{R^{a}_{n\ell}}{r}\,-\,m_{a}^{2}\,\frac{R^{a}_{n\ell}}{r}\,=\,0\,.
\eeq
Taking the derivative of the second term, we obtain therefore a Schr\"odinger like radial equation for the function:
\beq\label{AppenB9}
\frac{d}{dr}\!\Le\chi\,\frac{dR_{n\ell}^{a}}{dr}\Ra\,
+\,\left[\,\frac{\om^{a 2}_{n\ell}}{\chi}\,-\,\frac{\ell(\ell+1)}{r^{2}}\,-\,m_{a}^{2}\,
-\,\frac{\chi'(r)}{r}\,\right]\,R_{n\ell}^{a}\,=\,0\,.\;
\eeq
The last term in the brakets is constant in the case of static AdS solution we discuss and we can define an effective mass
\beq\label{AppenB10}
m_{a,{\rm eff}}^{2}(r)\,=\,m_{a}^{2}\,+\,\frac{\chi'(r)}{r}\,
=\,m_{a}^{2}\,+\,\frac{2}{L_{B}^{2}}\,.
\eeq
with correction which identically vanishes in the Minkowski limit
$L_{B}\to\infty$. For the quasi-static case we mind, we assume that $m_{\epsilon}^{2}\,\gg\,1/L_{B}^{2}$, i.e.
the  $2/L_{B}^{2}$ correction is subleading and does not change physical result. So, further, for sake of simplicity, we keep the notation $m_{a}^{2}$
for the \eq{AppenB10} shifted mass. We notice also, that
the excitations $\zeta,\,\epsilon$ exist only in the boundary--core
shell when $r_{*}\,\leq\,r\,\leq\,R_{-}$ and
$\phi\,\sim\,\phi_{0}$. So, the boundary conditions we impose are of the Dirichlet type at both walls where
\beq\label{AppenB11}
R_{n\ell}^{a}(r_{*})\,=\,R_{n\ell}^{a}(R_{-})\,=\,0\,,
\eeq
The energy eigenvalues of the problem $\om_{n\ell}^{a 2}$ are real
and discrete, they are labelled by $n\,=\,1,2,\ldots$ for each
$(\ell,a)$. Corresponding eigenfunctions with different $n$ at fixed
$(\ell,a)$ are orthogonal with a proper weight taken for orthogonality conditions of the K-G operator eigenfunctions.
Namely, an inner product on the eigenfunctions on the static 3-d hypersurface is defined as usual:
\beq\label{AppenB11001}
(u_{1},u_{2})\,=\,-\,i\,\int_{\Sigma_{t}}\!d\Sigma^{\mu}\,
\Le u_{1}\,\D_{\mu}u_{2}^{*}\,-\,u_{2}^{*}\,\D_{\mu}u_{1}\Ra\,,
\eeq
with $d\Sigma^{\mu}\,=\,n^{\mu}\,\sqrt{h}\,d^{3}x$, where $n^{\mu}$ is
the future--directed unit normal of $\Sigma_{t}$ and $h_{ij}$ the
induced spatial metric. For our metric we have:
\beq\label{AppenB12}
n^{\mu}\,=\,\Le\frac{1}{\sqrt{\chi}},0,0,0\Ra\,,\qquad
\sqrt{h}\,=\,\frac{r^{2}\sin\theta}{\sqrt{\chi}}\,,\qquad
\sqrt{h}\,n^{0}\,=\,\frac{r^{2}\sin\theta}{\chi}\,,
\eeq
see \eq{SC44}-\eq{SC50}.
Inserting the ansatz \eq{AppenB7} and evaluating the expression at fixed $t$, we obtain:
\beqar\label{AppenB13}
u_{1}\,\D_{t}u_{2}^{*}\,-\,u_{2}^{*}\,\D_{t}u_{1}\,&=&\,
i\,(\om_{1}+\om_{2})\,e^{-i(\om_{1}-\om_{2})t}\,
\frac{R_{1}\,R_{2}}{r^{2}}\,Y_{1}\,Y_{2}^{*}\,,
\eeqar
so that, using $\int d\Omega\,Y_{\ell m}Y_{\ell'm'}^{*}\,=\,
\delta_{\ell\ell'}\delta_{mm'}$, we have:
\beq\label{AppenB14}
(u_{n\ell m}^{a},\,u_{n'\ell'm'}^{a})\,=\,
(\om_{n\ell}^{a}+\om_{n'\ell'}^{a})\,
\delta_{\ell\ell'}\,\delta_{mm'}\,
e^{-i(\om_{n\ell}^{a}-\om_{n'\ell'}^{a})t}\,
\int_{r_{*}}^{R_{-}}\!\frac{R_{n\ell}^{a}\,R_{n'\ell}^{a}}{\chi(r)}\,dr\,.
\eeq
The radial functions function then should be defined as polynomials orthogonal on corresponding interval with  weight function
$1/\chi$. Namely, the orthonormality condition of the eigenfunctions
\beq\label{AppenB15}
(u_{n\ell m},u_{n'\ell'm'})\,=\,\delta_{nn'}\delta_{\ell\ell'}\delta_{mm'}
\eeq
fixes the orthogonality condition of their radial parts:
\beq\label{AppenB16}
\int_{r_{*}}^{R_{-}}\!\frac{R_{n\ell}^{a}\,R_{n'\ell}^{a}}{\chi(r)}\,dr\,=\,
\frac{\delta_{n n'}}{2\,\om_{n\ell}^{a}}\,.
\eeq
Next, we can define the free quantized modes of the fields we introduced expanding the field operators with the help of the \eq{AppenB7} normalized free field solutions:
\beqar\label{AppenB17}
\hat\zeta(x)\,&=&\,\sum_{n\ell m}\,\Le
\hat a_{n\ell m}^{\zeta}\,u_{n\ell m}^{(\zeta)}(x)\,+\,\hat a_{n\ell m}^{\zeta\dagger}\,u_{n\ell m}^{(\zeta)*}(x)\Ra\,,
\\
\hat\epsilon(x)\,&=&\,\sum_{n\ell m}\!\Le
\hat a_{n\ell m}^{\epsilon}\,u_{n\ell m}^{(\epsilon)}(x)\,+\,\hat a_{n\ell m}^{\epsilon\dagger}\,u_{n\ell m}^{(\epsilon)*}(x)\Ra\,,
\label{AppenB1701}
\eeqar
with the canonical commutation relations
\beq\label{AppenB18}
[\hat a_{n\ell m}^{a},\,\hat a_{n'\ell'm'}^{a'\dagger}]\,=\,
\delta^{aa'}\,\delta_{nn'}\,\delta_{\ell\ell'}\,\delta_{mm'}\,,
\eeq
all other commutators vanishing. 
The thermal Gibbs state at inverse temperature $\beta\,=\,1/T$,
\beq\label{AppenB19}
\hat\rho_{\beta}\,=\,\frac{e^{-\beta\,\hat H_{0}}}{Z}\,,\qquad
\hat H_{0}\,=\,\sum_{a,n\ell m}\,\om_{n\ell}^{(a)}\,\hat a_{n\ell m}^{a\dagger}\,\hat a_{n\ell m}^{a}\,,
\eeq
gives the Bose--Einstein occupation
\beq\label{AppenB20}
\langle\hat a_{n\ell m}^{a\dagger}\,\hat a_{n'\ell'm'}^{a'}\rangle_{\beta}\,=\,
n_{B}(\om_{n\ell}^{(a)})\,\delta^{aa'}\delta_{nn'}\delta_{\ell\ell'}\delta_{mm'}\,,
\quad
n_{B}(\om)\,=\,\frac{1}{e^{\om/T}-1}\,.
\eeq
The conjugate ordering follows from the \eq{AppenB18} commutator,
\beq\label{AppenB21}
\langle\hat a_{n\ell m}^{a}\,\hat a_{n'\ell'm'}^{a'\dagger}\rangle_{\beta}\,=\,
[1+n_{B}(\om_{n\ell}^{(a)})]\,\delta^{aa'}\delta_{nn'}\delta_{\ell\ell'}\delta_{mm'}\,,
\eeq
while $\langle aa\rangle_{\beta}\,=\,\langle a^{\dagger}a^{\dagger}\rangle_{\beta}\,=\,0$.

 Now, we have to define a local energy-momentum tensor in terms of the free field solutions, i.e. we have to define properly a 
kinetic and potential part of the EMT obtained from \eq{AppenB3} Lagrangian in terms of the new solutions at coinciding points,
see for example \cite{Birrell, Wald, Fulling}; for the thermal (Hartle--Hawking) state and the renormalized stress tensor near a black hole see also \cite{HartleHawking1976,Christensen1976,Candelas1980,Page1982}
We have:
\beq\label{AppenB21001}
\langle\hat\phi^{a}(x)\,\hat\phi^{a}(y)\rangle_{\beta}\,=\,
\sum_{n\ell m}\!\left\{[1+n_{B}(\om_{n\ell}^{a})]\,u_{n\ell m}^{a}(x)\,u_{n\ell m}^{a*}(y)\,
+\,n_{B}(\om_{n\ell}^{a})\,u_{n\ell m}^{a*}(x)\,u_{n\ell m}^{a}(y)\right\}\,
\eeq
obtained by use of 
\eq{AppenB17}, \eq{AppenB20}, \eq{AppenB21}\,.
For the sake of shortness we will use further also the following notations:
\beq\label{AppenB22}
u_{n\ell m}^{a}\rightarrow\,u^{a}\,.
\eeq
At coincident points,
\beq\label{AppenB23}
\langle\hat\phi^{a\,2}(x)\rangle_{\beta}\,=\,
\sum_{n\ell m}\,2\,[n_{B}(\om_{n\ell}^{a})\,+\,\frac{1}{2}]\,|u_{n\ell m}^{a}(x)|^{2}\,.
\eeq
Differentiating \eq{AppenB21} with respect to $x^{\mu}$ and $y^{\nu}$
and then taking $y\,\to\,x$,
\beq\label{AppenB24}
\langle\D_{\mu}\hat\phi^{a}(x)\,\D_{\nu}\hat\phi^{a}(x)\rangle_{\beta}\,=\,
\sum_{n\ell m}\!\Le[1\,+\,n_{B}]\,\D_{\mu}u^{a}\,\D_{\nu}u^{a*}\,
+\,n_{B}\,\D_{\mu}u^{a*}\,\D_{\nu}u^{a}\Ra\,.
\eeq
Symmetrizing the expression in respect to $\mu$ and $\nu$ we obtain:
\beq\label{AppenB25}
\langle\D_{\mu}\hat\phi^{a}\,\D_{\nu}\hat\phi^{a}\rangle_{\beta}^{\rm sym}\,=\,
\sum_{n\ell m}\,[n_{B}(\om_{n\ell}^{(a)})\,+\,\frac{1}{2}]\,
\Le\D_{\mu}u^{a}\,\D_{\nu}u^{a*}\,+\,\D_{\mu}u^{a*}\,\D_{\nu}u^{a}\Ra\,.
\eeq
Next,
the classical EMT for our modes, which has the \eq{AppenB7} form in the case of the free modes Lagrangian, we may obtain by the simple substitution:
\beq\label{AppenB26}
\D_{\mu}\phi^{a}\,\D_{\nu}\phi^{a}\,\to\,\frac{1}{2}\,\Le \D_{\mu}u^{a}\,\D_{\nu}u^{a*}\,+\,\D_{\mu}u^{a*}\,\D_{\nu}u^{a}\Ra\,;\qquad\,
\phi^{a 2}\,\to\,|u|^{a 2}\,.
\eeq
The EMT then acquires the following form:
\beqar
T^{a}_{\mu\nu}[u^{a},u^{a*}]\,&=&\,
\frac{1}{2}\,\Le\D_{\mu}u^{a}\,\D_{\nu}u^{a*}\,+\,\D_{\mu}u^{a*}\,\D_{\nu}u^{a}\Ra\,\nn\\
&&\,-\,\frac{1}{2}\,g_{\mu\nu}\!\Le g^{\al\beta}\,\D_{\al}u^{a}\,\D_{\beta}u^{a*}\,-\,m_{a}^{2}\,|u^{a}|^{2}\Ra\,,
\label{AppenB27}
\eeqar
and then we obtain
the thermal expectation value of the EMT:
\beq\label{AppenB28}
\langle T_{\mu\nu}^{a}(x)\rangle_{\beta}\,=\,
\sum_{n\ell m}\,[n_{B}(\om_{n\ell}^{a})\,+\,\frac{1}{2}]\,2\,
T^{a}_{\mu\nu}[u_{n\ell m}^{a},\,u_{n\ell m}^{a*}](x)\,,
\eeq
which is simply a consequence of the \eq{AppenB23} and \eq{AppenB25} expressions and \eq{AppenB26} substitutions.
Next we renormalize the obtained EMT by the
standard Hadamard prescription, see again \cite{Birrell,Wald,Fulling} and the covariant point-separation method of \cite{Christensen1976}, extracting from it a zero modes contribution and obtaining
\beq\label{AppenB29}
\langle T_{\mu\nu}^{a}\rangle_{\beta}^{\rm th}\,\equiv\,
\langle T_{\mu\nu}^{a}\rangle_{\beta}\,-\,\langle T_{\mu\nu}^{a}\rangle_{\rm vac}\,=\,
2\,\sum_{n\ell m}\,n_{B}(\om_{n\ell}^{a})\,
T^{a}_{\mu\nu}[u_{n\ell m}^{a},\,u_{n\ell m}^{a*}](x)\,,
\eeq
which is UV finite of course. An important property of the obtained thermal EMT is that 
the off--diagonal components of it are
vanishing when summed in the thermal state.

 Indeed, consider firstly the possible-time dependent non-diagonal terms.
We have for the classical solutions:
\beq\label{AppenB30}
u\,=\,e^{-i\om t}\,\frac{R}{r}\,Y_{\ell m}\,,\quad
\D_{t}u\,=\,-\,i\,\om\,u\,,\quad
\D_{i}u\,=\,e^{-i\om t}\,\D_{i}\!\Le\frac{R}{r}\,Y_{\ell m}\Ra\,
\eeq
where $i\,\in\,\{r,\theta,\ph\}$, then
\beq\label{AppenB31}
\frac{1}{2}\Le\D_{t}u\,\D_{i}u^{*}\,+\,\D_{t}u^{*}\,\D_{i}u\Ra\,=\,
\frac{i\om}{2}\,\Le u^{*}\,\D_{i}u\,-\,u\,\D_{i}u^{*}\Ra\,=\,\frac{i\om}{2}\,\D_{i}\Le\frac{R^{2}}{r^{2}}\Ra\,
\sum_{l}\,\sum_{m=-l}^{l}\Le Y^{*}_{\ell m}\,Y_{\ell m}\,-\,
Y_{\ell m}\,Y^{*}_{\ell m}\Ra\,=\,0
\eeq
because of the
\beq\label{AppenB32}
\sum_{m=-l}^{l}\Le Y^{*}_{\ell m}\,Y_{\ell m}\,-\,
Y_{\ell m}\,Y^{*}_{\ell m}\Ra\,=\,0
\eeq
property of the spherical functions which follows from
\beq\label{AppenB33}
\sum_{m=-\ell}^{\ell}\,Y_{\ell m}\,Y_{\ell m}^{*}\,=\,\frac{2\ell+1}{4\pi}\,
\eeq
identity.
Therefore we obtain that the diagonal $g_{ti}\,=\,0$ and this component contributes nothing to $T^{a}_{ti}$ either and so
$\langle T_{ti}^{(a)}\rangle_{\beta}^{\rm th}\,=\,0$ as well.
Similarly it can be shown that  the  spatial non-diagonal entries
$\sum_{m}T_{r\theta}^{a},\,\sum_{m}T_{r\ph}^{a},\,\sum_{m}T_{\theta\ph}^{a}$
vanish identically  and this shows that
\beq\label{AppenB34}
\langle T_{\mu\nu}^{a}\rangle_{\beta}^{\rm th}\,=\,
{\rm diag}\!\Le\langle T_{00}^{a}\rangle,\,\langle T_{rr}^{a}\rangle,\,
\langle T_{\theta\theta}^{a}\rangle,\,\langle T_{\ph\ph}^{a}\rangle\Ra^{\rm th}\,.
\eeq
Next we compute the components of the tensor and related to it quantities.

 The first quantity we need is a energy density corresponds to the calculated EMT, so firstly, using  \eq{AppenB27} with $\mu=\nu=0$, 
for the $T^{a}_{00}[u^{a},u^{a*}]$
we obtain:
\beq\label{AppenB35}
T^{a}_{00}\,=\,\frac{1}{2}\Le|\D_{t}u^{a}|^{2}+|\D_{t}u^{a*}|^{2}\Ra\,-\,
\frac{\chi}{2}\!\left[g^{\al\beta}\D_{\al}u^{a}\,\D_{\beta}u^{a*}\,-\,m_{a}^{2}|u^{a}|^{2}\right]\,.
\eeq
Taking now $\D_{t}u\,\D_{t}u^{*}\,=\,\om^{2}|u|^{2}$ and \eq{AppenB1} metric we write then:
\beqar\label{AppenB36}
g^{\al\beta}\D_{\al}u\,\D_{\beta}u^{*}\,&=&\,
\frac{\om^{2}}{\chi}\,|u|^{2}\,-\,\chi\,|\D_{r}u|^{2}\,\nn\\
&&\,-\,\frac{1}{r^{2}}\,|\D_{\theta}u|^{2}\,-\,\frac{1}{r^{2}\sin^{2}\theta}\,|\D_{\ph}u|^{2}\,
\eeqar
that provides finally
\beqar
T^{a}_{00}\,&=&\,\om^{2}|u^{a}|^{2}\,-\,\frac{\chi}{2}\!\left[
\frac{\om^{2}}{\chi}|u^{a}|^{2}\,-\,\chi|\D_{r}u^{a}|^{2}\,-\,\frac{|\D_{\theta}u^{a}|^{2}}{r^{2}}
\,-\,\frac{|\D_{\ph}u^{a}|^{2}}{r^{2}\sin^{2}\theta}\,-\,m_{a}^{2}|u|^{2}\right]
\nn\\
&=&\,\frac{\om^{2}}{2}\,|u^{a}|^{2}\,+\,\frac{\chi^{2}}{2}\,|\D_{r}u^{a}|^{2}\,+\,
\frac{\chi}{2r^{2}}\!\Le|\D_{\theta}u^{a}|^{2}\,+\,\frac{|\D_{\ph}u^{a}|^{2}}{\sin^{2}\theta}\Ra\,+\,
\frac{\chi\,m_{a}^{2}}{2}|u^{a}|^{2}\,.
\label{AppenB37}
\eeqar
Therefore, using \eq{SC4}, \eq{AppenB12} and \eq{AppenB29} definitions, we obtain:
\beqar
\rho^{a}_{\rm th}\,&=&\,\frac{1}{\chi}\,\langle T_{00}^{a}\rangle_{\beta}^{\rm th}\,=\,
\frac{2}{\chi}\sum_{n\ell m}\,n_{B}\,T^{a}_{00}\,=\,
\nn\\
&=&\,\sum_{n\ell m}\,n_{B}(\om_{n\ell}^{a})\!\left[
\frac{\om^{2}}{\chi}|u^{a}|^{2}\,+\,\chi|\D_{r}u^{a}|^{2}\,
\,+\,\frac{1}{r^{2}}\!\Le|\D_{\theta}u^{a}|^{2}\,+\,\frac{|\D_{\ph}u^{a}|^{2}}{\sin^{2}\theta}\Ra\,+\,
m_{a}^{2}|u^{a}|^{2}\right]\,.
\label{AppenB38}
\eeqar
Then angular part of the expression, summed over $m$, is fixed by the
spectral relation $\nabla_{S^{2}}^{2}Y_{\ell m}\,=\,-\,\ell(\ell+1)Y_{\ell m}$ in the angular integration through
\beq\label{AppenB39}
\int\!d\Omega\,\Le|\D_{\theta}Y_{\ell m}|^{2}+\frac{|\D_{\ph}Y_{\ell m}|^{2}}{\sin^{2}\theta}\Ra\,=\,
\int\!d\Omega\,Y_{\ell m}^{*}\Le-\nabla_{S^{2}}^{2}Y_{\ell m}\Ra\,=\,\ell(\ell+1)\,,
\eeq
equality. 
After  the $m$ summation performed, the integrand becomes to be angle independent by the
spherical function property: 
\beq\label{AppenB40}
\sum_{m=-\ell}^{\ell}\!\Le|\D_{\theta}Y_{\ell m}|^{2}+\frac{|\D_{\ph}Y_{\ell m}|^{2}}{\sin^{2}\theta}\Ra\,=\,
\frac{\ell(\ell+1)(2\ell+1)}{4\pi}\,=\,\ell(\ell+1)\sum_{m}|Y_{\ell m}|^{2}\,
\eeq
and hence the $m$ sum converts to the following answer:
\beq\label{AppenB41}
\sum_{m}\frac{1}{r^{2}}\!\Le|\D_{\theta}u|^{2}+\frac{|\D_{\ph}u|^{2}}{\sin^{2}\theta}\Ra\,=\,
\frac{\ell(\ell+1)}{r^{2}}\sum_{m}|u|^{2}\,,
\eeq
where the $u\,=\,e^{-i\om t}(R/r)Y_{\ell m}$ was used.
Taking all together, we obtain the following expression
\beq\label{AppenB42}
\rho^{a}_{\rm th}(r)\,=\,\sum_{n\ell m}\,n_{B}(\om_{n\ell}^{a})\!\left[
\frac{(\om_{n\ell}^{a})^{2}}{\chi(r)}\,|u^{a}|^{2}\,+\,\chi(r)\,|\D_{r}u^{a}|^{2}\,+\,\frac{\ell(\ell+1)}{r^{2}}\,|u^{a}|^{2}\,+\,m_{a}^{2}\,|u^{a}|^{2}\right]\,
\eeq
for the energy density of the boundary core region.

 Next we compute an answer for the radial pressure $p_{r,\rm th}$ quantity. For that we need $T^{a}_{rr}[u^{a},u^{a*}]$
EMT component:
\beq\label{AppenB43}
T^{a}_{rr}\,=\,|\D_{r}u^{a}|^{2}\,-\,\frac{1}{2}(-\chi^{-1})\!
\left[g^{\al\beta}\D_{\al}u^{a}\,\D_{\beta}u^{a*}\,-\,m_{a}^{2}|u^{a}|^{2}\right]\,.
\eeq
with $g_{rr}\,=\,-\,\chi^{-1}$ used.
Performing the same steps as above, we have:
\beq\label{AppenB44}
T^{a}_{rr}\,=\,|\D_{r}u^{a}|^{2}\,+\,\frac{1}{2\chi}\!\left[
\frac{(\om_{n\ell}^{a})^{2}}{\chi}|u^{a}|^{2}\,-\,\chi|\D_{r}u^{a}|^{2}\,
-\,\frac{|\D_{\theta}u^{a}|^{2}}{r^{2}}-\frac{|\D_{\ph}u^{a}|^{2}}{r^{2}\sin^{2}\theta}-m_{a}^{2}|u^{a}|^{2}\right]\,.
\eeq
Therefore
\beqar
p_{r,\rm th}^{a}\,&=&\,\chi\,\langle T_{rr}^{(a)}\rangle_{\beta}^{\rm th}\,=\,2\,\chi\sum_{n\ell m}\,n_{B}\,T^{a}_{rr}\,=\,
\nn\\
&=&\,
\sum_{n\ell m}\,n_{B}\!\left[2\,\chi|\D_{r}u^{a}|^{2}\,+\,\frac{\om^{2}}{\chi}|u^{a}|^{2}\,-\,\chi|\D_{r}u^{a}|^{2}\,
-\,\frac{1}{r^{2}}\!\Le|\D_{\theta}u^{a}|^{2}+\frac{|\D_{\ph}u^{a}|^{2}}{\sin^{2}\theta}\Ra\,-\,m_{a}^{2}|u^{a}|^{2}\right]\,=\,
\nn\\
&=&\,\sum_{n\ell m}\,n_{B}\!\left[\frac{(\om_{n\ell}^{a})^{2}}{\chi}|u^{a}|^{2}\,+\,\chi|\D_{r}u^{a}|^{2}\,-\,\frac{\ell(\ell+1)}{r^{2}}|u^{a}|^{2}\,-\,
m_{a}^{2}|u^{a}|^{2}\right]\,,
\label{AppenB45}
\eeqar
where again \eq{AppenB41} was used in the last line.
Comparing the answer with the with energy density \eq{AppenB42} expression, 
we see that the radial pressure
differs from it only in the sign of the
centrifugal barrier term $\ell(\ell+1)/r^{2}|u^{a}|^{2}$ and the mass
term $m_{a}^{2}|u^{a}|^{2}$, while the "radial kinetic" piece
$\frac{\om^{2}}{\chi}|u^{a}|^{2}\,+\,\chi|\D_{r}u^{a}|^{2}$ in the expressions is the same, it is a standard difference of course, see \eq{SD5} and \eq{SDD7}.

 For the tangential pressure $p_{\perp,\rm th}$ we consider $\mu=\nu=\theta$ component of the EMT with $g_{\theta\theta}\,=\,-\,r^{2}$ obtaining
\beq\label{AppenB46}
T^{a}_{\theta\theta}\,=\,|\D_{\theta}u^{a}|^{2}\,-\,\frac{1}{2}(-r^{2})\!
\left[g^{\al\beta}\D_{\al}u^{a}\,\D_{\beta}u^{a*}\,-\,m_{a}^{2}|u^{a}|^{2}\right]\,.
\eeq
Using again \eq{AppenB30} free field solution, we write:
\beqar\label{AppenB47}
T^{a}_{\theta\theta}\,&=&\,|\D_{\theta}u^{a}|^{2}\,+\,\frac{r^{2}}{2}\!\left[
\frac{\om^{2}}{\chi}|u^{a}|^{2}\,-\,\chi|\D_{r}u^{a}|^{2}
\,-\,\frac{|\D_{\theta}u^{a}|^{2}}{r^{2}}\,-\,\frac{|\D_{\ph}u^{a}|^{2}}{r^{2}\sin^{2}\theta}\,-\,m_{a}^{2}|u|^{2}\right]\,=\,
\nn\\
&=&\,\frac{|\D_{\theta}u^{a}|^{2}}{2}\,-\,\frac{|\D_{\ph}u^{a}|^{2}}{2\sin^{2}\theta}\,+\,
\frac{r^{2}}{2}\!\left[\frac{\om^{2}}{\chi}|u^{a}|^{2}-\chi|\D_{r}u^{a}|^{2}-m_{a}^{2}|u^{a}|^{2}\right]\,
\eeqar
obtaining
\beqar\label{AppenB48}
p_{\perp,\rm th}^{a}\,&=&\,\frac{1}{r^{2}}\,\langle T_{\theta\theta}^{a}\rangle_{\beta}^{\rm th}\,=\,
\frac{2}{r^{2}}\sum_{n\ell m}\,n_{B}\,T^{a}_{\theta\theta}\,=\,
\nn\\
&=&\,\sum_{n\ell m}\,n_{B}\!\left[
\frac{|\D_{\theta}u^{a}|^{2}}{r^{2}}\,-\,\frac{|\D_{\ph}u^{a}|^{2}}{r^{2}\sin^{2}\theta}\,
\,+\,\frac{\om^{2}}{\chi}|u^{a}|^{2}-\chi|\D_{r}u^{a}|^{2}-m_{a}^{2}|u^{a}|^{2}\right]\,.
\eeqar
In the expression the first two terms appear with opposite signs and their $m$
sum is symmetric under
$|\D_{\theta}|^{2}\,\leftrightarrow\,|\D_{\ph}|^{2}/\sin^{2}\theta$.
After the summation the angular distribution is uniform, it means simply that
\beq\label{AppenB49}
\sum_{m}\,|\D_{\theta}Y_{\ell m}|^{2}\,=\,\sum_{m}\,\frac{|\D_{\ph}Y_{\ell m}|^{2}}{\sin^{2}\theta}\,=\,
\frac{\ell(\ell+1)(2\ell+1)}{8\pi}\,.
\eeq
Therefore the two angular derivative terms in the \eq{AppenB48} are canceled
exactly after the summation over $m$ taken:
\beq\label{AppenB50}
\sum_{m}\!\left[\frac{|\D_{\theta}u|^{2}}{r^{2}}\,-\,\frac{|\D_{\ph}u|^{2}}{r^{2}\sin^{2}\theta}\right]\,=\,0\,.
\eeq
and the $\ell(\ell+1)/r^{2}$ term is not appearing in the answer. We are left with the radial
and mass pieces alone then:
\beq\label{AppenB51}
p^{a}_{\perp,\rm th}(r)\,=\,\sum_{n\ell m}\,n_{B}(\om_{n\ell}^{(a)})\!\left[
\frac{(\om_{n\ell}^{a})^{2}}{\chi(r)}\,|u^{a}|^{2}\,-\,\chi(r)\,|\D_{r}u^{a}|^{2}\,-\,m_{a}^{2}|u^{a}|^{2}\right]\,.
\eeq
We see, also that 
\beq\label{B52}
p_{r,\rm th}^{a}-p_{\perp,\rm th}^{a}\,=\,
\sum_{n\ell m}n_{B}\!\left[2\,\chi|\D_{r}u^{a}|^{2}
\,-\,\frac{\ell(\ell+1)}{r^{2}}|u^{a}|^{2}\right]\,\neq\,0\,,
\eeq
i.e  the pressure tensor is anisotropic, mode by mode, and the isotropy, even approximate, can emerge only after the spectral sum performed.

 Now we can introduce the four scalar densities that build all three components of the EMT. For
the boundary core they are the following discrete sums:
\beqar\label{B53}
K(r)\,&\equiv&\,\sum_{n\ell m}n_{B}(\om_{n\ell}^{a})\,
\frac{(\om_{n\ell}^{a})^{2}}{\chi}\,|u^{a}|^{2}\,,\nn\\
G^{\rm rad}(r)\,&\equiv&\,\sum_{n\ell m}n_{B}(\om_{n\ell}^{a})\,
\chi\,|\D_{r}u^{a}|^{2}\,,\\
G^{\rm ang}(r)\,&\equiv&\,\sum_{n\ell m}n_{B}(\om_{n\ell}^{a})\,
\frac{\ell(\ell+1)}{r^{2}}\,|u^{a}|^{2}\,,\nn\\
m^{2}\Pi(r)\,&\equiv&\,\sum_{n\ell m}n_{B}(\om_{n\ell}^{a})\,
m_{a}^{2}\,|u^{a}|^{2}\,,\nn
\eeqar
which are respectively the temporal kinetic, radial gradient,
tangential (centrifugal) gradient and mass densities. In terms of
these quantities, the pressure reads as
\beqar\label{B54}
\rho^{a}_{\rm th}\,&=&\,K\,+\,G^{\rm rad}\,+\,G^{\rm ang}\,+\,m^{2}\Pi\,
-\,\tfrac12\!\Le K-G^{\rm rad}-G^{\rm ang}-m^{2}\Pi\Ra ,\nn
\eeqar
or, in terms of the three independent components:
\beqar\label{B55}
\rho^{a}_{\rm th}\,&=&\,\tfrac12 K+\tfrac12 G^{\rm rad}
+\tfrac12 G^{\rm ang}+\tfrac12 m^{2}\Pi\,,\nn\\
p_{r,\rm th}^{a}\,&=&\,\tfrac12 K+\tfrac12 G^{\rm rad}
-\tfrac12 G^{\rm ang}-\tfrac12 m^{2}\Pi\,,\\
p_{\perp,\rm th}^{a}\,&=&\,\tfrac12 K-\tfrac12 G^{\rm rad}
-\tfrac12 m^{2}\Pi\,,\nn
\eeqar
We note that the tangential pressure $p_{\perp,\rm th}^{a}$ contains no $G^{\rm ang}$ term by construction, the centrifugal contribution cancels in the $m$-sum, see \eq{AppenB50}, rather than being omitted.
A case of a perfect fluid when $p_{r}=p_{\perp}$ requires
\beq\label{B56}
G^{\rm rad}\,=\,\tfrac12\,G^{\rm ang}\,,
\eeq
which fails mode by mode and could be restored only by the WKB sum.

 For a  boundary core region we assume that the thermally populated modes provide
$\om_{n\ell}^{a}\,(R_{-}-r_{*})\gg1$ and the dominant angular momenta then is
reach $\ell_{\max}\sim\om\,r/\sqrt{\chi}\gg1$. The WKB counting of the modes in a shell bounded by Dirichlet walls, together with the associated local density of states, is the method of the "brick-wall" model \cite{tHooft1985,MannTarasovZelnikov1992,MukohyamaIsrael1998}.
In this regime the
curvature term $\chi'/r$ in the Schr\"odinger form \eq{AppenB9} is
negligible relative to $\om^{2}/\chi$ and the radial equation is solved
by the geometrical optic (WKB) form for the radial function:
\beq\label{B57}
R_{n\ell}^{a}(r)\,\simeq\,\frac{{\cal N}}{\sqrt{k(r)}}\,
\cos\Theta(r)\,;\,
\Theta(r)\,=\,\int^{r}\!k(r')\,dr'\,+\,\delta\,;\,
\Theta'(r)\,=\,k(r)\,;\,
k(r)\,\equiv\,\sqrt{\frac{1}{\chi}\!\Le\frac{(\om_{n\ell}^{a})^{2}}{\chi}
-\frac{\ell(\ell+1)}{r^{2}}-m_{a}^{2}\Ra}\,;
\eeq
this is a standing wave combination which we consider 
together with the \eq{AppenB7} mode, $u\,=\,e^{-i\om t}(R/r)Y_{\ell m}$,
and the exact radial orthonormality:
\beq\label{B57001}
\int_{r_{*}}^{R_{-}}\!\frac{R_{n\ell}^{a}\,R_{n'\ell}^{a}}{\chi(r)}\,dr
\,=\,\frac{\delta_{nn'}}{2\,\om_{n\ell}^{a}}\,,
\eeq
see above; the WKB quantization
\beq\label{B58}
\int_{r_{*}}^{R_{-}}k\,dr=(n-\tfrac12)\pi
\eeq
fixes the discrete spectrum in the case.

Next, firstly, we perform the rapid phase averaging. In the geometric optics regime we discuss, the
amplitude ${\cal N}/\sqrt{k}$ and the metric function $\chi$ vary slowly
over one radial wavelength $2\pi/k$, while $\Theta$ varies rapidly, so
that $\langle\cos^{2}\Theta\rangle\,=\,\langle\sin^{2}\Theta\rangle\,=\,
\tfrac12$ and $\langle\sin\Theta\cos\Theta\rangle\,=\,0$. Hence
\beq\label{B58001}
\overline{R^{2}}\,=\,\frac{{\cal N}^{2}}{k}\,\langle\cos^{2}\Theta\rangle
\,=\,\frac{{\cal N}^{2}}{2\,k}\,.
\eeq
Considering the derivative parts of the modes, as always for the WKB approximation,there are the rapidly oscillating piece which survive and provide
leading WKB order contributions, we have then:
\beq\label{B59}
\frac{dR}{dr}\,\simeq\,-\,{\cal N}\,\sqrt{k}\,\sin\Theta\,,\qquad
\overline{\Le\frac{dR}{dr}\Ra^{2}}\,=\,{\cal N}^{2}\,k\,
\langle\sin^{2}\Theta\rangle\,=\,\frac{{\cal N}^{2}\,k}{2}\,
\eeq
and so 
\beq\label{B60}
\frac{\chi\,\overline{|\D_{r}u^{a}|^{2}}}{\overline{|u^{a}|^{2}}}
\,=\,\chi\,k^{2}\,\equiv\,k_{\rm rad}^{2}\,,
\eeq
the 
equality is the definition $k_{\rm rad}^{2}\,\equiv\,\chi\,k^{2}$ which appears as well in
\eq{B58001}. 

 Secondly we fix the normalization ${\cal N}$ by \eq{B57001} expression. Inserting in it the \eq{B58001}  we have:
\beq\label{B61}
\int_{r_{*}}^{R_{-}}\!\frac{\overline{R^{2}}}{\chi}\,dr
\,=\,\frac{{\cal N}^{2}}{2}\int_{r_{*}}^{R_{-}}\!\frac{dr}{k\,\chi}
\,=\,\frac{1}{2\,\om_{n\ell}^{a}}\,.
\eeq
With the proper radial length $d\gamma\,=\,dr/\sqrt{\chi}$ and
$k\,=\,k_{\rm rad}/\sqrt{\chi}$, see \eq{B60}, one has
$dr/(k\chi)\,=\,d\gamma/k_{\rm rad}$, hence
\beq\label{B62}
{\cal N}^{2}\,\int\!\frac{d\gamma}{k_{\rm rad}}
\,=\,\frac{1}{\om_{n\ell}^{a}}\,.
\eeq
The WKB quantization \eq{B58}, in proper length
\beq\label{B63}
\int k_{\rm rad}\,d\gamma\,=\,(n-\tfrac12)\pi\,,
\eeq
provides correspondingly the local radial density of states 
\beq\label{B64}
dn\,=\,(1/\pi)\,k_{\rm rad}\,d\gamma
\eeq
so that \eq{B62} sets the coarse--grained value of ${\cal N}^{2}$ per unit proper length.

 Next we perform an averaging of the free field modes over the phase replaces the oscillating
profiles by their local mean values obtaining:
\beq\label{B59001}
\overline{|u^{a}|^{2}}\,\to\,\frac{1}{2\,\om_{\rm loc}\,k_{\rm loc}}
\cdot\frac{1}{\chi}\,,\qquad
\chi\,\overline{|\D_{r}u^{a}|^{2}}\,\to\,
\frac{k_{\rm rad}^{2}}{2\,\om_{\rm loc}\,k_{\rm loc}}\cdot\frac{1}{\chi}\,,
\eeq
where
\beq\label{B60001}
k_{\perp}^{2}\,\equiv\,\frac{\ell(\ell+1)}{r^{2}}\,,\qquad
k_{\rm rad}^{2}\,\equiv\,\chi\,k^{2}\,,\qquad
k_{\rm loc}^{2}\,\equiv\,k_{\rm rad}^{2}+k_{\perp}^{2}\,=\,\frac{\ell(\ell+1)}{r^{2}}\,+\,\chi\,k^{2}\,,\qquad
\om_{\rm loc}\,\equiv\,\frac{\om_{n\ell}^{a}}{\sqrt{\chi}}\,.
\eeq
We see now that then the WKB dispersion relation in \eq{B57} takes
the flat local Minkowski form:
\beq\label{B61001}
\om_{\rm loc}^{2}\,=\,k_{\rm rad}^{2}\,+\,k_{\perp}^{2}\,+\,m_{a}^{2}\,
=\,k_{\rm loc}^{2}\,+\,m_{a}^{2}\,,
\eeq
i.e.\ the boundary--core excitations propagate, in the local frame, as
ordinary relativistic Bose quanta of mass $m_{a}$. The Bose weight is
correspondingly local, since
$\om_{n\ell}^{a}/T_{\infty}=\om_{\rm loc}/T_{\rm loc}$,
\beq\label{B62001}
n_{B}(\om_{n\ell}^{a})\,=\,\frac{1}{e^{\om_{n\ell}^{a}/T_{\infty}}-1}\,
=\,\frac{1}{e^{\om_{\rm loc}/T_{\rm loc}}-1}\,.
\eeq
At the next step  we pass from the discrete spectrum to the contonious one, we define:
\beq\label{B63001}
\sum_{\ell}\sum_{m}(\cdots)\,\to\, \sum_{\ell}(2\ell+1)\,\to\,\int d[\ell(\ell+1)]=\int 2k_{\perp}r^{2}dk_{\perp},
\eeq
see first term of \eq{B60}.
Combining the radial and angular factors and passing to proper
volume, the discrete triple sum collapses to the isotropic local
phase space integral therefore; explicitly, the radial measure \eq{B64}, $dn\,=\,(1/\pi)\,k_{\rm rad}\,d\gamma$, together with the angular measure \eq{B63001}, 
$\sum_{\ell}(2\ell+1)\,\to\,\int 2k_{\perp}r^{2}\,dk_{\perp}$, 
combine into $dn\,d[\ell(\ell+1)]\,\to\,dV_{\rm proper}\,d^{3}k_{\rm loc}/(2\pi)^{3}$ with $dV_{\rm proper}\,=\,4\pi\,r^{2}\,dr/\sqrt{\chi}$. We have:
\beq\label{B64001}
\sum_{n\ell m}(\cdots)\,\longrightarrow\,
\int dV_{\rm proper}\!\int\!\frac{d^{3}k_{\rm loc}}{(2\pi)^{3}}\,(\cdots)\,,
\eeq
in which the integrand depends on $k_{\rm loc}$ and
$\om_{\rm loc}=\sqrt{k_{\rm loc}^{2}+m_{a}^{2}}$ only. Per unit proper
volume the four blocks \eq{B53} therefore acquire the following form:
\beqar\label{B65}
K\,&=&\,\int\!\frac{d^{3}k_{\rm loc}}{(2\pi)^{3}}\,
\frac{\om_{\rm loc}}{e^{\om_{\rm loc}/T_{\rm loc}}-1}\,,
\nn\\
G^{\rm rad}\,&=&\,\int\!\frac{d^{3}k_{\rm loc}}{(2\pi)^{3}}\,
\frac{\langle k_{\rm rad}^{2}\rangle/\om_{\rm loc}}
{e^{\om_{\rm loc}/T_{\rm loc}}-1}
\,=\,\frac{1}{3}\!\int\!\frac{d^{3}k_{\rm loc}}{(2\pi)^{3}}\,
\frac{k_{\rm loc}^{2}/\om_{\rm loc}}{e^{\om_{\rm loc}/T_{\rm loc}}-1}\,,
\\
G^{\rm ang}\,&=&\,\int\!\frac{d^{3}k_{\rm loc}}{(2\pi)^{3}}\,
\frac{\langle k_{\perp}^{2}\rangle/\om_{\rm loc}}
{e^{\om_{\rm loc}/T_{\rm loc}}-1}
\,=\,\frac{2}{3}\!\int\!\frac{d^{3}k_{\rm loc}}{(2\pi)^{3}}\,
\frac{k_{\rm loc}^{2}/\om_{\rm loc}}{e^{\om_{\rm loc}/T_{\rm loc}}-1}\,,
\nn\\
m^{2}\Pi\,&=&\,\int\!\frac{d^{3}k_{\rm loc}}{(2\pi)^{3}}\,
\frac{m_{a}^{2}/\om_{\rm loc}}{e^{\om_{\rm loc}/T_{\rm loc}}-1}\,,
\nn
\eeqar
where the $\tfrac13$ and $\tfrac23$ weights follow from the isotropic
angular averages
$\langle k_{\rm rad}^{2}\rangle=\tfrac13 k_{\rm loc}^{2}$ (one radial
direction of three) and
$\langle k_{\perp}^{2}\rangle=\tfrac23 k_{\rm loc}^{2}$ (the two
tangential directions). The averaging 
converts the centrifugal sum $G^{\rm ang}$ into exactly twice the
radial gradient $G^{\rm rad}$, so that the perfect fluid condition
\eq{B56} is now satisfied identically:
\beq\label{B66}
G^{\rm rad}\,=\,\tfrac13\,G\,,\qquad
G^{\rm ang}\,=\,\tfrac23\,G\,,\qquad
G\,\equiv\,G^{\rm rad}+G^{\rm ang}
\,=\,\int\!\frac{d^{3}k_{\rm loc}}{(2\pi)^{3}}\,
\frac{k_{\rm loc}^{2}/\om_{\rm loc}}{e^{\om_{\rm loc}/T_{\rm loc}}-1}\,.
\eeq
So, using \eq{B66}, we obtain for the the boundary core thermal tensor the answer similar to the
isotropic Bose gas:
\beqar\label{B67}
\rho^{a}_{\rm th}(r)\,&=&\,\tfrac12 K\,+\,\tfrac12 G\,+\,\tfrac12 m^{2}\Pi
\,=\,\int\!\frac{d^{3}k_{\rm loc}}{(2\pi)^{3}}\,
\frac{\om_{\rm loc}}{e^{\om_{\rm loc}/T_{\rm loc}}\,-\,1}\,,
\nn\\
p_{r,\rm th}^{a}\,&=&\,p_{\perp,\rm th}^{a}\,\equiv\,p^{a}_{\rm th}(r)\,=\,
\tfrac12 K\,-\,\tfrac16 G\,-\,\tfrac12 m^{2}\Pi\,=\,
\frac{1}{3}\!\int\!\frac{d^{3}k_{\rm loc}}{(2\pi)^{3}}\,
\frac{k_{\rm loc}^{2}/\om_{\rm loc}}{e^{\om_{\rm loc}/T_{\rm loc}}\,-\,1}\,,
\eeqar
where the on shell identity
$K\,=\,G\,+\,m^{2}\Pi$ was used, it follows directly from $\om_{\rm loc}^{2}\,=\,k_{\rm loc}^{2}+m_{a}^{2}$ under the integral. The
mode by mode anisotropy \eq{B52} has been canceled then,
the radial excess $2\chi|\D_{r}u|^{2}$ and the centrifugal deficit
$\ell(\ell+1)|u|^{2}/r^{2}$  by averaging of the WKB sum reduce to the single
isotropic gradient pressure $\tfrac13 G$ in both transverse and radial channels. 

  More precisely, we obtained that the difference between the two pressures is
\beq\label{B68}
p_{r,\rm th}^{a}(r)\,-\,p_{\perp,\rm th}^{a}(r)\,=\,
G^{\rm rad}\,-\,\tfrac12\,G^{\rm ang}\,,
\eeq
which vanishes identically using the WKB approximation. The leading corrections then arise, from, first of all, the curvature/centrifugal potential term
$\chi'/r$ dropped in the approximation together with the boundary
reflection at the Dirichlet walls $r_{*},R_{-}$, and, secondly, from the radial
variation of $T_{\rm loc}(r)$ and of $\chi(r)$ across a dominant
thermal wavelength $\lambda_{T}$. Both are controlled by
$(\lambda_{T}/\Delta R)^{2}$ with $\Delta R=R_{-}-r_{*}$ the shell
thickness, so
\beq\label{Banisoest}
\frac{p_{r,\rm th}^{a}-p_{\perp,\rm th}^{a}}{p^{a}_{\rm th}}\,\sim\,
\Le\frac{\lambda_{T}}{\Delta R}\Ra^{2}\,\ll\,1\,,
\eeq
with $\lambda_{T}\sim(m_{a}T_{\rm loc})^{-1/2}$ in the
non--relativistic (cold, massive) regime and
$\lambda_{T}\sim T_{\rm loc}^{-1}$ in the relativistic (hot) regime.
For a thin boundary shell of macroscopic radius this ratio is small,
so the free thermal scalars $\zeta,\epsilon$ on the AdS patch
\eq{AppenB1} are isotropic up to short--wavelength corrections, and
the perfect--fluid Bose description \eq{B66} is the correct
leading answer for the EMT components.

We notice finally that in the cold limit $T_{\infty}\,\ll\,m_{\epsilon}$ adopted in Section~\ref{SecBB}, see \eq{BB9}, the massive mode $\epsilon$ is exponentially suppressed and only the massless Goldstone mode $\zeta$ survives; the perfect-fluid result \eq{B67} then reduces to the radiation expressions $\rho_{\zeta}\,=\,\tfrac{\pi^{2}}{30}\,T_{\rm loc}^{4}$, $p_{\zeta}\,=\,\tfrac13\,\rho_{\zeta}$ used in the main text, see \eq{BBA12} and \eq{BB10}.


\newpage
\section{Thermal excitations in the crust layer}\label{AppenC}
\renewcommand{\theequation}{C.\arabic{equation}}
\setcounter{equation}{0}

 The form of the metric we consider in the crust layer is
\beq\label{C1}
ds^{2}\,=\,F(r)\,dt^{2}\,-\,\chi^{-1}(r)\,dr^{2}\,-\,r^{2}\,d\Omega^{2}\,,\qquad\,\sqrt{-g}\,=\,r^{2}\sin\theta\,\sqrt{F/\chi}\,;
\eeq
the $\epsilon$ field is defined in the crust and it is taken as
a free massive scalar which obeys usual K-G equation and it's free feld modes can be found in the way similar to the described to the previous Appendix. 
The simplest way to deal with the metric is to consider local tetrad (vierbein) frame
\beq\label{C1001}
\{\hat e_{0},\hat e_{1},\hat e_{2},\hat e_{3}\}
\eeq 
with
\beq\label{C1002}
\hat e_{0}\,=\,e^{-\Phi/2}\chi^{-1/2}\,\D_{t}\,,\,\,\,\hat e_{1}\,=\,\sqrt{\chi}\,\D_{r}\,\,\,\,
\hat e_{2}\,=\,r^{-1}\D_{\theta}\,\,\,\,\hat e_{3}\,=\,(r\sin\theta)^{-1}\D_{\varphi}
\eeq,
and $\tau^{\mu}\,=\,(1,0,0,0)$. Define the following thermally averaged quantities then
\beqar\label{C1003}
K_{T}\,&\equiv&\,\langle\,(\hat e_{0}\hat\epsilon)^{2}\,\rangle_{\beta}
\,=\,\langle\,e^{-\Phi}\chi^{-1}\,(\D_{t}\hat\epsilon)^{2}\,\rangle_{\beta}\,,
\\
G_{T}\,&\equiv&\,\sum_{i=1}^{3}\,\langle\,(\hat e_{i}\hat\epsilon)^{2}\,\rangle_{\beta}\,,
\nn\\
\Phi_{T}\,&\equiv&\,\langle\hat\epsilon^{2}\rangle_{\beta}\,,
\nn
\eeqar
which are manifestly positive. An spatial isotropy of the thermal state in
this frame gives
\beq\label{C1004}
\langle(\hat e_{i}\hat\epsilon)^{2}\rangle_{\beta}\,=\,G_{T}/3
\eeq 
for each $i\,=\,1,2,3$, and 
\beq\label{C1005}
\eta^{\alpha\beta}\langle\hat e_{\alpha}\epsilon\,
\hat e_{\beta}\epsilon\rangle_{\beta}\,=\,K_{T}\,-\,G_{T}\,.
\eeq
The EMT operator in this local orthonormal frame reads therefore as
\beq\label{C1006}
\hat T_{\mu\nu}\,=\,\hat e_{\mu}\hat\epsilon\,\hat e_{\nu}\hat\epsilon\,
-\,\tfrac{1}{2}\,\eta_{\mu\nu}\,
\Le\eta^{\al\beta}\hat e_{\al}\hat\epsilon\,\hat e_{\beta}\hat\epsilon\,-\,m^{2}\hat\epsilon^{2}\Ra\,.
\eeq 
Taking thermal averages we obtain
\beqar\label{C1007}
\langle\hat T_{0 0}\rangle_{\beta}\,&=&\,K_{T}\,-\,\tfrac{1}{2}\Le K_{T}\,-\,G_{T}\,-\,m^{2}\Phi_{T}\Ra
\,=\,\tfrac{1}{2}\Le K_{T}\,+\,G_{T}\,+\,m^{2}\Phi_{T}\Ra\,,\\
\langle\hat T_{i i}\rangle_{\beta}\,&=&\,\frac{G_{T}}{3}\,+\,\tfrac{1}{2}\Le K_{T}\,-\,G_{T}\,-\,m^{2}\Phi_{T}\Ra
\,=\,\tfrac{1}{2}K_{T}\,-\,\tfrac{1}{6}G_{T}\,-\,\tfrac{1}{2}m^{2}\Phi_{T}\,\,\,\,(\mbox{no sum})\,,\nn
\eeqar
where $\eta_{0 0}\,=\,+1$ and $\eta_{i i}\,=\,-1$ were used.

  The covariant d'Alembertian for the \eq{C1} metric is
\beq\label{C2}
\Box\,\epsilon\,=\,\frac{1}{\sqrt{-g}}\,\D_{\mu}\!\Le\sqrt{-g}\,
g^{\mu\nu}\,\D_{\nu}\epsilon\Ra\,.
\eeq
Using $g^{tt}=1/F$, $g^{rr}=-\chi$, $g^{\theta\theta}=-1/r^{2}$,
$g^{\varphi\varphi}=-1/(r^{2}\sin^{2}\theta)$ and the volume element from \eq{C1} we have:
\beqar\label{C3}
\Box\,\epsilon\,&=&\,\frac{1}{F}\,\D_{t}^{2}\epsilon\,
-\,\frac{1}{r^{2}}\,\sqrt{\frac{\chi}{F}}\,
\D_{r}\!\Le r^{2}\,\sqrt{F\,\chi}\,\D_{r}\epsilon\Ra\,
+\,\frac{1}{r^{2}}\,\Delta_{S^{2}}\,\epsilon\,,
\eeqar
where $\Delta_{S^{2}}$ is the  Laplacian on the
unit sphere, $\Delta_{S^{2}}Y_{\ell m}\,=\,-\ell(\ell+1)\,Y_{\ell m}$, the same as above.
The equation allows a varible separation by the use of the following free modes ansatz:
\beq\label{C4}
\epsilon_{\om\ell m}(x)\,=\,e^{-i\om t}\,
\frac{u_{\om\ell}(r)}{r}\,Y_{\ell m}(\theta,\varphi)\,,
\eeq
where $\om$ is the conserved Killing (Tolman) frequency conjugate to
$t$. Inserting the ansatz into K-G equation  and multiplying the expression by
$-r/\sqrt{F\chi}$, we obtain a radial equation for the radial function:
\beq\label{C5}
\sqrt{\frac{F}{\chi}}\,\frac{d}{dr}\!\Le\sqrt{F\,\chi}\,
\frac{d}{dr}\!\Le\frac{u}{r}\Ra\Ra\,r\,
+\,\Le\frac{\om^{2}}{F}\,-\,\frac{\ell(\ell+1)}{r^{2}}\,-\,m^{2}\Ra u\,=\,0\,.
\eeq
In general, it is convenient to rewrite the equation in terms of the generalized tortoise coordinate then:
\beq\label{C6}
\frac{dr_{*}}{dr}\,=\,\frac{1}{\sqrt{F\,\chi}}\,,
\eeq
so that radial derivatives combine into a one dimensional kinetic
operator. Writing $u=u(r_{*})$ and performing some algebra, the
\eq{C5} reduces to the Schr\"odinger--like form:
\beq\label{C7}
\frac{d^{2}u}{dr_{*}^{2}}\,+\,\Le\om^{2}\,-\,V_{\ell}(r)\Ra\,u\,=\,0\,,
\eeq
with the effective potential
\beq\label{C8}
V_{\ell}(r)\,=\,F(r)\Le\frac{\ell(\ell+1)}{r^{2}}\,+\,m^{2}\,
+\,\frac{1}{r}\,\frac{d(F\chi)^{1/2}}{dr_{*}}\,\frac{1}{(F\chi)^{1/2}}\Ra
\,=\,F\Le\frac{\ell(\ell+1)}{r^{2}}\,+\,m^{2}\,+\,\frac{(F\chi)'}{2r}\Ra\,,
\eeq
where in the last step $d/dr_{*}=\sqrt{F\chi}\,d/dr$ was used and the
prime denotes $d/dr$. The term $\propto(F\chi)'/(2r)$ is the
gravitational scattering ("curvature"') potential. This term is
subleading at large $\ell$ and it is dropped in the
geometric optics (WKB) approximation we use below. 

 The modes \eq{C4} are normalized through the Klein--Gordon inner
product on a constant $t$ slice as following:
\beq\label{C9}
(\epsilon_{1},\epsilon_{2})\,=\,i\!\int_{\Sigma_{t}}\!d^{3}x\,
\sqrt{-g}\,g^{tt}\,\Le\epsilon_{1}^{*}\,\D_{t}\epsilon_{2}\,
-\,\epsilon_{2}\,\D_{t}\epsilon_{1}^{*}\Ra\,.
\eeq
With $\sqrt{-g}\,g^{tt}=r^{2}\sin\theta/(F\sqrt{F\chi})\cdot F
=r^{2}\sin\theta/\sqrt{F\chi}\cdot(1/\sqrt{F})\,$, and using
$\int d\Omega\,Y_{\ell m}^{*}Y_{\ell'm'}=\de_{\ell\ell'}\de_{mm'}$,
the angular and time factors dissappear in the norm and \eq{C9} becomes a
one dimensional norm in $r_{*}$,
\beq\label{C10}
(\epsilon_{\om\ell m},\epsilon_{\om'\ell m})\,=\,
2\om\!\int\! dr_{*}\,u_{\om\ell}^{*}(r_{*})\,u_{\om'\ell}(r_{*})\,
=\,2\pi\,\de(\om-\om')\,,
\eeq
fixing the normalization
$u_{\om\ell}\sim(2\om)^{-1/2}\,(\cdots)$ of the modes. The free field operator expansion then acquires the following form:
\beq\label{C11}
\hat\epsilon(x)\,=\,\sum_{\ell m}\!\int_{0}^{\infty}\!\frac{d\om}{2\pi}\,
\Le \epsilon_{\om\ell m}(x)\,\hat a_{\om\ell m}\,+\,
\epsilon_{\om\ell m}^{*}(x)\,\hat a_{\om\ell m}^{\dagger}\Ra\,,
\eeq
with 
\beq\label{C12}
[\hat a_{\om\ell m},\hat a_{\om'\ell'm'}^{\dagger}]
=2\pi\,\de(\om-\om')\,\de_{\ell\ell'}\de_{mm'}
\eeq 
For the thermal static (Hartle--Hawking/Tolman) state the only non-vanishing
quadratic average is the following one correspondingly:
\beq\label{C13}
\langle \hat a_{\om\ell m}^{\dagger}\,\hat a_{\om'\ell'm'}\rangle_{\beta}
\,=\,\frac{2\pi\,\de(\om-\om')\,\de_{\ell\ell'}\de_{mm'}}
{e^{\om/T_{\infty}}-1}\,,
\eeq
where $T_{\infty}=\sqrt{F}\,T_{\rm loc}$ is the (constant) Tolman
temperature conjugate to the Killing time $t$; the locally measured
frequency is $\om_{\rm loc}=\om/\sqrt{F}$ and the local occupation is
$(e^{\om_{\rm loc}/T_{\rm loc}}-1)^{-1}=(e^{\om/T_{\infty}}-1)^{-1}$.

  Now we can compute thermally averaged components of the EMT $\langle\hat T^{\mu}{}_{\nu}\rangle_{\beta}$ term by term, using above \eq{C4},
\eq{C12}-\eq{C13}. Because the stationarity of the solution we llok for, the terms
$\langle\D_{t}\hat\epsilon\,\D_{r}\hat\epsilon\rangle_{\beta}$ are absent; in turn, the
$\sum_{m}$ summation together with the following spherical functions properties
\beq\label{C14}
\sum_{m=-\ell}^{\ell}|Y_{\ell m}|^{2}\,=\,\frac{2\ell+1}{4\pi}\,,\quad
\sum_{m}|\D_{\theta}Y_{\ell m}|^{2}\,+\,\frac{1}{\sin^{2}\theta}\,
\sum_{m}|\D_{\varphi}Y_{\ell m}|^{2}\,=\,\frac{2\ell+1}{4\pi}\,
\frac{\ell(\ell+1)}{1}
\eeq
restores spherical symmetry of the averaged EMT tensor. Namely, the result should depend on $r$ only, and the two angular pressures are equal, i.e.
$p_{\theta}=p_{\varphi}\equiv p_{\perp}$. So, next for each mode, the
local orthonormal-frame "squared-derivative" densities summed over $\ell$ and integrated over $\om$ together with the thermal weight
\eq{C13} are defined. Performing the calculations, the diagonal components in the mixed index form we obtain are the following:
\beqar\label{C15}
\rho(r)\,\equiv\,\langle\hat T^{t}{}_{t}\rangle_{\beta}
\,&=&\,\sum_{\ell}\frac{2\ell+1}{8\pi}\!\int_{0}^{\infty}\!\frac{d\om}{2\pi}\,
\frac{1}{e^{\om/T_{\infty}}-1}\,\frac{1}{r^{2}}
\nn\\
&&\times\Le\frac{\om^{2}}{F}\,|u|^{2}\,+\,\chi\,|u'_{\rm rad}|^{2}\,
+\,\frac{\ell(\ell+1)}{r^{2}}\,|u|^{2}\,+\,m^{2}\,|u|^{2}\Ra\,,
\\
-\,p_{r}(r)\,\equiv\,\langle\hat T^{r}{}_{r}\rangle_{\beta}
\,&=&\,\sum_{\ell}\frac{2\ell+1}{8\pi}\!\int\!\frac{d\om}{2\pi}\,
\frac{1}{e^{\om/T_{\infty}}-1}\,\frac{1}{r^{2}}\nn\\
&&\times\Le\frac{\om^{2}}{F}\,|u|^{2}\,+\,\chi\,|u'_{\rm rad}|^{2}\,
-\,\frac{\ell(\ell+1)}{r^{2}}\,|u|^{2}\,-\,m^{2}\,|u|^{2}\Ra\,,
\nn\\
-\,p_{\perp}(r)\,\equiv\,\langle\hat T^{\theta}{}_{\theta}\rangle_{\beta}
\,&=&\,\sum_{\ell}\frac{2\ell+1}{8\pi}\!\int\!\frac{d\om}{2\pi}\,
\frac{1}{e^{\om/T_{\infty}}-1}\,\frac{1}{r^{2}}\nn\\
&&\times\Le\frac{\om^{2}}{F}\,|u|^{2}\,-\,\chi\,|u'_{\rm rad}|^{2}\,
-\,m^{2}\,|u|^{2}\Ra\,,
\nn
\eeqar
where $u=u_{\om\ell}(r)$, $u'_{\rm rad}\equiv\frac{d}{dr}(u/r)$ is the
properly $1/r$--stripped radial derivative entering
$\langle(\hat e_{1}\hat\epsilon)^{2}\rangle$, and the four bracketed
terms are respectively the temporal kinetic, radial gradient,
tangential gradient, and mass contributions. These terms are the
exact, mode by mode, components defined by the  \eq{C1} metric and they are manifestly anisotropic because the tangential
gradient term $\ell(\ell+1)|u|^{2}/r^{2}$ enters $p_{r}$ and
$p_{\perp}$ with opposite signs.
Equivalently, in the vierbein frame, using \eq{C1003} definitions, we obtain:
\beqar\label{C16}
K_{T}(r)\,&=&\,\sum_{\ell}\frac{2\ell+1}{8\pi r^{2}}\!\int\!\frac{d\om}{2\pi}\,
\frac{\om^{2}/F}{e^{\om/T_{\infty}}-1}\,|u|^{2}\,,\nn\\
G_{T}^{\rm rad}(r)\,&=&\,\sum_{\ell}\frac{2\ell+1}{8\pi r^{2}}\!\int\!\frac{d\om}{2\pi}\,
\frac{\chi\,|u'_{\rm rad}|^{2}}{e^{\om/T_{\infty}}-1}\,,\\
G_{T}^{\rm ang}(r)\,&=&\,\sum_{\ell}\frac{2\ell+1}{8\pi r^{2}}\!\int\!\frac{d\om}{2\pi}\,
\frac{\ell(\ell+1)\,|u|^{2}/r^{2}}{e^{\om/T_{\infty}}-1}\,,\nn\\
m^{2}\Phi_{T}(r)\,&=&\,\sum_{\ell}\frac{2\ell+1}{8\pi r^{2}}\!\int\!\frac{d\om}{2\pi}\,
\frac{m^{2}\,|u|^{2}}{e^{\om/T_{\infty}}-1}\,,\nn
\eeqar
with $G_{T}=G_{T}^{\rm rad}+2\,(\tfrac{1}{2}G_{T}^{\rm ang})
=G_{T}^{\rm rad}+G_{T}^{\rm ang}$ once the two equal tangential
directions are summed. In terms of these we have:
\beqar\label{C17}
\rho\,&=&\,\tfrac{1}{2}\,K_{T}\,+\,\tfrac{1}{2}\,G_{T}^{\rm rad}\,
+\,\tfrac{1}{2}\,G_{T}^{\rm ang}\,+\,\tfrac{1}{2}\,m^{2}\Phi_{T}\,,\nn\\
p_{r}\,&=&\,\tfrac{1}{2}\,K_{T}\,+\,\tfrac{1}{2}\,G_{T}^{\rm rad}\,
-\,\tfrac{1}{2}\,G_{T}^{\rm ang}\,-\,\tfrac{1}{2}\,m^{2}\Phi_{T}\,,\\
p_{\perp}\,&=&\,\tfrac{1}{2}\,K_{T}\,-\,\tfrac{1}{2}\,G_{T}^{\rm rad}\,
-\,\tfrac{1}{2}\,m^{2}\Phi_{T}\,,\nn
\eeqar
which reduces to a perfect fluid  answer $p_{r}=p_{\perp}$ only when
$G_{T}^{\rm rad}=\tfrac{1}{3}G_{T}$ and
$G_{T}^{\rm ang}=\tfrac{2}{3}G_{T}$, i.e. when the exact isotropy condition $G_{T}^{\rm rad}=\tfrac{1}{2}G_{T}^{\rm ang}$ is satisfied. As we see, this condintion is not
true mode by mode resukt but it holds only after the WKB approximation applied.

 As in the previous Appendix we assume that for a shell placed between the macroscopic radiuses $R_{+}$ and $R_{S}$, the dominant thermal
modes have $\om R_{+}\gg1$ and the relevant angular momenta reach up
to $\ell_{\max}\sim\om R_{+}/\sqrt{F}\gg1$. This is again the WKB "brick-wall" counting of the thermal atmosphere near the surface 
\cite{tHooft1985,MukohyamaIsrael1998}. In this regime the defined above
curvature term in \eq{C8} is negligible and the radial equation \eq{C7} can be solved by the use of WKB form for the ftree modes:
\beq\label{C18}
u_{\om\ell}(r)\,\simeq\,\frac{1}{\sqrt{2\om}}\,
\frac{e^{\pm i\int^{r_{*}}\!k\,dr_{*}'}}{\sqrt{k}}\,,\quad
k(r)\,\equiv\,\sqrt{\om^{2}\,-\,V_{\ell}(r)}\,
\simeq\,\sqrt{\om^{2}\,-\,F\Le\frac{\ell(\ell+1)}{r^{2}}+m^{2}\Ra}\,.
\eeq
The rapidly oscillating phase averages providing 
\beq\label{C19}
|u|^{2}\to1/(2\om k)\,,
\eeq 
also we have 
\beq\label{C20}
|u'_{\rm rad}|^{2}\to k^{2}/(2\om k)\cdot\chi^{-1}\cdot\,.
(\cdots)
\eeq
Correspondingly, the orthonormal radial gradient density we obtain is 
\beq\label{C21}
\chi|u'_{\rm rad}|^{2}\to k_{\rm rad}^{2}/(2\om k)
\eeq
where 
\beq\label{C22}
k_{\rm rad}^{2}\equiv\chi\,k^{2}
\eeq 
is the proper radial wave-number squared. Now convert the discrete sum over $\ell$ to an
integral over the proper tangential wave-number. We define the local proper momenta as
\beq\label{C23}
k_{\perp}^{2}\,\equiv\,\frac{\ell(\ell+1)}{r^{2}}\,,\qquad
k_{\rm rad}^{2}\,\equiv\,\chi\,k^{2}\,,\qquad
k_{\rm loc}^{2}\,\equiv\,k_{\rm rad}^{2}\,+\,k_{\perp}^{2}\,,\qquad
\om_{\rm loc}\,\equiv\,\frac{\om}{\sqrt{F}}\,,
\eeq
so that the dispersion relation in \eq{C18} becomes a flat local like one:
\beq\label{C24}
\om_{\rm loc}^{2}\,=\,k_{\rm rad}^{2}\,+\,k_{\perp}^{2}\,+\,m^{2}\,
=\,k_{\rm loc}^{2}\,+\,m^{2}\,.
\eeq
Using therefore
\beq\label{C25}
\sum_{\ell}(2\ell+1)\to\int d[\ell(\ell+1)]=\int 2\,k_{\perp}\,r^{2}\,dk_{\perp}\cdot(\cdots)
\eeq 
and
\beq\label{C26}
\int_{0}^{\infty}\!d\om/(2\pi)\to\sqrt{F}\!\int d\om_{\rm loc}/(2\pi)\,,
\eeq
the discrete measure in the expressions can written as an isotropic local phase-space integral
\beq\label{C27}
\sum_{\ell}\frac{2\ell+1}{4\pi r^{2}}\!\int\!\frac{d\om}{2\pi}\,
(\cdots)\,\longrightarrow\,
\int\!\frac{d^{3}k_{\rm loc}}{(2\pi)^{3}}\,(\cdots)\,,
\eeq
with the integrands now defined as a functions of $k_{\rm loc}$ and
$\om_{\rm loc}=\sqrt{k_{\rm loc}^{2}+m^{2}}$ only. Applying the \eq{C27} to the \eq{C16} expressions and using the averaged
$|u|^{2}\to1/(2\om k)$, $\,\chi|u'_{\rm rad}|^{2}\to k_{\rm rad}^{2}/(2\om k)$, we obtain:
\beqar\label{C28}
K_{T}\,&=&\,\int\!\frac{d^{3}k_{\rm loc}}{(2\pi)^{3}}\,
\frac{\om_{\rm loc}}{e^{\om_{\rm loc}/T_{\rm loc}}-1}\,,
\nn\\
G_{T}^{\rm rad}\,&=&\,\int\!\frac{d^{3}k_{\rm loc}}{(2\pi)^{3}}\,
\frac{\langle k_{\rm rad}^{2}\rangle/\om_{\rm loc}}{e^{\om_{\rm loc}/T_{\rm loc}}-1}
\,=\,\frac{1}{3}\int\!\frac{d^{3}k_{\rm loc}}{(2\pi)^{3}}\,
\frac{k_{\rm loc}^{2}/\om_{\rm loc}}{e^{\om_{\rm loc}/T_{\rm loc}}-1}\,,
\\
G_{T}^{\rm ang}\,&=&\,\int\!\frac{d^{3}k_{\rm loc}}{(2\pi)^{3}}\,
\frac{\langle k_{\perp}^{2}\rangle/\om_{\rm loc}}{e^{\om_{\rm loc}/T_{\rm loc}}-1}
\,=\,\frac{2}{3}\int\!\frac{d^{3}k_{\rm loc}}{(2\pi)^{3}}\,
\frac{k_{\rm loc}^{2}/\om_{\rm loc}}{e^{\om_{\rm loc}/T_{\rm loc}}-1}\,,
\nn\\
m^{2}\Phi_{T}\,&=&\,\int\!\frac{d^{3}k_{\rm loc}}{(2\pi)^{3}}\,
\frac{m^{2}/\om_{\rm loc}}{e^{\om_{\rm loc}/T_{\rm loc}}-1}\,,
\nn
\eeqar
where the factors $\tfrac13$ and $\tfrac23$ follow from the isotropic
angular averages $\langle k_{\rm rad}^{2}\rangle=\tfrac13 k_{\rm loc}^{2}$
(one of three directions) and
$\langle k_{\perp}^{2}\rangle=\tfrac23 k_{\rm loc}^{2}$ (the two
tangential directions). The answers reproduce the regular ideal Boze gas expressions with the 
$k\equiv k_{\rm loc}$, $\om_{k}\equiv\om_{\rm loc}$ replacements made, and
also make manifest the isotropy condition provided by the WKB approximation:
\beq\label{C29}
G_{T}^{\rm rad}\,=\,\tfrac{1}{3}\,G_{T}\,,\qquad
G_{T}^{\rm ang}\,=\,\tfrac{2}{3}\,G_{T}\,,\qquad
G_{T}\,\equiv\,G_{T}^{\rm rad}+G_{T}^{\rm ang}
\,=\,\int\!\frac{d^{3}k_{\rm loc}}{(2\pi)^{3}}\,
\frac{k_{\rm loc}^{2}/\om_{\rm loc}}{e^{\om_{\rm loc}/T_{\rm loc}}-1}\,.
\eeq
Substituting the \eq{C29} answers into the  \eq{C17} gives, in the WKB limit, the following energy density and pressure expressions:
\beqar\label{C30}
\rho_{\rm crust}(r)\,&=&\,\tfrac12 K_{T}+\tfrac12 G_{T}+\tfrac12 m^{2}\Phi_{T}
\,=\,\int\!\frac{d^{3}k_{\rm loc}}{(2\pi)^{3}}\,
\frac{\om_{\rm loc}}{e^{\om_{\rm loc}/T_{\rm loc}}-1}\,,\nn\\
p_{r}\,=\,p_{\perp}\,=\,p_{\rm crust}(r)\,&=&\,
\tfrac12 K_{T}-\tfrac16 G_{T}-\tfrac12 m^{2}\Phi_{T}
\,=\,\frac{1}{3}\int\!\frac{d^{3}k_{\rm loc}}{(2\pi)^{3}}\,
\frac{k_{\rm loc}^{2}/\om_{\rm loc}}{e^{\om_{\rm loc}/T_{\rm loc}}-1}\,,
\eeqar
where the algebraic identity 
$K_{T}\,=\,G_{T}+m^{2}\Phi_{T}$, which follows directly from the local dispersion relation
$\om_{\rm loc}^{2}=k_{\rm loc}^{2}+m^{2}$ under the integral, was used.

  We notice again, that the exact difference between the two pressures is
\beq\label{C31}
p_{r}(r)\,-\,p_{\perp}(r)\,=\,G_{T}^{\rm rad}\,-\,\tfrac{1}{2}\,G_{T}^{\rm ang}\,,
\eeq
is vanishing  only when the WKB isotropy conditions \eq{C29} holds. The leading corrections, which violates the isotropy, comes, first of all, from the curvature potential term $(F\chi)'/(2r)$ dropped in
the WKB approximation, and, secondly through the radial variation of $T_{\rm loc}(r)$ across
the dominant thermal wavelength $\lambda_{T}$. Both are controlled by $(\lambda_{T}/R_{+})^{2}$ quantity, so
\beq\label{C32}
\frac{p_{r}-p_{\perp}}{p}\,\sim\,
\Le\frac{\lambda_{T}}{R_{+}}\Ra^{2}\,\ll\,1\,,
\eeq
with $\lambda_{T}\sim(m T_{\rm loc})^{-1/2}$ in the cold regime and
$\lambda_{T}\sim T_{\rm loc}^{-1}$ in the hot regime, see corresponding Section. So, the quantitative statement then is that
the EMT of the free thermal scalar is isotropic
up to exponentially / power--law small short--wavelength corrections to the field
and that the perfect fluid description of the crust thermal system is the correct leading one.


\newpage
\section{Einstein equations for the crust layer}\label{AppenD}
\renewcommand{\theequation}{D.\arabic{equation}}
\setcounter{equation}{0}

  The static, spherically symmetric metric we use is \eq{CC6} metric defined as
\beq\label{D1}
ds^{2}\,=\,e^{\Phi(r)}\chi(r)\,dt^{2}\,-\,\chi^{-1}(r)\,dr^{2}\,
-\,r^{2}\,d\theta^{2}\,-\,r^{2}\sin^{2}\theta\,d\varphi^{2}\,,
\eeq
i.e.\ $g_{00}\,=\,e^{\Phi}\chi\,\equiv\,F$ and $g_{rr}\,=\,-\chi^{-1}$,
with inverse metric components
\beq\label{D2}
g^{tt}\,=\,e^{-\Phi}\chi^{-1}\,,\,\,\,g^{rr}\,=\,-\chi\,,\,\,\,
g^{\theta\theta}\,=\,-\frac{1}{r^{2}}\,,\,\,\,
g^{\varphi\varphi}\,=\,-\frac{1}{r^{2}\sin^{2}\theta}\,.
\eeq
The Christoffel symbols computed computed by the metric are
\beqar\label{D3}
\Gamma^{t}_{tr}\,&=&\,\frac{1}{2}\Le\Phi'\,+\,\frac{\chi'}{\chi}\Ra\,,\quad
\Gamma^{r}_{tt}\,=\,\frac{1}{2}\,\chi\,e^{\Phi}\Le\chi\,\Phi'\,+\,\chi'\Ra\,,\quad
\Gamma^{r}_{rr}\,=\,-\frac{\chi'}{2\chi}\,,
\nn\\
\Gamma^{r}_{\theta\theta}\,&=&\,-r\,\chi\,,\quad
\Gamma^{r}_{\varphi\varphi}\,=\,-r\,\chi\,\sin^{2}\theta\,,
\\
\Gamma^{\theta}_{r\theta}\,&=&\,\Gamma^{\varphi}_{r\varphi}\,=\,\frac{1}{r}\,,\quad
\Gamma^{\theta}_{\varphi\varphi}\,=\,-\sin\theta\,\cos\theta\,,\quad
\Gamma^{\varphi}_{\theta\varphi}\,=\,\cot\theta\,,
\nn
\eeqar
with all others vanishing, as usual the prime denotes $d/dr$. We have also
$\Gamma^{t}_{tr}\,=\,\tfrac{1}{2}F'/F\,=\,\tfrac12(\Phi'+\chi'/\chi)$
exhibits the split of the $00$ redshift slope into a genuine
redshift part $\Phi'$ and the radial part $\chi'/\chi$.

The Einstein tensor in mixed--index form
$G^{\mu}_{\nu}\,=\,R^{\mu}_{\nu}\,-\,\tfrac{1}{2}\de^{\mu}_{\nu}\,R$,
computed for the metric \eq{D1}, has nonzero components
\beqar\label{D4}
G^{t}_{t}\,&=&\,\frac{1\,-\,\chi\,-\,r\,\chi'}{r^{2}}\,,
\\
G^{r}_{r}\,&=&\,\frac{1\,-\,\chi\,-\,r\,\chi'}{r^{2}}\,-\,\frac{\chi\,\Phi'}{r}\,
=\,\frac{1\,-\,\chi}{r^{2}}\,-\,\frac{\chi}{r}\Le\Phi'\,+\,\frac{\chi'}{\chi}\Ra\,,
\nn\\
G^{\theta}_{\theta}\,&=&\,G^{\varphi}_{\varphi}\,=\,
-\,\frac{\chi}{4}\Le\Phi'^{2}\,+\,2\Phi''\,+\,3\,\Phi'\frac{\chi'}{\chi}\,+\,2\frac{\chi''}{\chi}\Ra\,-\,\frac{\chi\,\Phi'}{2r}\,-\,\frac{\chi'}{r}\,.
\nn
\eeqar
We see that the $G^{t}_{t}$ depends on $\chi$ only and an energy density $\rho$ fixes the $\chi$ as usual, while the combination
\beq\label{D5}
G^{r}_{r}\,-\,G^{t}_{t}\,=\,-\,\frac{\chi\,\Phi'}{r}
\eeq
determines the redshift function $\Phi$ with known $\chi$, it is sourced by
$\rho+p$ sum and is independent of $\chi'$. The mixed index form for the equations is used because the $T^{\mu}_{\nu}$ of a perfect fluid has a diagonal form
$T^{\mu}_{\nu}\,=\,\mathrm{diag}(\rho,-p,-p,-p)$. 

 The Einstein equation $G^{\mu}_{\nu}\,=\,8\pi G\,T^{\mu}_{\nu}$ for the $00$ component provides:
\beq\label{D6}
\frac{1\,-\,\chi\,-\,r\,\chi'}{r^{2}}\,=\,8\pi G\,\rho\,.
\eeq
Defining the Misner--Sharp mass function $m(r)$ as
\beq\label{MSdef}
\chi(r)\,=\,1\,-\,\frac{2\,G\,m(r)}{r}\,,
\eeq
we obtain then
\beq\label{D7}
\chi'(r)\,=\,-\,\frac{2\,G\,m'(r)}{r}\,+\,\frac{2\,G\,m(r)}{r^{2}}\,,
\eeq
and the l.h.s of \eq{D6} acquires an usual form:
\beq\label{D8}
\frac{1\,-\,\chi\,-\,r\,\chi'}{r^{2}}\,=\,\frac{2\,G\,m'(r)}{r^{2}}\,
\eeq
Therefore, from \eq{D6}, we obtain
\beq\label{D9}
\frac{dm}{dr}\,=\,4\pi\,r^{2}\,\rho(r)\,,
\eeq
or in an integrated form 
\beq\label{D10}
m(r)\,=\,C\,+\,4\pi\int_{R_{+}}^{r}r'^{2}\,\rho(r')\,dr'\,.
\eeq
with $C$ as yet undefined integration constant.

 Next we discuss the $rr$ Einstein equation component $G^{r}_{r}\,=\,8\pi G\,T^{r}{}_{r}\,=\,-8\pi G\,p$, which 
reads as
\beq\label{D11}
\frac{1\,-\,\chi}{r^{2}}\,-\,\frac{\chi}{r}\Le\Phi'\,+\,\frac{\chi'}{\chi}\Ra\,=\,
-\,8\pi G\,p\,.
\eeq
Subtracting the $00$ equation from the given one we arrive to \eq{D5} expression which has the following form then: 
\beq\label{D12}
\frac{d\Phi}{dr}\,=\,8\pi G\,\frac{r}{\chi(r)}\,\Le\rho(r)\,+\,p(r)\Ra.
\eeq
Next we notice that since $F\,=\,e^{\Phi}\chi$, the redshift slope of the $00$ component
splits as
\beq\label{D13}
\frac{F'}{F}\,=\,\Phi'\,+\,\frac{\chi'}{\chi}\,.
\eeq
Solving \eq{D11} for this combination, the familiar answer is reproducing:
\beq\label{D14}
\frac{F'}{F}\,=\,2\,G\,\frac{1}{r^{2}\,\chi}\Le m\,+\,4\pi\,r^{3}\,p\Ra\,
\eeq
or
\beq\label{D1501}
\frac{dF}{dr}\,=\,2\,G\,\frac{F(r)}{r^{2}\,\chi(r)}\!\Le m(r)\,+\,4\pi\,r^{3}\,p(r)\Ra\,.
\eeq
In the case of the static spherically symmetric ansatz, the covariant conservation equation $\nabla_{\mu}T^{\mu}{}_{\nu}\,=\,0$
has nontrivial component only for $\nu\,=\,r$:
\beq\label{D16}
\nabla_{\mu}T^{\mu}{}_{r}\,=\,\D_{\mu}T^{\mu}{}_{r}\,
+\,\Gamma^{\mu}{}_{\mu\al}\,T^{\al}{}_{r}\,
-\,\Gamma^{\al}{}_{\mu r}\,T^{\mu}{}_{\al}\,=\,0\,,
\eeq
and substituting $T^{\mu}_{\nu}\,=\,\mathrm{diag}(\rho,-p,-p,-p)$ together with the \eq{D3} expressions provides after simplification:
\beq\label{D17}
-\,\frac{dp}{dr}\,=\,\Le \rho\,+\,p\Ra\,\Gamma^{t}_{tr}\,=\,\frac{1}{2}\,\Le \rho\,+\,p\Ra\,\Le\Phi'+\frac{\chi'}{\chi}\Ra\,=\,
\Le \rho\,+\,p\Ra\,\frac{F'}{2F}\,.
\eeq
so using \eq{D14} it gives
\beq\label{dpdrApp}
\frac{dp}{dr}\,=\,-\,(\rho\,+\,p)\,\frac{F'}{2F}\,
=\,-\,G\,\frac{\Le m+4\pi r^{3}p\Ra}{r^{2}\chi}\,\Le \rho\,+\,p\Ra\,
\eeq
and finally
\beq\label{D15}
\frac{dp}{dr}\,=\,
-\,\frac{G}{r^{2}}\,\frac{\Le m\,+\,4\pi\,r^{3}\,p\Ra }{1\,-\,2\,G\,m(r)/r}\,\Le \rho\,+\,p\Ra\,,
\eeq
which is the standard TOV equation for the metric of interest of the crust. We notice finally that, as usual, the other components of the Einstein equations are no independent but depend on the 
considered ones.

 Now we have a system of three first order ODEs which requires for a solution properly defined boundary conditions, so we have the following.
At the inner edge of the crust at $r\,=\,R_{+}$ we defiine $m(R_{+})\,=\,C\,=\,0$ requesting that the total mass of the core is equal to
$M_{core}\,=\,M_{\rm ADM}$. The overall BH mass then is defined in a way which as well  
provides the correct mass of the black hole through $M_{BH}\,=\,M_{\rm ADM}\,+\,|m_{-}|\,+\,m_{-}\,=\,M_{\rm ADM}$ sum where $m_{-}\,<\,0$ 
is a negative mass of the whole core and $|m_{-}|$ is the positive mass belongs to the transition layer placed between the core and crust,
see corresponding Section~\ref{SecDD}.
There the redshift zero point $\Phi(R_{+})\,=\,0$ as well and it is fixed by continuity with the wall metric.
At the outer edge when $r\,=\,R_{S}$ the metric must be matched to the Schwarzschild one that requests to provide the $\Phi(R_{S})$
boundary value, see discussion in the corresponding Section~\ref{SecCC}. The last ingredient we need is 
an equation of state $p\,=\,p(\rho)$ of the thermal excitations in the crust, the equation is derived in the previous Appendix.


\newpage
\section{ Linear stability analysis of zero gravity hypersurface}\label{AppenE}
\renewcommand{\theequation}{E.\arabic{equation}}
\setcounter{equation}{0}

  The vanishing of the force on $\mathcal{Z}$, \eq{DD48}, makes $r=R_{+}$ an
equilibrium radius for a static particle. In this Appendix we analyze the linear stability of
that equilibrium, separately for positive- and negative-mass test bodies, by
perturbing the radial geodesic motion.

  For the radial motion of a test body the conserved energy is $E=F\,\dot t$ (dot
$\equiv d/d\tau$), and the interval
$ds^{2}\,=\,F\dot t^{2}\,-\,\chi^{-1}\dot r^{2}=1$ gives the first integral of motion:
\beq\label{E1}
\dot r^{2}\,=\,\chi\Le\frac{E^{2}}{F}-1\Ra\,.
\eeq
The radial acceleration is defined by the differentiating of the expression, it is
\beq\ddot r=\tfrac12\,d(\dot r^{2})/dr\,,
\eeq 
for a body at rest at some moment at time and at radius
$r$, when $E^{2}=F(r)$, it reduces to the force per unit mass expression:
\beq\label{E2}
f(r)\,\equiv\,\ddot r\big|_{\dot r=0}\,=\,-\,\frac{1}{2}\,\chi\,\frac{F'}{F}\,=\,-\,a^{r}(r)\,, 
\eeq
see \eq{DD47}. Introducing a small radial displacement
\beq\label{E3}
x\,\equiv\, r\,-\,R_{+}
\eeq
and expanding \eq{E2} at vicinity of
$\mathcal{Z}$, where $f(R_{+})=0$, the linearized equations of motion we obtain is
\beq\label{E4}
\ddot x\,=\,f'(R_{+})\,x\,+\,O(x^{2})\,,\qquad
f'(R_{+})\,=\,-\,\frac{1}{2}\,\frac{\chi(R_{+})}{F(R_{+})}\,F''(R_{+})\,.
=\,-\,\frac{1}{2}\,F''(R_{+})\,,
\eeq
Here, in the last equality, we used $\chi(R_{+})\,=\,F(R_{+})\,=\,1$ and $F'(R_{+})\,=\,0$, see \eq{DD26} and \eq{DD48}. There are no 
terms contain $F'$ or $\chi'$ in the expressions therefore. We define the effective constant appears in the linearized equation as
\beq\label{E5}
k_{\rm eff}\,\equiv\,-\,f'(R_{+})\,=\,\frac{1}{2}\,F''(R_{+})\,
=\,\frac{G\,\mathcal{M}_{\rm Komar}'(R_{+})}{R_{+}^{2}}\,,
\eeq
the second equality following from \eq{DD50}  which provides $F''(R_{+})\,=\,2G\,\mathcal{M}_{\rm Komar}'(R_{+})/R_{+}^{2}$ after differentiation. The sign of
$k_{\rm eff}$ is therefore the sign of the gradient of the Tolman--Komar active mass through the surface.
Namely, the force on a static test body is set by the enclosed Tolman--Komar active
mass 
\beq\label{E6}
\mathcal{M}_{\rm Komar}(r)=m_{w}(r)+4\pi r^{3}p_{w}(r)
\eeq
through 
\beq\label{E7}
f(r)\,=\,-a^{r}\,=\,-G\,\mathcal{M}_{\rm Komar}(r)/(r^{2}\sqrt{F})
\eeq
where $f\,<\,0$ is inward with $\mathcal{M}_{\rm Komar}\,>\,0$, and $f\,>\,0$ is outward with $\mathcal{M}_{\rm Komar}\,<\,0$.
Important that it is the enclosed mass and not the local density. Although the wall
density is positive in the outer sub-shell where$\rho_{w}>0$ for $r_{0}<r<R_{+}$, the
mass enclosed within such a radius is still negative, because it contains
the whole negative-energy core together with the negative wall tail,
\beq\label{E8}
m_{w}(r)\,<\,0\quad (r\,<\,R_{+})\,,\qquad m_{w}(R_{+})\,=\,0\,.
\eeq
Since $\mathcal{M}_{\rm Komar}(R_{+})\,=\,0$ is a simple zero, see \eq{DD50}, the active mass changes sign across the zero gravity hypersurface $\mathcal{Z}$:
\beqar\label{E9}
\mathcal{M}_{\rm Komar}(r)\,&<&\,0\,,\qquad r\,<\,R_{+}\,,
\\
\mathcal{M}_{\rm Komar}(r)\,&>&\,0\,,\qquad r\,>\,R_{+}\,.
\label{E9001}
\eeqar
Consequently, for the ordinary matter, the static force is repulsive (outward) just inside$\mathcal{Z}$ and attractive (inward) just outside it. 
In these cases both cases it is directed back toward $\mathcal{Z}$, and, correspondingly, the surface is therefore a genuine two-sided normal matter
attractor where $\mathcal{M}_{\rm Komar}$ rises through zero with positive slope:
\beq\label{E10}
\mathcal{M}_{\rm Komar}'(R_{+})\,>\,0
\quad\rightarrow\quad
F''(R_{+})\,=\,\frac{2G\,\mathcal{M}_{\rm Komar}'(R_{+})}{R_{+}^{2}}\,>\,0\,,
\eeq
so $F$ has a minimum at $\mathcal{Z}$, the redshift function forms a
potential well, with $\mathcal{Z}$ at its bottom. The restoring force is two-sided, there are repulsion from the net-negative enclosed mass on the inner
side and attraction from the net-positive enclosed mass on the outer side, therefore the
equilibrium we obtain is stable.

 Now we can discuss a behavior of the ordinary matter particles and negative energy quasi-particles at vicinity of the hypersurface, see
\cite{Bondi,My,My1,My2,My3} for additional discussions.
For an ordinary positive mass test particles, the motion is geodesic and, with
$k_{\rm eff}>0$ from \eq{E5}, the linearised equation \eq{E4} is simple harmonic,
\beq\label{E11}
\ddot x\,=\,-\,k_{\rm eff}\,x\,,\qquad
\om_{+}\,=\,\sqrt{k_{\rm eff}}\,=\,\sqrt{\tfrac12 F''(R_{+})}\,
=\,\frac{1}{R_{+}}\sqrt{G\,\mathcal{M}_{\rm Komar}'(R_{+})}\,,
\eeq
in proper time. Since $dt/d\tau\,=\,E/F\,=\,1$ at $\mathcal{Z}$, the 
frequency coincides with $\om_{+}$ at leading order. So the picture is that the positive mass particle
displaced inward is repelled by the net-negative enclosed mass and pushed back out
to $\mathcal{Z}$. The same particle displaced outward it is attracted by the net-positive enclosed
mass and pulled back in. It therefore oscillates about $R_{+}$, we obatin that the
zero-gravity hypersurface is a stable equilibrium for ordinary matter, we can identify the $\mathcal{Z}$ then as
maximal regular-matter density.

 For the marginal case of negative mass when $m_{g}\,=\,m_{i}\,<\,0$,
the mass cancels from $m_{i}\ddot x\,=\,m_{g}g(x)$ equation and the body follows the same geodesic  as positive matter,
\beq\label{E12}
\ddot x\,=\,-\,k_{\rm eff}\,x\,,\qquad \om\,=\,\om_{+}\,,
\eeq
so this is equally stable situation. This kind of matter will oscillate with the same frequencyas the regular one, displaced
inward or outward it is pushed back to $\mathcal{Z}$ as well.

 An another marginal negative mass definition is an anomalous negative mass when $m_{i}\,<\,0$ and $m_{g}\,>\,0$.
In this case the response to the restoring field reverses sign,
\beq\label{E13}
\ddot x\,=\,\frac{m_{g}}{m_{i}}\,g(x)\,=\,\frac{m_{g}}{m_{i}}\,(-k_{\rm eff}\,x)\,
=\,+\,k_{\rm eff}\,x\,,
\eeq
whose solutions $x\propto e^{\pm\sqrt{k_{\rm eff}}\,\tau}$ grow exponentially, the
$\mathcal{Z}$ is unstable for such a body, which runs away on the proper
time scale $\tau_{\rm run}\sim k_{\rm eff}^{-1/2}
=R_{+}/\sqrt{G\,\mathcal{M}_{\rm Komar}'(R_{+})}$. This particle's bejaviour is anomalous one, the gravitational pull is
ordinary, i.e. toward the well, but the negative inertia drives the response in a wrong direction.

 The mostly acceptable and physical definition oof the negative mass particle is a situation with a negative gravitational charge but positive inertia mass when
$m_{i}\,>\,0$ and $m_{g}\,<\,0$.
This is the physically most natural realization of a negative energy core quasi-particles, the inertia is ordinary but an interaction with ordinary matter is repulsive. Then
\beq\label{E14}
\ddot x\,=\,\frac{m_{g}}{m_{i}}\,g(x)\,=\,\frac{m_{g}}{m_{i}}\,(-k_{\rm eff}\,x)\,
=\,+\,\frac{|m_{g}|}{m_{i}}\,k_{\rm eff}\,x\,,
\eeq
again exponentially growing, with rate
\beq\label{E15}
\Om\,=\,\sqrt{\frac{|m_{g}|}{m_{i}}\,k_{\rm eff}}\,
=\,\frac{1}{R_{+}}\sqrt{\frac{|m_{g}|}{m_{i}}\,G\,\mathcal{M}_{\rm Komar}'(R_{+})}\,.
\eeq
So $\mathcal{Z}$ is again unstable, a quasi-particle displaced inward is pushed further inward toward the
core, one displaced outward is pushed further outward toward the horizon $R_{S}$.

 Formulating an unifying criterion, we see that near the well all cases share the linear law with a sign set only by the ratio
$m_{g}/m_{i}$:
\beq\label{E16}
\ddot x\,=\,-\,\frac{m_{g}}{m_{i}}\,k_{\rm eff}\,x\,,\qquad k_{\rm eff}>0\,,
\qquad
\begin{cases}
m_{g}/m_{i}\,>\,0\ \rightarrow\ \text{stable},\ \om\,=\,\sqrt{(m_{g}/m_{i})\,k_{\rm eff}}\,,
\\
m_{g}/m_{i}\,<\,0\ \rightarrow\ \text{unstable},\ \Om\,=\,\sqrt{|m_{g}/m_{i}|\,k_{\rm eff}}\,.
\end{cases}
\eeq
Stability depends only on the sign of the ratio $m_{g}/m_{i}$. The ordinary
body $(+,+)$ and the equivalence-principle negative mass $(-,-)$ have
$m_{g}/m_{i}\,>\,0$ and oscillate stably in the well; both mixed-sign realisations, the
negative inertia case $(-,+)$ and the negative gravitational charge case 
$(+,-)$, have $m_{g}/m_{i}\,<\,0$ and run away.



\newpage
\begin{thebibliography}{99}


\bibitem{Beken1}
J.~D.~Bekenstein,
Lett. Nuovo Cim. \textbf{4} (1972), 737-740.

\bibitem{Beken2}
J.~D.~Bekenstein,
Phys. Rev. D \textbf{7} (1973), 2333-2346.



\bibitem{Hawk1}
S.~Hawking, 
  Phys.Rev.Lett. \textbf{26} (1971) 1344--1346.

\bibitem{Hawk2}
J.~M. Bardeen, B.~Carter, and S.~Hawking, 
Commun.Math.Phys. \textbf{\bf 31} (1973) 161--170.

\bibitem{Hawk3}
S.~Hawking, 
Commun.Math.Phys. \textbf{43} (1975) 199--220.



\bibitem{Strom1}
A.~Strominger and C.~Vafa, 
Phys.Lett. \textbf{B379} (1996) 99--104.


\bibitem{Frol1}
V.~P.~Frolov and I.~Novikov,
Phys. Rev. D \textbf{48} (1993), 4545-4551.

\bibitem{Frol2}
I.~D.~Novikov and V.~P.~Frolov,
"Physics of black holes",
Kluwer Academic, 1989.




\bibitem{Page1}
D.~N.~Page, 
Phys.Rev.Lett. \textbf{71} (1993) 3743--3746.

\bibitem{Page2}
D.~N.~Page, 
Phys.Rev.Lett. \textbf{71} (1993) 1291--1294.
	
\bibitem{Page3}
D.~N.~Page, 
JCAP \textbf{1309} (2013) 028.	



\bibitem{BondBH}
S.~Bondarenko, D.~Cheskis and R.~Singh,
Annals Phys. \textbf{490} (2026), 170500.


\bibitem{Int}
E.~Poisson and W.~Israel,
Phys. Rev. D \textbf{41} (1990), 1796-1809.

\bibitem{Int1}
G.~Magli,
Rept. Math. Phys. \textbf{44} (1999), 407-412.

\bibitem{Int2}
K.~Papadodimas and S.~Raju,
Phys. Rev. Lett. \textbf{112} (2014) no.5, 051301.

\bibitem{Int3}
P.~O.~Mazur and E.~Mottola,
Class. Quant. Grav. \textbf{32} (2015) no.21, 215024

\bibitem{Int4}
G.~Penington, S.~H.~Shenker, D.~Stanford and Z.~Yang,
JHEP \textbf{03} (2022), 205.




\bibitem{Nov1}
Y.~B.~Zel'dovich, I.~D.~Novikov, "Relativistic astrophysics", Vol.1, University of Chicago Press, 1971.

\bibitem{Nov2}
Y.~B.~Zel'dovich, I.~D.~Novikov, "Relativistic astrophysics", Vol.2, University of Chicago Press, 1971.


\bibitem{Neut1}
J.~M.~Lattimer and M.~Prakash,
Astrophys. J. \textbf{550} (2001), 426.

\bibitem{Neut2}
F.~Douchin and P.~Haensel,
Astron. Astrophys. \textbf{380} (2001), 151.

\bibitem{Neut3}
N.~K.~Glendenning,
Phys. Rept. \textbf{342} (2001), 393-447.

\bibitem{Neut4}
A.~Kurkela, E.~S.~Fraga, J.~Schaffner-Bielich and A.~Vuorinen,
Astrophys. J. \textbf{789} (2014), 127.




\bibitem{FB}
V.~A.~Khodel, V.~R.~Shaginyan,
Jetp Lett, \textbf{51}, no. 9 (1990), 553.

\bibitem{FB1}
G.~E.~Volovik, 
JETP Lett \textbf{53}, no. 4 (1991), 222.

\bibitem{FB2}
N.~B.~Kopnin, T.~T.~Heikkila and G.~E.~Volovik,
Phys. Rev. B \textbf{83} (2011), 220503.

\bibitem{FB3}
D.~Leykam, A.~Andreanov, S.~Flach, 
Advances in Physics: X, \textbf{3.1} (2018), 1473052.




\bibitem{Zeld1}
Ya. B. Zel’dovich, 
Zh. Eksp. Teor. Fiz. \textbf{42} (1962), 641. 

\bibitem{Zeld2}
Y.~B.~Zeldovich,
Usp. Fiz. Nauk \textbf{123} (1977), 487-503



\bibitem{Wil1}
M.~G.~Alford, K.~Rajagopal and F.~Wilczek,
Phys. Lett. B \textbf{422} (1998), 247-256.

\bibitem{Wil2}
M.~G.~Alford, K.~Rajagopal and F.~Wilczek,
Nucl. Phys. B \textbf{537} (1999), 443-458.

\bibitem{Wil3}
M.~G.~Alford, K.~Rajagopal, S.~Reddy and F.~Wilczek,
Phys. Rev. D \textbf{64} (2001), 074017.

\bibitem{Wil4}
M.~G.~Alford,
Ann. Rev. Nucl. Part. Sci. \textbf{51} (2001), 131-160.



\bibitem{Nozi}
P.~Nozieres, "Theory of interacting Fermi systems"', CRC Press, 2018.



\bibitem{Interior}
S.~Bondarenko,
Mod. Phys. Lett. A \textbf{34} (2019) no.11, 1950084.


\bibitem{Latt1}
A.~F.~Kapustinskii, "Lattice energy of ionic crystals." Quarterly Reviews, Chemical Society 10, no. \textbf{3} (1956), 283-294.

\bibitem{Latt2}
C.~Kittel, ``Introduction to Solid State Physics'', 8th ed., Wiley, New York (2005).


\bibitem{FalB}
S.~R.~Coleman,
Phys. Rev. D \textbf{15} (1977), 2929-2936 [erratum: Phys. Rev. D \textbf{16} (1977), 1248].

\bibitem{FalB1}
C.~G.~Callan and S.~R.~Coleman,
Phys. Rev. D \textbf{16} (1977), 1762-1768.

\bibitem{FalB2}
S.~K.~Blau, E.~I.~Guendelman and A.~H.~Guth,
Phys. Rev. D \textbf{35} (1987), 1747-1766.

\bibitem{NEM}
S.~Bondarenko and R.~Singh,
Annals Phys. \textbf{487} (2026), 170365.


\bibitem{YounLaplaceEq}
L.~D.~Landau and E.~M.~Lifshitz, "Fluid Mechanics", Course of Theoretical
Physics Vol.~6, 2nd ed., Pergamon Press, Oxford (1987).


\bibitem{TwoT1}
I.~B.~Rumer and M.~S.~Ryvkin, "Thermodynamics, Statistical Physics and
Kinetics", Mir Publishers, Moscow (1980).

\bibitem{TwoT2}
Y.~B.~Zel'dovich and Y.~P.~Raizer, "Physics of Shock Waves and
High-Temperature Hydrodynamic Phenomena", Academic Press, New York (1966).



\bibitem{Zelmanov}
A.~L.~Zelmanov,
Dover (translation), originally Dissertation, Moscow (1944).

\bibitem{Vladinirov}
Yu.~S.~Vladimirov, ``Reference Frames in Gravitation Theory'' (in Russian),
Energoizdat, Moscow (1982).


\bibitem{Alt1}
P.~O.~Mazur and E.~Mottola,
Universe \textbf{9} (2023) no.2, 88.

\bibitem{Alt2}
L.~J.~Garay, J.~R.~Anglin, J.~I.~Cirac and P.~Zoller,
Phys. Rev. Lett. \textbf{85} (2000), 4643-4647.

\bibitem{Alt3}
P.~O.~Mazur and E.~Mottola,
Proc. Nat. Acad. Sci. \textbf{101} (2004), 9545-9550.

\bibitem{Alt4}
C.~Barcelo, S.~Liberati and M.~Visser,
Living Rev. Rel. \textbf{8} (2005), 12.

\bibitem{Alt5}
S.~D.~Mathur,
Fortsch. Phys. \textbf{53} (2005), 793-827.

\bibitem{Alt6}
G.~Dvali and C.~Gomez,
Eur. Phys. J. C \textbf{74} (2014), 2752.


\bibitem{Ho1}	
F.~Hoyle,
Mon. Not. Roy. Astron. Soc. \textbf{108} (1948), 372-382.

\bibitem{Ho2}	
F. Hoyle and J. V. Narlikar,
Proc. R. Soc. Lond. A 1963 273, 1-11.

\bibitem{Ho3}	
Hoyle, F., and J. V. Narlikar,
Proceedings of the Royal Society of London. Series A, Mathematical and Physical Sciences 273, no. 1352 (1963): 1–11.

\bibitem{Ho4}	
Hoyle, Fred, and J. V. Narlikar,
Proceedings of the Royal Society of London. Series A. Mathematical and Physical Sciences 282.1389 (1964): 178-183.

\bibitem{Ho5}	
Hoyle, Fred, and Jayant Vishnu Narlikar,
Proceedings of the Royal Society of London. Series A. Mathematical and Physical Sciences 278.1375 (1964): 465-478.



\bibitem{Zub}
D.~N.~Zubarev and V.~P.~Kalashnikov,
Teor. Mat. Fiz. \textbf{5} (1970), 406-416.

\bibitem{Zub1}
D.~N.~Zubarev, A.~V.~Prozorkevich and S.~A.~Smolyanskii,
Theor. Math. Phys. \textbf{40} (1979) no.3, 821-831.

\bibitem{Zub2}
D.~N.~Zubarev and M.~V.~Tokarchuk,
Teor. Mat. Fiz. \textbf{88N2} (1991), 286-310



\bibitem{Birrell}
N.~D.~Birrell and P.~C.~W.~Davies, ``Quantum Fields in Curved Space'',
Cambridge University Press, Cambridge (1982).

\bibitem{Wald}
R.~M.~Wald, ``Quantum Field Theory in Curved Spacetime and Black Hole
Thermodynamics'', University of Chicago Press, Chicago (1994).

\bibitem{Fulling}
S.~A.~Fulling, ``Aspects of Quantum Field Theory in Curved Space-Time'',
London Math. Soc. Student Texts Vol.~17, Cambridge University Press (1989).



\bibitem{HawkingHartle1972}
S.~W.~Hawking and J.~B.~Hartle,
Commun.\ Math.\ Phys.\ \textbf{27} (1972), 283.

\bibitem{Hartle1973}
J.~B.~Hartle,
Phys.\ Rev.\ D \textbf{8} (1973), 1010.

\bibitem{PoissonSasaki1995}
E.~Poisson and M.~Sasaki,
Phys.\ Rev.\ D \textbf{51} (1995), 5753.

\bibitem{Datta2020}
S.~Datta, R.~Brito, S.~Bose, P.~Pani, and S.~A.~Hughes,
Phys.\ Rev.\ D \textbf{101} (2020), 044004.

\bibitem{ReggeWheeler1957}
T.~Regge and J.~A.~Wheeler,
Phys.\ Rev.\ \textbf{108} (1957), 1063.

\bibitem{Zerilli1970}
F.~J.~Zerilli,
Phys.\ Rev.\ Lett.\ \textbf{24} (1970), 737.

\bibitem{KokkotasSchmidt1999}
K.~D.~Kokkotas and B.~G.~Schmidt,
Living Rev.\ Rel.\ \textbf{2} (1999), 2.

\bibitem{CardosoFranzinPani2016}
V.~Cardoso, E.~Franzin, and P.~Pani,
Phys.\ Rev.\ Lett.\ \textbf{116} (2016), 171101
[Erratum: \textbf{117}, 089902 (2016)].

\bibitem{Hinderer2008}
T.~Hinderer,
Astrophys.\ J.\ \textbf{677} (2008), 1216.

\bibitem{BinningtonPoisson2009}
T.~Binnington and E.~Poisson,
Phys.\ Rev.\ D \textbf{80} (2009), 084018.

\bibitem{Charalambous2021}
P.~Charalambous, S.~Dubovsky, and M.~M.~Ivanov,
JHEP \textbf{05} (2021), 038.

\bibitem{Hawking1975}
S.~W.~Hawking,
Commun.\ Math.\ Phys.\ \textbf{43} (1975), 199
[Erratum: \textbf{46}, 206 (1976)].

\bibitem{Page1976}
D.~N.~Page,
Phys.\ Rev.\ D \textbf{13} (1976), 198.

\bibitem{Damour1978}
T.~Damour,
Phys.\ Rev.\ D \textbf{18} (1978), 3598.

\bibitem{ThornePriceMacDonald1986}
K.~S.~Thorne, R.~H.~Price, and D.~A.~MacDonald (eds.),
Yale University Press, New Haven, 1986.

\bibitem{CardosoHopper2016}
V.~Cardoso, S.~Hopper, C.~F.~B.~Macedo, C.~Palenzuela, and P.~Pani,
Phys.\ Rev.\ D \textbf{94} (2016), 084031.

\bibitem{CardosoPani2019}
V.~Cardoso and P.~Pani,
Living Rev.\ Rel.\ \textbf{22} (2019), 4.

\bibitem{Maggio2020}
E.~Maggio, L.~Buoninfante, A.~Mazumdar, and P.~Pani,
Phys.\ Rev.\ D \textbf{102} (2020), 064053.


\bibitem{tHooft1985}
G.~'t~Hooft,
Nucl.\ Phys.\ B \textbf{256} (1985), 727-745.

\bibitem{MannTarasovZelnikov1992}
R.~B.~Mann, L.~Tarasov, and A.~Zelnikov,
Class.\ Quantum Grav.\ \textbf{9} (1992), 1487-1494.

\bibitem{MukohyamaIsrael1998}
S.~Mukohyama and W.~Israel,
Phys.\ Rev.\ D \textbf{58} (1998), 104005.


\bibitem{HartleHawking1976}
J.~B.~Hartle and S.~W.~Hawking,
Phys.\ Rev.\ D \textbf{13} (1976), 2188-2203.

\bibitem{Christensen1976}
S.~M.~Christensen,
Phys.\ Rev.\ D \textbf{14} (1976), 2490-2501.

\bibitem{Candelas1980}
P.~Candelas,
Phys.\ Rev.\ D \textbf{21} (1980), 2185-2202.

\bibitem{Page1982}
D.~N.~Page,
Phys.\ Rev.\ D \textbf{25} (1982), 1499-1509.



\bibitem{Bondi}
H.~Bondi,
Rev. Mod. Phys. \textbf{29} (1957), 423-428.

\bibitem{My}
S.~Bondarenko,
Mod. Phys. Lett. A \textbf{34} (2019) no.11, 1950084.

\bibitem{My1}
S.~Bondarenko,
Universe \textbf{6} (2020) no.8, 121.

\bibitem{My2}
S.~Bondarenko,
Eur. Phys. J. C \textbf{81} (2021) no.3, 253.

\bibitem{My3}
S.~Bondarenko and V.~De La Hoz-Coronell,
Class. Quant. Grav. \textbf{41} (2024) no.7, 075001.






\end{thebibliography}
\end{document}